\documentclass[CRPHYS,Unicode,manuscript,french]{cedram}

\usepackage{amssymb,amsmath,amsbsy}
\hypersetup{breaklinks=true}
\hypersetup{colorlinks=true}
\hypersetup{urlcolor=blue}
\hypersetup{citecolor=blue}
\hypersetup{linkcolor=blue}

\usepackage{comment}
\addto\captionsfrench{\def\tablename{\textsc{Tableau}}}
\addto\captionsfrench{\renewcommand{\contentsname}{Sommaire}}
\DeclareMathOperator\re{Re}
\DeclareMathOperator\im{Im}
\newcommand{\zero}{\mathbf{0}}
\newcommand{\kk}{\mathbf{k}}
\newcommand{\KK}{\mathbf{K}}
\newcommand{\qq}{\mathbf{q}}
\newcommand{\pp}{\mathbf{p}}

\newcommand{\vv}{\mathbf{v}}
\newcommand{\rr}{\mathbf{r}}
\newcommand{\RR}{\mathbf{R}}
\newcommand{\CC}{\mathbf{C}}
\newcommand{\g}[1]{«~#1~»}
\newcommand{\be}{\begin{equation}}
\newcommand{\ee}{\end{equation}}
\newcommand{\ii}{\mathrm{i}}
\newcommand{\eee}{\mathrm{e}}
\newcommand{\dd}{\mathrm{d}}
\newcommand{\veps}{\varepsilon}
\newcommand\raz{\par
  \setcounter{section}{0}%
  \setcounter{subsection}{0}%
  \setcounter{equation}{0}
  \setcounter{figure}{0}
  \setcounter{footnote}{0}
  \setcounter{table}{0}
  \gdef\thefigure{\arabic{figure}}%
  \gdef\thetable{\arabic{table}}%
  \gdef\thesection{\arabic{section}}%
  \gdef\theequation{\arabic{equation}}%
}

\title{Questions ouvertes pour les gaz de fermions en interaction forte et de portée nulle}

\alttitle{Open questions for strongly interacting Fermi gases with zero-range interactions}

\author{\firstname{Yvan} \lastname{Castin}}
\address{Laboratoire Kastler Brossel, ENS-Université PSL, CNRS, Université Sorbonne et Collège de France, 24 rue Lhomond, 75231 Paris, France}
\email[Y. Castin]{yvan.castin@lkb.ens.fr}

\begin{abstract}
Nous passons en revue quelques questions théoriques non résolues dans les gaz tridimensionnels de fermions à deux composantes, en nous inspirant des expériences réalisées récemment sur les atomes froids dans des pièges immatériels près d'une résonance de Feshbach magnétique. Nous distinguons successivement (i) les questions ouvertes apparaissant dans le problème à petit nombre de corps avec interactions de contact dites de Wigner-Bethe-Peierls - essentiellement la stabilité du gaz vis-à-vis de l'effet Efimov et le calcul des coefficients d'amas (ou du viriel), (ii) celles relevant de la théorie effective de basse énergie dite hydrodynamique quantique de Landau et Khalatnikov - essentiellement l'amortissement des modes de phonons et le temps de cohérence du condensat de paires liées, et enfin (iii) les questions nécessitant une résolution complète, microscopique, du problème à $N$ corps, comme les propriétés précises de la branche d'excitation sonore (de Goldstone) du condensat de paires, ou de sa branche d'excitation collective (de Higgs) dans le continuum de paire brisée.
\end{abstract}

\begin{altabstract}
We review some unresolved theoretical issues in three-dimensional two-component Fermi gases, drawing on recent experiments on cold atoms in immaterial traps close to a magnetic Feshbach resonance. We distinguish successively (i) the open questions arising in the few-body problem with Wigner-Bethe-Peierls contact interactions - essentially the stability of the gas with respect to the Efimov effect and the calculation of the cluster (or virial) coefficients, (ii) those arising in the effective low-energy theory of Landau and Khalatnikov quantum hydrodynamics - essentially the damping of phonon modes and the coherence time of the condensate of pairs, and finally (iii) questions requiring a complete, microscopic solution of the many-body problem, such as the specific properties of the acoustic excitation branch (Goldstone) of the condensate of pairs, or its collective excitation branch (Higgs) in the broken-pair continuum.
\end{altabstract}

\altkeywords{Fermi gases, quantum hydrodynamics, Efimov effect, unitary limit, contact interactions, cluster or virial expansion, Higgs mode}

\begin{document}

\maketitle

\tableofcontents


\section*{Version fran\c{c}aise (English version starts on page \pageref{debe})}
\label{debf}

\section{Introduction et présentation générale}
\label{sec1}

Ce texte est essentiellement la retranscription de notre exposé de 90 minutes au colloque de prospective \g{Questions ouvertes dans le problème quantique à $N$ corps} qui s'est tenu à l'Institut Henri Poincaré à Paris, du 8 au 12 juillet 2024, d'où son style et son niveau de précision différents de ceux d'un article de recherche habituel. Il est plus complet que l'exposé sur la section \ref{sec4} (traitée rapidement à l'oral) et sur la section \ref{sec5} (omise à l'oral par manque de temps). Les notes en bas de page peuvent être ignorées en première lecture. L'exposé a été enregistré et est disponible en ligne sur la chaîne Carmin de l'IHP (cliquer \href{https://www.carmin.tv/fr/collections/symposium-open-questions-in-the-quantum-many-body-problem}{ici}).

Le système considéré est inspiré des expériences sur les atomes froids: il s'agit d'un gaz tridimensionnel de fermions à deux composantes (comprendre deux états internes $\uparrow$ et $\downarrow$) dans un piège immatériel - fait de lumière, à des températures très basses de l'ordre du microkelvin.  C'est le digne descendant des gaz d'atomes refroidis par laser (dans les fameuses \g{mélasses optiques}, voir le prix Nobel de physique 1997 décerné à Steven Chu, William Phillips et Claude Cohen-Tannoudji) puis des condensats de Bose-Einstein atomiques gazeux refroidis par évaporation (voir le prix Nobel de physique 2001 décerné à Eric Cornell, Carl Wieman et Wolfgang Ketterle). 

Par rapport à leurs illustres prédécesseurs, les gaz d'atomes froids fermioniques ont l'avantage (i) d'être composés de fermions, ce qui permet de couvrir les deux statistiques possibles (on peut toujours \g{bosoniser} le gaz en formant des paires fortement liées $\uparrow\downarrow$) et de faire un lien direct avec les systèmes d'électrons (des fermions!) de la physique du solide, (ii) de rester collisionnellement stables (peu de pertes à trois corps par recombinaison vers des états moléculaires profonds) même dans le régime d'interaction forte comme dans la fameuse \g{limite unitaire} décrite plus bas (au contraire pour l'instant des gaz d'atomes froids bosoniques), et (iii) de constituer dans ledit régime des systèmes modèles, beaux, simples et universels, grâce à la portée négligeable des interactions de van der Waals entre $\uparrow$ et $\downarrow$ (plus précisément, la longueur de van der Waals associée est négligeable); comme nous le verrons, ceci autorise à remplacer l'interaction par des conditions de contact sur la fonction d'onde à $N$ corps dépendant de la seule longueur de diffusion $a$ dans l'onde $s$, longueur que les expérimentateurs ajustent à volonté au moyen d'une résonance de Feshbach, par simple application d'un champ magnétique uniforme bien choisi.

Notre système n'est pas sans rapport avec ceux d'autres exposés du colloque. Le lien est évident avec la contribution de Tilman Enss sur la viscosité des gaz de fermions en interaction forte \cite{Enss}, complémentaire de la nôtre. Mais si l'on place nos fermions dans un réseau optique, à raison d'environ une particule par site (près du demi-remplissage), on retombe sur les problèmes de fermions fortement corrélés et de supraconductivité à haute température critique discutés par Antoine Georges. Dans un régime qui plus est d'interaction sur site $U_{\uparrow\downarrow}$ forte devant le couplage tunnel $t$ entre sites voisins, $U_{\uparrow\downarrow}\gg t$, le système est décrit par un hamiltonien modèle de spins de type Heisenberg, avec un couplage magnétique $J\propto t^2/U_{\uparrow\downarrow}$, ce qui fait le lien avec l'exposé de Sylvain Capponi \cite{Cappo}.  En revenant à un système uniforme (sans réseau) mais en appliquant un champ de jauge artificiel (un champ magnétique fictif) à nos atomes froids fermioniques pourtant neutres, ce que les expérimentateurs savent faire, voir l'exposé de Sylvain Nascimbène \cite{Nascim}, on tombe sur des problématiques proches des exposés de Thierry Jolicœur \cite{Jolic} (sur l'effet Hall quantique fractionnaire à 2D) et de Carlos Sá de Melo \cite{SadeM} (couplage spin-orbite à une dimension d'espace). Tous ces ponts vers la physique du solide ne sont cependant pas si faciles que cela à emprunter, à cause d'effets parasites non conservatifs, de la taille finie des échantillons et d'une difficulté à descendre à suffisamment basse température (en unités de la température de Fermi $T_{\rm F}$ ou de couplage magnétique $J/k_{\rm B}$), voir les exposés de Wolfgang Ketterle, de Sylvain Nascimbène et d'Antoine Georges. 

Terminons par le plan de notre contribution.  Dans la section \ref{sec2}, nous partons du réel en esquissant le cheminement des expériences sur les atomes froids depuis les années 1980 et la situation atteinte dans le cas des fermions.  Dans la section \ref{sec3}, nous adoptons un point de vue microscopique, d'interactions remplacées par des conditions de contact, et passons en revue quelques questions ouvertes dans le problème à petit nombre de fermions. Dans la section \ref{sec4}, nous adoptons au contraire un point de vue macroscopique, celui d'une théorie effective de basse énergie (l'hydrodynamique quantique), et passons en revue quelques questions ouvertes liées à l'interaction entre les phonons (les quanta des ondes sonores) dans la phase superfluide.  Enfin, dans la courte section \ref{sec5}, nous croisons les points de vue, en listant quelques questions ouvertes requérant un traitement théorique microscopique du problème à $N$ corps complet. 

\section{Un système physique assez récent}
\label{sec2}

Commençons par une mise en contexte de nos gaz de fermions, au moyen d'un bref historique des atomes froids.  

L'aventure commence au début des années 1980 par le refroidissement laser des alcalins.  Les basses températures atteintes sont spectaculaires lorsqu'on les exprime en kelvins, $T\approx 1\,\mu{\rm K}$, mais les densités spatiales sont malheureusement très faibles, $\rho\lesssim 10^{10}{\rm at/cm}^3$,  si bien que les gaz ont une très faible dégénérescence quantique, c'est-à-dire une très faible densité dans l'espace des phases, $\rho\lambda^3\ll 1$, où $\lambda$ est la longueur d'onde thermique de de Broglie des atomes de masse $m$:
\be
\lambda=\left(\frac{2\pi\hbar^2}{m k_{\rm B} T}\right)^{1/2}
\ee
Les effets de statistique quantique (bosoniques ou fermioniques) sont imperceptibles.  

Tout change en 1995, lorsqu'Eric Cornell et Carl Wieman au JILA \cite{CBEJila},  suivis de peu par Wolfgang Ketterle au MIT \cite{CBEMit},  atteignent la condensation de Bose-Einstein (CBE), évidemment sur des isotopes bosoniques,  grâce au refroidissement par évaporation dans des potentiels de piégeage non dissipatifs à fond harmonique.\footnote{La référence \cite{Schreck} a réussi plus tard, au moyen d'astuces bien trouvées, à obtenir un condensat de Bose-Einstein sans évaporation, par le seul refroidissement laser (voir aussi la référence \cite{Vuletic}) ;  pour cela, il a fallu en particulier (i) utiliser une raie atomique étroite à faible saturation pour rendre aussi basse que possible la température limite du refroidissement laser \cite{josa} et (ii) réussir à éviter que les photons d'émission spontanée, qui emportent une partie de l'énergie du mouvement des atomes, ne la redéposent par réabsorption dans le gaz.}  Les températures de transition restent dans la gamme du refroidissement laser, $T_{\rm c}^{\rm CBE}\simeq 0,\!1\ \mbox{à}\ 1\,\mu{\rm K}$, mais les densités spatiales sont considérablement plus élevées, $\rho=10^{12}\ \mbox{à}\ 10^{15}\mbox{at/cm}^3$, ce qui permet l'atteinte de la dégénérescence quantique $\rho\lambda^3\gtrsim 1$. 

Enfin, en 2004, le refroidissement par évaporation est étendu avec succès aux isotopes fermioniques jusqu'à la température de transition \cite{Jin,Zwier}; les gaz à deux états internes $\uparrow$ et $\downarrow$  ne forment plus des condensats de Bose-Einstein mais condensent par paires $\uparrow\downarrow$ par le mécanisme BCS \cite{revuefer}: les interactions de van der Waals attractives entre $\uparrow$ et $\downarrow$ conduisent, en présence d'une mer de Fermi dans chaque état interne, à la formation de paires liées, les fameuses paires de Cooper, des \g{bosons composites}, qui peuvent former un condensat à suffisamment basse température, $T<T_{\rm c}^{\rm BCS}$. Les températures les plus basses accessibles expérimentalement sont de l'ordre de $0,\!1 T_{\rm F}$, où la température de Fermi $T_{\rm F}$ reste de l'ordre du microkelvin; ceci suffit néanmoins à franchir $T_{\rm c}^{\rm BCS}$ car les interactions entre $\uparrow$ et $\downarrow$ sont rendues très fortes au moyen d'une résonance de diffusion à deux corps (résonance de Feshbach magnétique): la température de transition $T_{\rm c}^{\rm BCS}$ est alors une fraction de $T_{\rm F}$ et l'on évite la situation extrême des supraconducteurs BCS, pour lesquels $T_{\rm c}^{\rm BCS}\lll T_{\rm F}$ par plusieurs ordres de grandeur. 

\begin{figure}[tb]
\includegraphics[width=8cm,clip=]{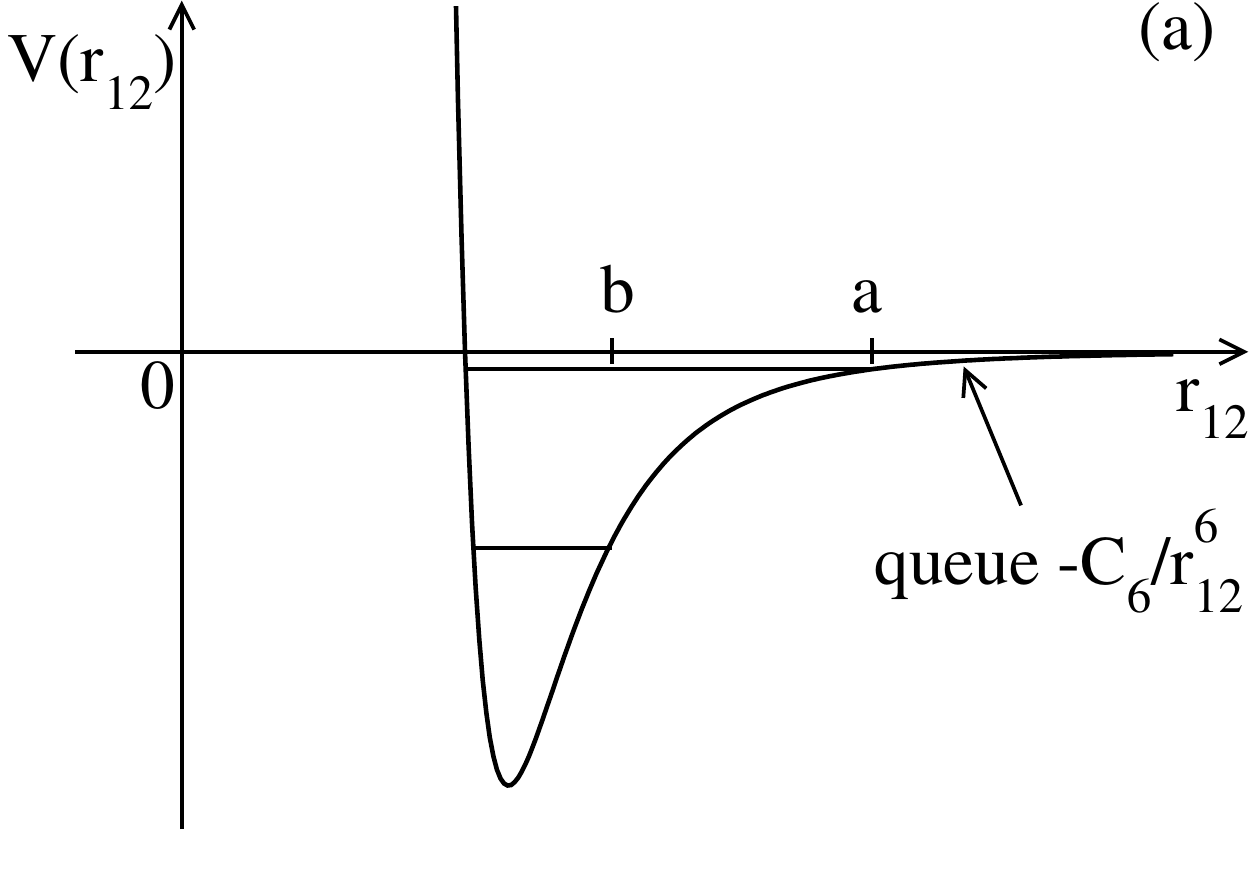}
\includegraphics[width=8cm,clip=]{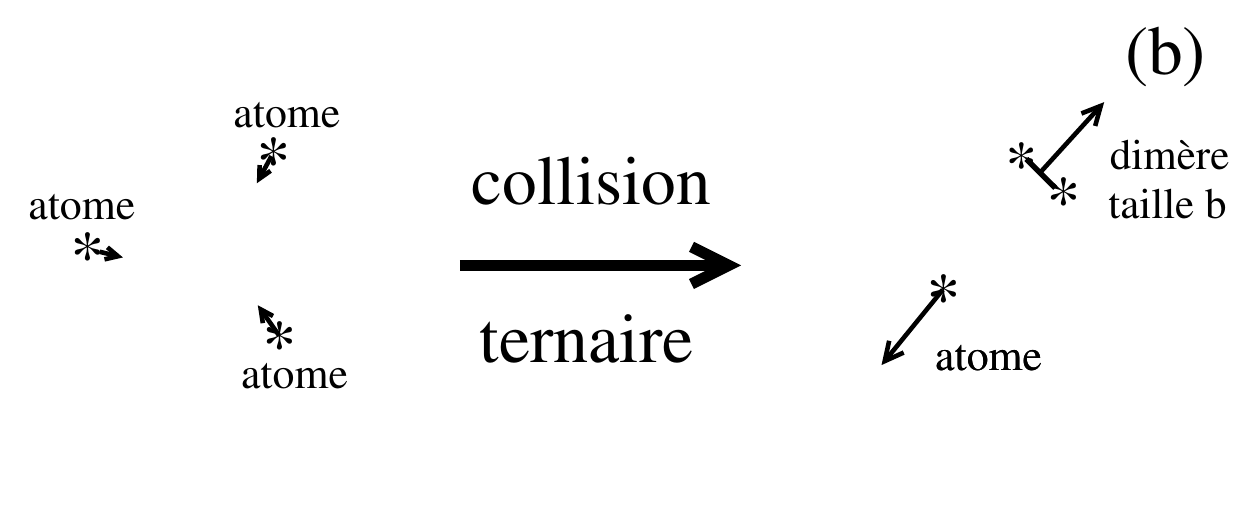}
\caption{(a) Représentation schématique de l'interaction de van der Waals (plus précisément de Lennard-Jones) résonnante ($|a|\gg b$) entre les fermions $\uparrow$ et $\downarrow$ en fonction de leur distance relative.  Le potentiel admet au moins un état fortement lié d'extension de l'ordre de la longueur de van der Waals $b\simeq (m C_6/\hbar^2)^{1/4}$ donc d'énergie de liaison $\approx \hbar^2/mb^2$ et, dans le cas d'une longueur de diffusion $a>0$ comme sur la figure, un dernier niveau d'énergie d'extension $a$ (d'énergie de liaison $\hbar^2/ma^2$) sur \g{le point de disparaître} (ici $m_\uparrow=m_\downarrow=m$ comme dans les expériences); si $a$ était grande mais négative ($|a|\gg b, a<0$), cet état faiblement lié serait sur \g{le point d'apparaître}. (b) L'état dimère fortement lié peut être peuplé par des collisions à trois corps, ce qui est à l'origine de pertes de particules dans le gaz de fermions, dites pertes à trois corps (les produits de la collision emportent l'énergie de liaison $\approx \hbar^2/m b^2$ considérable sous forme d'énergie cinétique et quittent le piège). Les flèches représentent les quantités de mouvement avant et après la collision.}
\label{fig1}
\end{figure}

Décrivons maintenant notre système d'atomes froids fermioniques dans ses grandes lignes, dans un début d'idéalisation de la réalité expérimentale. (i) Les fermions sont à deux états internes $\uparrow$ et $\downarrow$; comme nous n'envisageons pas ici de couplage de Rabi interconvertissant $\uparrow$ et $\downarrow$, nos considérations s'appliquent aussi au cas d'un mélange de deux espèces chimiques de fermions formellement sans spin; pour cette raison, nous ne supposons pas que les masses $m_\sigma$ des particules sont égales dans les deux états internes\footnote{Dans le cas où $\uparrow$ et $\downarrow$  sont deux états de spin d'une même espèce chimique, on a naturellement $m_\uparrow=m_\downarrow$ dans l'expérience.  On pourrait cependant, par application d'un réseau optique se couplant différemment aux deux états internes (dans une limite de faible taux de remplissage), produire des masses effectives $m_\sigma$ différentes.  Ceci reste à faire.} et nous considérons le rapport $m_{\uparrow}/m_{\downarrow}$ comme un paramètre libre. (ii) Les fermions sont piégés, soit dans des potentiels harmoniques isotropes de même pulsation de piégeage $\omega$ pour les deux composantes $\sigma$,
\be
\label{eq1_5}
U_\sigma(\rr)=\frac{1}{2} m_\sigma\omega^2 r^2
\ee
où $\rr$ est le vecteur position à 3D, soit dans la boîte de quantification cubique $[0,L]^3$ commune aux deux composantes, avec les habituelles conditions aux limites périodiques.\footnote{Expérimentalement, on sait réaliser des boîtes de potentiel à fond plat au moyen de faisceaux de Laguerre-Gauss ou de Bessel-Gauss et de nappes de lumière laser, après compensation de la pesanteur (mise en lévitation des atomes) par un gradient de champ magnétique \cite{Zoran,buxida,revbox}.} (iii) L'interaction de van der Waals entre les deux états internes $\uparrow$ et $\downarrow$, représentée schématiquement sur la figure \ref{fig1}a, est rendue de manière effective très forte (résonnante) dans l'onde $s$ (moment cinétique orbital relatif $l=0$) par application d'un champ magnétique idoine\footnote{\label{note2_5}Sans entrer dans les détails, signalons que, pour comprendre cette résonance, il faut tenir compte de la structure interne des atomes et décrire leur interaction binaire a minima par un modèle à deux voies, une voie ouverte de potentiel d'interaction $V_{\rm o}(r_{12})$ et une voie fermée de potentiel d'interaction $V_{\rm f}(r_{12})$ -- on pourra penser aux potentiels d'interaction singulet et triplet de deux fermions de spin 1/2.  Lors d'une collision,  les atomes $\uparrow$ et $\downarrow$ entrent par la voie ouverte et, par conservation de l'énergie, sortent aussi par la voie ouverte car leur énergie cinétique relative incidente est inférieure à la différence des limites de dissociation $V_{\rm f}(+\infty)-V_{\rm o}(+\infty)>0$.  Comme il existe un couplage entre les deux voies, les atomes peuplent cependant virtuellement la voie fermée pendant la collision.  Le champ magnétique $B$ appliqué induit un déplacement Zeeman différent dans les deux voies.  Il suffit alors de choisir $ B$ astucieusement pour que l'énergie d'un état lié dans $V_{\rm f}(r_{12})$ - pas l'énergie nue mais l'énergie déplacée par le couplage - coïncide presque avec la limite de dissociation $V_{\rm o}(+\infty)$, ce qui induit une résonance de diffusion (ou de collision) à deux corps dans la voie ouverte et fait diverger la longueur de diffusion $a$.} si bien que la longueur de diffusion $a$ entre deux atomes $\uparrow$ et $\downarrow$ (définie mathématiquement dans la section \ref{sec3.1}) est suffisamment grande en valeur absolue (elle peut être positive ou négative) pour que
\be
\label{eq1_6}
\rho^{1/3}|a|\gtrsim 1
\ee
On rappelle que la théorie des gaz de bosons en interaction faible fait usage du petit paramètre $(\rho a^3)^{1/2}\ll 1$, voir la contribution de Jan Solovej \cite{Solov} aux actes du colloque; la condition (\ref{eq1_6}) est donc au contraire la marque d'un gaz en interaction forte.
La longueur du diffusion $a$ est également beaucoup plus grande en  valeur absolue que la portée $b$ de l'interaction, définie sur la figure \ref{fig1}a,
\be
|a|\gg b
\ee
 ce qui est bien la marque d'une résonance de diffusion à deux corps.  Comme $b$ est de l'ordre de quelques nanomètres dans les expériences, on a aussi
\be
b\ll \rho^{-1/3},\lambda
\ee
 ce qui donne l'idée de construction d'un système modèle, par passage à la limite $b\to 0$ à $a$ fixé d'une interaction de portée nulle, caractérisée seulement par la longueur algébrique $a$.  Cette idée sera mise en œuvre dans la section \ref{sec3.1}.  (iv) En revanche, l'interaction n'est pas résonnante dans l'onde $p$ (moment cinétique orbital relatif $l=1$) donc les interactions $\uparrow\uparrow$ et $\downarrow\downarrow$, qui se produisent de façon prédominante dans cette onde à basse énergie (antisymétrie fermionique oblige), sont négligeables. 

Comme nous le verrons dans la section \ref{sec3}, l'existence d'un modèle bien défini (d'énergie bornée inférieurement lorsque $b\to 0$) constitue un problème mathématiquement non trivial.  On peut déjà en proposer une condition nécessaire, inspirée de la réalité expérimentale.  On l'aura en effet bien compris sur la figure \ref{fig1}a: puisque l'interaction de van der Waals admet (au moins) un état lié à deux corps de taille $\approx b$, la phase gazeuse considérée jusqu'à présent et vue dans les expériences n'est qu'une phase métastable, échappant temporairement à la solidification prédite par les lois de la physique à l'équilibre, solidification dont les pertes à trois corps sont les précurseurs (voir la figure \ref{fig1}b).  Ces pertes se produisent avec un taux estimé comme suit dans la référence \cite{Shlyap} pour des masses égales:
\be
\Gamma_{\rm pertes}^{\rm 3\, corps}\propto \frac{\hbar}{mb^2} \ \mbox{Proba}(\mbox{3 fermions $\uparrow\uparrow\downarrow$ ou $\downarrow\downarrow\uparrow$ dans une même boule de rayon $b$})
\label{eq1_7}
\ee
Le premier facteur représente l'échelle d'énergie pertinente de ce processus de recombinaison: c'est l'énergie de liaison du dimère fortement lié formé, et l'échelle de longueur $|a|\gg b$ ne peut intervenir.  Le second facteur tient compte du fait que le processus à trois fermions ne peut pas se produire si l'un des fermions est séparé des deux autres par une distance $\gg b$, par quasi-localité dans l'espace des positions: en effet, la portée des interactions et la taille du dimère fortement lié sont toutes deux de l'ordre de $b$.  Le coefficient de proportionnalité dans l'équation (\ref{eq1_7}) dépend des détails de la physique microscopique.  Nous aboutissons ainsi à une condition de stabilité expérimentale du gaz de fermions dans la limite $b\to 0$ d'une interaction de contact:
\be
\Gamma_{\rm pertes}^{\rm 3\, corps}\underset{b\to 0}{\to} 0
\ee
L'étude de ce système, pourtant gazeux, est rendue non triviale par la force des interactions.  Par exemple,  puisque $k_{\rm F}|a|\approx 1$, où $k_{\rm F}=(3\pi^2\rho)^{1/3}$ est le nombre d'onde de Fermi, la température de transition superfluide est a priori de l'ordre de la température de Fermi $T_{\rm F}=E_{\rm F}/k_{\rm B}$ (il n'y a pas d'autre échelle  disponible que l'énergie de Fermi $E_{\rm F}=\hbar^2k_{\rm F}^2/2m$) et sera difficile à calculer avec précision: la théorie BCS sera au mieux qualitative, et les méthodes de Monte-Carlo quantique sont difficiles à appliquer aux fermions; le défi a cependant été relevé par la référence \cite{SvisunovTc}, il est vrai dans le cas symétrique de masses et de potentiels chimiques égaux dans les deux composantes, où existent des méthodes de Monte-Carlo exemptes du fameux \g{problème de signe}.

\section{Questions ouvertes dans un point de vue microscopique}
\label{sec3}

Dans cette section, les interactions entre fermions sont remplacées dans une limite de portée nulle par des conditions de contact sur la fonction d'onde à $N$ corps, l'opérateur  hamiltonien se réduisant alors à celui du gaz parfait (modèle de Wigner-Bethe-Peierls \cite{Wigner,BethePeierls}). 

\subsection{Définition du modèle de Wigner-Bethe-Peierls}
\label{sec3.1}

\begin{figure}[t]
\includegraphics[width=12cm,clip=]{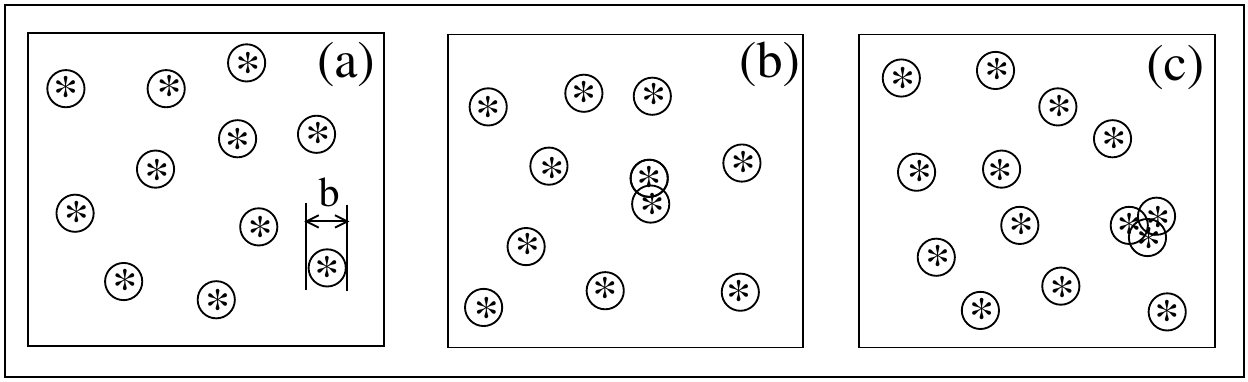}
\caption{Photographie du gaz montrant les positions (étoiles) des $N$ fermions, comme le ferait un microscope à gaz quantique, dans la limite d'une interaction de portée $b\to 0$ (à titre indicatif, nous avons entouré chaque étoile d'un cercle de diamètre $b$). (a)  Cas typique: les particules sont séparées deux à deux par une distance $\gg b$ et n'interagissent pas.  Ceci fixe l'opérateur hamiltonien (\ref{eq2_9}) du modèle de Wigner-Bethe-Peierls.  (b) Cas où deux particules $\uparrow$ et $\downarrow$, bien séparées des autres, subissent une collision binaire.  Ceci fixe les conditions de contact (\ref{eq1_12}) du modèle.  (c) Cas d'une collision ternaire isolée.  Ceci interroge sur la nécessité de conditions de contact à trois corps.}
\label{fig2}
\end{figure}

Pour construire le modèle, partons de la perception simple que nous en donnerait une photographie du gaz, c'est-à-dire une mesure des positions des $N$ fermions comme les microscopes à gaz quantique permettent de le faire depuis peu dans le cas homogène \cite{Tarik}.  Dans la limite où la portée $b$ de l'interaction tend vers zéro, la photo typique ressemble à la figure \ref{fig2}a: les fermions sont séparés deux à deux par une distance $\gg b$ et le potentiel d'interaction $V(\rr_i-\rr_j)$ est négligeable.  La fonction d'onde à $N$ corps  obéit dans ce cas à l'équation de  Schrödinger stationnaire
\be
\label{eq1_9}
E\psi = H_{\rm gaz\, parfait}\psi
\ee
avec l'opérateur hamiltonien du gaz parfait, somme des termes d'énergie cinétique $\pp^2/2m_\sigma$ et de piégeage $U_\sigma(\rr)$ dans chaque état interne $\sigma$:
\be
\label{eq2_9}
H_{\rm gaz\, parfait}=\sum_{i=1}^{N_\uparrow}\left(\frac{\pp_i^2}{2m_\uparrow} + U_\uparrow(\rr_i)\right) + \sum_{j=N_\uparrow+1}^{N} \left(\frac{\pp_j^2}{2m_\downarrow} + U_\downarrow(\rr_j)\right)
\ee
On convient ici de numéroter les particules  de façon que les $N_\uparrow$ premières soient dans l'état interne $\uparrow$ et les $N_\downarrow$ dernières soient dans l'état interne $\downarrow$; la fonction d'onde $\psi(\rr_1,\ldots,\rr_N)$ est alors une fonction antisymétrique des $N_\uparrow$ premières positions et une fonction antisymétrique des $N_\downarrow$ dernières positions. 

Certaines photos ressembleront cependant à la figure \ref{fig2}b: deux fermions $i$ et $j$, d'états internes différents,  respectivement $\uparrow$ et $\downarrow$,  sont séparés des autres par une distance $\gg b$ mais sont séparés entre eux d'une distance $\approx b$ donc subissent l'effet du potentiel d'interaction $V(\rr_i-\rr_j)$. La bonne façon de voir est de dire que $i$ et $j$ sont en train de subir dans le gaz une diffusion à deux corps isolée, ce qui a deux conséquences, l'une qualitative, l'autre quantitative. 

Qualitativement, on comprend qu'il vaut mieux,  dans notre gaz très peu dense (au sens où $\rho b^3\ll 1$), caractériser l'interaction entre $\uparrow$ et $\downarrow$ par son amplitude de  diffusion à deux corps,  plus généralement par un opérateur de transmission dit matrice $T$, que par la fonction $V(\rr)$ elle-même;  comme l'interaction se produit dans l'onde $s$, l'amplitude de diffusion $f$ est isotrope et ne dépend que du nombre d'onde relatif $k_{\rm rel}$ des deux particules; dans la limite $b\to 0$ à longueur de diffusion $a$ fixée, on dispose alors du développement de basse énergie\footnote{Si $V(\rr)$ décroît plus vite que $1/r^7$ à l'infini, on peut mettre un $O(k_{\rm rel}^4b^3)$ au dénominateur.}
\be
\label{eqfk}
f_{k_{\rm rel}} = \frac{-1}{a^{-1}+\ii k_{\rm rel}-(1/2) k_{\rm rel}^2 r_{\rm e} + O(k_{\rm rel}^3b^2)}
\ee
Nous supposons dans la suite que la portée effective de l'interaction $r_{\rm e}$ est un $O(b)$ donc devient négligeable lorsque $b\to 0$.\footnote{\label{note1_10}Expérimentalement, il existe cependant des résonances de Feshbach magnétiques dites étroites, pour lesquelles $r_{\rm e}$, négatif, est gigantesque à l'échelle atomique et peut être de l'ordre de $1/k_{\rm F}$, à cause d'un couplage anormalement faible entre les voies ouverte et fermée de notre note \ref{note2_5}, voir la référence \cite{Gur}. Ces résonances sont difficiles à utiliser car elles nécessitent un très bon contrôle du champ magnétique. L'existence d'une portée effective de limite non nulle lorsque $b\to 0$ a cependant l'avantage de stabiliser le gaz dans le régime instable de la section \ref{sec3.2} (le spectre reste borné inférieurement et le taux de pertes à trois corps (\ref{eq1_7}) tend vers zéro), et devrait permettre la préparation et l'observation d'états liés efimoviens de longue durée de vie, pour peu que le rapport de masse $m_\uparrow/m_\downarrow$ soit assez grand. L'expérience reste à faire.} Aussi l'amplitude de diffusion se réduit-elle à la forme universelle pour une interaction de contact
\be
\label{eq2_10}
f_{k_{\rm rel}}^{\rm contact}=\frac{-1}{a^{-1}+\ii k_{\rm rel}}
\ee

Quantitativement, on s'attend à ce que les deux fermions $\uparrow$ et $\downarrow$ proches se découplent des $N-2$ autres dans la fonction d'onde à $N$ corps, au sens où
\be
\label{eq3_10}
\psi(\rr_1,\ldots,\rr_N) \underset{r_{ij}=O(b)}{\simeq} \phi(\rr_i-\rr_j) A_{ij}(\RR_{ij}; (\rr_k)_{k\neq i,j})
\ee
où $\RR_{ij}=(m_\uparrow\rr_i+m_\downarrow \rr_j)/(m_\uparrow+m_\downarrow)$ est la position du centre de masse des particules $i$ et $j$, $\rr_{ij}=\rr_i-\rr_j$ est leur position relative, $(\rr_k)_{k\neq i,j}$ est le $(N-2)$-uplet  des positions des autres particules, la fonction $A_{ij}$ n'est en général pas connue  mais $\phi(\rr)$ est un état de diffusion à deux corps, solution de l'équation de Schrödinger
\be
\label{eq1_11}
\veps\phi(\rr)= -\frac{\hbar^2}{2 m_{\rm rel}} \Delta \phi(\rr) + V(\rr)\phi(\rr)
\ee
pour le mouvement relatif de masse $m_{\rm rel}=m_\uparrow m_\downarrow/(m_\uparrow+m_\downarrow)$ à une énergie $\veps$ dont l'expression formelle est donnée dans la référence \cite{Gen} (voir son équation (85)) mais dont nous retiendrons seulement qu'elle est $\approx \hbar^2 k_{\rm typ}^2/2 m_{\rm rel}$, où le nombre d'onde typique $k_{\rm typ}$ dans le gaz est de l'ordre de $k_{\rm F}$ pour $T=O(T_{\rm F})$. Dans la limite $b\to 0$, il suffit en fait d'analyser l'équation (\ref{eq1_11}) dans l'intervalle
\be
\label{eq2_11}
b\ll r \ll k_{\rm typ}^{-1}
\ee
le cas $r\leq b$ n'apportant que des détails non universels sur l'interaction et le cas $r>k_{\rm typ}^{-1}$ invalidant la factorisation (\ref{eq3_10}) (la paire $ij$ n'est plus bien isolée comme sur la photo de la figure \ref{fig2}b).  La première inégalité dans l'équation (\ref{eq2_11}) permet de mettre $V(\rr)$ à zéro au second membre, la seconde permet d'assimiler l'énergie $\veps$ à zéro au premier membre de l'équation (\ref{eq1_11}), d'où l'équation de Schrödinger simplifiée
\be
0=-\frac{\hbar^2}{2 m_{\rm rel}} \Delta \phi(\rr)
\ee
Sa solution générale dans l'onde $s$ (invariante par rotation) est combinaison linéaire de la solution constante $1$ (l'onde incidente d'énergie nulle) et de la solution de Coulomb (l'onde  diffusée) avec une amplitude relative fixée par $V(\rr)$ aux courtes distances:
\be
\label{eq4_11}
\phi(\rr)=\mathcal{N}\left(1-\frac{a}{r}\right) = \frac{1}{a}-\frac{1}{r}
\ee
Par définition, voir la contribution de Jan Solovej \cite{Solov}, la quantité $a$ est la longueur de diffusion du potentiel.  Au troisième membre, nous avons choisi la normalisation commode (facteur $\mathcal{N}$ pris égal à $1/a$ au second membre) pour avoir un résultat fini à la résonance de diffusion $a^{-1}=0$. 

Nous arrivons ainsi naturellement à la définition du modèle de Wigner-Bethe-Peierls pour notre système tridimensionnel de $(N_\uparrow,N_\downarrow)$ fermions à deux composantes en interaction de portée nulle et de longueur de diffusion $a\neq 0$  dans l'onde $s$:
\begin{enumerate}
\item  l'opérateur hamiltonien est le même que celui du gaz parfait, comme dans les équations (\ref{eq1_9},\ref{eq2_9})
\item il y a antisymétrie fermionique du vecteur d'état $\psi$ pour les $N_\uparrow$ premières et pour les $N_\downarrow$ dernières positions
\item  l'interaction est décrite non pas par un potentiel $V$ mais par les conditions de contact suivantes sur $\psi$: pour tout indice $i\in\{1,\ldots,N_\uparrow\}$ et tout indice $j\in\{N_\uparrow+1,\ldots,N=N_\uparrow+N_\downarrow\}$, il existe une fonction $A_{ij}$ telle que\footnote{\label{notedep}Les fonctions $A_{ij}$ ne sont pas indépendantes. L'antisymétrie fermionique impose que $A_{ij}(\RR_{ij},(\rr_k)_{k\neq i,j})=(-1)^{i-1} (-1)^{j-(N_\uparrow+1)} A_{1,N_\uparrow+1}(\RR_{ij},(\rr_k)_{k\neq i,j})$ (pour les amener en première position dans leur état interne respectif et faire ainsi apparaître la fonction $A_{1,N_\uparrow+1}$, on a dû faire passer $\rr_i$ à travers $i-1$ vecteurs positions de fermions $\uparrow$ et $\rr_j$ à travers $j-(N_\uparrow+1)$ vecteurs positions de fermions $\downarrow$, d'où les signes).}
\be
\label{eq1_12}
\psi(\rr_1,\ldots,\rr_N)\underset{r_{ij}\to 0}{=}A_{ij}(\RR_{ij},(\rr_k)_{k\neq i,j})\left(\frac{1}{r_{ij}}-\frac{1}{a}\right)+O(r_{ij})
\ee
où l'on fait tendre vers zéro la distance $r_{ij}$ entre les particules $i$ et $j$ à positions fixées de leur centre de masse $\RR_{ij}$ et des autres particules $\rr_k$, en imposant $\RR_{ij}\neq\rr_k \forall k\neq i,j$ et les $\rr_k$ deux à deux distincts (comme sur la figure \ref{fig2}b). 
\end{enumerate}
Mathématiquement, le point 3 signifie que le domaine de l'opérateur hamiltonien n'est pas le même que celui du gaz parfait: en l'absence d'interaction ($a=0$), on élimine à juste titre les solutions qui divergent en $1/r_{ij}$, comme il est dit dans tout bon ouvrage de mécanique quantique.  C'est la seule différence mais elle est de taille!\footnote{Un point clé est que l'état de diffusion $\phi(\rr)=1/r-1/a$ est bien de carré sommable sur un voisinage de l'origine, $\int_{r< r_{\rm max}}\dd^3r\, |\phi(\rr)|^2<\infty$: il n'y a donc pas de coupure à mettre à courte distance et $a$ est la seule longueur associée à l'interaction. C'est différent dans les ondes de moment cinétique $l>0$: $\phi(\rr)=Y_l^{m_l}(\theta,\varphi)\left(r^l+a_{\rm gen}^{2l+1}/r^{l+1}\right)$ (où le paramètre $a_{\rm gen}\neq 0$ généralisant $a$ est une longueur et $Y_l^{m_l}$ est une harmonique sphérique) n'est alors plus de carré sommable, et il faut introduire une coupure donc une seconde longueur pour caractériser l'interaction \cite{Ludoondel}.}

La figure \ref{fig2}c, qui montre un trio d'atomes proches, bien séparés des autres et en train de subir une diffusion à trois corps, fait naître une interrogation légitime: faut-il compléter le modèle par des conditions de contact à trois  corps? à quatre corps? etc.  Réponse dans la section suivante. 

\subsection{Questions d'existence}
\label{sec3.2}

Il n'est pas évident que le modèle de Wigner-Bethe-Peierls, tel que nous l'avons défini en page \pageref{eq1_12}, conduise à un hamiltonien auto-adjoint (sans conditions de contact supplémentaires) et, surtout, à un spectre d'énergie borné inférieurement.  En effet, nous avons quand même fait tendre une échelle d'énergie vers $-\infty$, celle $-\hbar^2/m_{\rm rel} b^2$ associée à la portée de l'interaction, en prenant la limite $b\to 0$ à longueur de diffusion $a$ fixée donc sans faire tendre la force des interactions vers zéro, ce qui pourrait provoquer un effondrement du système sur lui-même, comme dans l'effet Thomas bien connu en physique nucléaire \cite{Thom}! 

 La discussion s'éclaire dans le cas particulier $a^{-1}=0$, dit de la limite unitaire (l'amplitude de diffusion (\ref{eq2_10}) du modèle atteint en module la valeur maximale $k_{\rm rel}^{-1}$ autorisée dans l'onde $s$ par l'unitarité de la matrice de collision $S$), car les conditions de contact (\ref{eq1_12}) deviennent invariantes d'échelle (c'est aussi le régime le plus intéressant et le plus ouvert car d'interaction maximale en phase gazeuse).  Pour simplifier encore, limitons-nous aux états propres d'énergie $E=0$ dans l'espace libre, avec un centre de masse des $N$ fermions  au repos.  Comme  il n'y a alors ni énergie ni potentiel extérieur  pour introduire une échelle de longueur, on s'attend à ce que l'état propre $\psi$ lui-même soit invariant d'échelle, c'est-à-dire une fonction homogène des coordonnées (invariante à un facteur près par l'homothétie $\rr_i\to \lambda \rr_i$  de rapport $\lambda$ sur les $N$ positions), de la forme \cite{FWsep,livre}
\be
\label{eq1_14}
\psi(\rr_1,\ldots,\rr_N)=R^{s-\frac{3N-5}{2}}\Phi(\Omega)
\ee
où (i) $R$ est  l'hyperrayon interne, écart quadratique moyen des positions des $N$ particules à leur centre de masse $\CC$ pondérées par les masses,
\be
\label{eq2_14}
MR^2 = \sum_{i=1}^{N} m_i (\rr_i-\CC)^2
\ee
avec $M=\sum_{i=1}^{N} m_i$ la masse totale et $M\CC=\sum_{i=1}^{N} m_i \rr_i$; (ii) l'exposant d'échelle (le degré d'homogénéité) est commodément repéré par la quantité $s$ après translation de $(3N-5)/2$ - pour révéler une symétrie $s\leftrightarrow -s$; (iii) $\Phi$ est une fonction inconnue des $3N-4$ hyperangles complétant $R$ dans le paramétrage des $\rr_i-\CC$ en coordonnées hypersphériques.  Le report de l'ansatz (\ref{eq1_14})  dans l'équation de Schrödinger (\ref{eq1_9}) (avec $E=0$ et $U_\sigma\equiv 0$ comme il a été dit) donne une équation aux valeurs propres sur $\Phi$:
\be
\label{eq3_14}
\left[-\Delta_\Omega +\left(\frac{3N-5}{2}\right)^2\right]\Phi(\Omega) = s^2 \Phi(\Omega)
\ee
dont les valeurs propres ne sont autres que $s^2$! Comme le laplacien $\Delta_\Omega$ est pris sur un compact, l'hypersphère unité $S_{3N-4}$,  les valeurs possibles de $s^2$ forment un ensemble discret, dans $\mathbb{R}$ à supposer que l'hamiltonien soit hermitien; on ne sait en général pas les calculer, à cause des difficiles conditions de contact (\ref{eq1_12}) sur $\Phi(\Omega)$.\footnote{\label{note1_14}Les conditions de contact (\ref{eq1_12}) ne contraignent en revanche pas la dépendance de $\psi$ en l'hyperrayon.  C'est que, si $\psi$ obéit aux conditions de contact, $f(R)\psi$  y obéit aussi, pourvu que le facteur $f(R)$ soit une fonction régulière de $R$. En effet, dans la limite $r_{ij}\to 0$ à $\RR_{ij}$ et $(\rr_k)_{k\neq i,j}$ fixés, on a $MR^2=m_i(\rr_i-\CC)^2+m_j(\rr_j-\CC)^2+\mbox{cte}=m_i\rr_i^2+m_j\rr_j^2+\mbox{cte}=(m_i+m_j)\RR_{ij}^2+m_{\rm rel} r_{ij}^2+\mbox{cte}=O(r_{ij}^2)+\mbox{cte}$. Or, $(1/r_{ij}-1/a)O(r_{ij}^2)$ est un $O(r_{ij})$  négligeable.}

 Nous nous contenterons dans la suite d'écrire formellement que $s$ est la racine d'une fonction transcendante paire, dite fonction d'Efimov,
\be
\label{eq1_15}
\Lambda_{N_\uparrow,N_\downarrow}(s)=0
\ee
sans spécifier cette fonction (le plus direct pour l'obtenir est d'imposer les conditions de contact (\ref{eq1_12}) sur un ansatz de Faddeev écrit dans l'espace réciproque,\footnote{Rappelons brièvement la construction de l'ansatz. On écrit d'abord l'équation de Schrödinger à énergie nulle au sens des distributions, $H_{\rm gaz\, parfait}\psi= \sum_{i=1}^{N_\uparrow} \sum_{j=N_\uparrow+1}^{N} (2\pi\hbar^2/m_{\rm rel})\delta(\rr_{ij}) A_{ij}(\RR_{ij},(\rr_k)_{k\neq i,j})$ où les distributions de Dirac proviennent de l'action des opérateurs d'énergie cinétique sur les singularités en $1/r_{ij}$, en vertu de l'équation de Poisson $\Delta_\rr(1/r)=-4\pi\delta(\rr)$, et $m_{\rm rel}$ est la masse réduite de deux fermions de spins opposés comme nous l'avons dit. On en prend ensuite la transformée de Fourier ($\psi\to\tilde{\psi}$, $\Delta_\rr\to -k^2$). En tirant parti de l'antisymétrie fermionique comme dans la note \ref{notedep} et de l'invariance par translation spatiale (le centre de masse est au repos), on se réduit à $\tilde{\psi}(\kk_1,\ldots,\kk_N)=\frac{\delta(\kk_1+\ldots+\kk_N)}{\sum_{n=1}^{N} \hbar^2 k_n^2/2m_n} \sum_{i=1}^{N_\uparrow} \sum_{j=N_\uparrow+1}^{N} (-1)^{i+j} D((\kk_n)_{n\neq i,j})$ où $D$ est la seule fonction inconnue (chaque $A_{ij}$ est une fonction des $(\rr_k-\RR_{ij})_{k\neq i,j}$ dont $D((\kk_n)_{n\neq i,j})$ est la transformée de Fourier à un facteur près).}
ce qui mène à une équation intégrale dite de Skorniakov-Ter-Martirosian - ici à la limite unitaire et à énergie nulle, dans laquelle on reporte l'équivalent de Fourier de l'ansatz (\ref{eq1_14}); l'équation transcendante (\ref{eq1_15}) qui en résulte admet une écriture explicite pour $N=3$ \cite{Tignone}, et s'écrit comme le déterminant d'un opérateur pour $N>3$, cet opérateur étant donné explicitement pour $N=4$ dans le secteur $(3,1)$ par la référence \cite{MoraCastinPricoupenko} et dans le secteur $(2,2)$ par la référence \cite{Endopas4}).  Il faut maintenant distinguer deux cas. 

\paragraph{Premier cas: $s^2>0$} Il y a alors deux valeurs possibles correspondantes de l'exposant d'échelle, une valeur $>0$ que nous convenons d'appeler $s$, et  la valeur opposée $-s<0$.  Par un phénomène similaire à celui de l'équation (\ref{eq4_11}), $\psi$  est en général une combinaison linéaire de deux solutions, l'une contenant un facteur $R^s$, l'autre contenant un facteur $R^{-s}$, les amplitudes relatives étant fixées de manière univoque par une longueur $\ell$ (l'équivalent de $a$ dans l'équation (\ref{eq4_11})) déterminée par les détails microscopiques de l'interaction à courte distance $O(b)$:\footnote{\label{noteN2} L'état de diffusion (\ref{eq4_11}) correspond au cas $N=2$; alors $s=1/2$ et $(3N-5)/2=1/2$, et $\psi$ dans l'équation (\ref{eq2_15}) est bien combinaison linéaire de $R^0$ et $R^{-1}$; dans ce cas, $R\propto r_{12}$ et $\Phi(\Omega)=\mbox{cte}$ dans l'onde $s$. Le calcul explicite de l'expression (\ref{eq2_15}) pour $N=2$ donne en effet $\psi\propto (r_{12}/\bar{\ell})^{s-1/2}-(r_{12}/\bar{\ell})^{-s-1/2}$ avec $\bar{\ell}=(m_1+m_2)\ell/(m_1m_2)^{1/2}$, ce qui doit être proportionnel à (\ref{eq4_11}), d'où la valeur annoncée de l'exposant $s=1/2$; le paramètre à deux corps $\bar{\ell}$ n'est autre que la longueur de diffusion $a$.} \footnote{On a mis un signe moins entre les deux termes entre crochets dans l'équation (\ref{eq2_15}); un signe plus serait aussi possible, suivant le modèle microscopique.}
\be
\label{eq2_15}
\psi=\left[(R/\ell)^{s}-(R/\ell)^{-s}\right]R^{-\frac{3N-5}{2}}\Psi(\Omega)
\ee
Cependant, en l'absence de résonance de diffusion à $N$ corps, on s'attend à ce que $\ell=O(b)$, si bien que $\ell\to 0$ lorsque $b\to 0$: la solution en $R^{-s}$ devient négligeable, la longueur $\ell$ disparaît du problème et l'on garde la condition de contact à $N$ corps invariante d'échelle suivante dans la voie d'exposant d'échelle $s$ \cite{theseFelix}:
\be
\label{eq3_15}
\psi\underset{R\to 0}{\approx} R^{s-\frac{3N-5}{2}}
\ee
La fonction d'onde $\psi$, considérée comme une fonction de $R$, est sans nœud donc l'énergie $E=0$ correspond à l'état fondamental: il n'y a pas d'état lié, d'énergie $E<0$ pouvant tendre vers $-\infty$ lorsque $b\to 0$.\footnote{\label{note13} Le cas spécial d'une résonance de diffusion à $(N_\uparrow,N_\downarrow)$ corps, où $\ell$ reste non infinitésimal dans la limite $b\to 0$, est traité en détail dans la référence \cite{LudovicN}, qui explique quelle condition de contact à $N$ corps utiliser pour décrire correctement l'état faiblement lié qui en résulte. En effet, la condition (\ref{eq2_15}) déjà proposée dans \cite{FWsep} n'est satisfaisante que pour $s$ assez petit (pour $s>1$, on voit bien que l'état (\ref{eq2_15}) n'est plus de carré intégrable en $R=0$ et qu'une deuxième longueur - une coupure - doit être introduite).} \footnote{\label{note13bis} Sur une résonance de Feshbach étroite, voir la note \ref{note1_10}, la portée effective $r_{\rm e}$ est - pour $1/a=0$ - la seule échelle de longueur pertinente lorsque la portée vraie $b$ tend vers zéro, si bien que la longueur $\ell$ est de l'ordre de $|r_{\rm e}|\gg b$. La fonction d'onde $\psi$ dans l'équation (\ref{eq2_15}) admet alors un nœud très \g{en dehors} du potentiel d'interaction: la solution à $E=0$ ne serait pas d'énergie minimale, et le système admettrait un état lié à $(N_\uparrow,N_\downarrow)$ fermions (avec $N>2$)! Cependant, une étude spécifique du cas $(N_\uparrow=2,N_\downarrow=1)$ montre qu'il n'en est rien (tant que le rapport de masse $m_\uparrow/m_\downarrow$ reste assez faible pour que $s^2\geq 0$ bien entendu) \cite{Tignone,epls}. Fallait-il s'en étonner ? Raisonnons par l'absurde. S'il y avait vraiment un état lié, il conduirait à un nombre d'onde relatif $k_{\rm rel}\approx 1/|r_{\rm e}|$ entre les fermions, le terme de portée effective ne serait pas négligeable au dénominateur de l'amplitude de diffusion (\ref{eqfk}) et l'on perdrait l'invariance d'échelle donc la séparabilité en coordonnées hypersphériques. L'équation (\ref{eq2_15}) serait inapplicable et la prédiction d'un état lié caduque. Plus généralement, pour pouvoir croire à (\ref{eq2_15}) - c'est une condition nécessaire, il faut que l'hyperrayon $R$ soit beaucoup plus grand que toute échelle de longueur apparaissant dans tout sous-système $(n_\uparrow,n_\downarrow)$ [avec $n_\uparrow\leq N_\uparrow$, $n_\downarrow\leq N_\downarrow$ et $n_\uparrow+n_\downarrow<N_\uparrow+N_\downarrow=N$], en particulier $R\gg b$ et $R\gg |r_{\rm e}|$ pour $(n_\uparrow=1,n_\downarrow=1)$. Faut-il le préciser, la résonance à $(N_\uparrow,N_\downarrow)$ corps de la note \ref{note13} ne remet pas (\ref{eq2_15}) et l'existence d'un nœud à l'hyperdistance $\ell$ en question puisque la longueur $\ell\gg b$ anormalement grande qui apparaît ne préexiste dans aucun sous-système.}
\paragraph{Deuxième cas: $s^2<0$} Il y a là aussi deux valeurs possibles de l'exposant d'échelle, l'une $s=\ii|s|$ dans $\ii\mathbb{R}^+$ que nous convenons d'appeler $s$, et l'autre, son complexe conjugué $-\ii |s|$, ou encore son opposé $-s$, dans $\ii\mathbb{R}^-$.  Comme dans le premier cas, on conclut qu'il existe une longueur $\ell$, fonction des détails microscopiques de l'interaction, réglant l'amplitude relative des deux solutions:
\be
\label{eq1_16}
\psi=\left[(R/\ell)^{\ii|s|}-(R/\ell)^{-\ii|s|}\right]R^{-\frac{3N-5}{2}}\Psi(\Omega)=2\ii\sin[|s|\ln(R/\ell)]R^{-\frac{3N-5}{2}}\Psi(\Omega)
\ee
Cette fois, les deux solutions sont de même module donc il faut les garder  toutes les deux (aucune ne l'emporte sur l'autre dans la limite $b\to 0$)! La longueur $\ell$ ne disparaît pas du problème mais définit dans la limite $b\to 0$ une condition de contact à $N$ corps (\ref{eq1_16}) qui brise explicitement l'invariance d'échelle continue de la limite unitaire. Comme $\psi$ comporte un nombre infini de nœuds à une hyperdistance $R$ arbitrairement grande - arbitrairement plus grande que la portée $b$ de l'interaction - (voir l'écriture du troisième membre de l'équation (\ref{eq1_16})), il existe un nombre infini d'états liés à $N$ corps sous la solution d'énergie $E=0$; comme la condition aux limites (\ref{eq1_16}) est invariante par changement de $\ell$ en $\exp(\pm\pi/|s|)\ell$, on passe d'un état $N$-mère à l'autre par une homothétie de rapport $\exp(\pi/|s|)$, le spectre correspondant  formant une suite géométrique de limite nulle mais non bornée  inférieurement dans la limite de portée nulle:\footnote{L'expression précise de $E_{\rm glob}$ en fonction de $s$ et $\ell$, et de l'ordre de $\hbar^2/Mb^2$ pour $\ell\approx b$, figure par exemple dans la référence \cite{virtrans}. Dans l'équation (\ref{eq2_16}) nous prenons $n\geq 1$ (étant admis que $\exp(-2\pi/|s|)\ll 1$ - sinon le spectre ne serait pas entièrement géométrique \cite{Tignone}) car le modèle de Wigner-Bethe-Peierls  ne peut s'appliquer qu'à un état lié de taille $\gg b$. On peut toutefois avoir $\ell\approx |r_{\rm e}|\gg b$ et $E_{\rm glob}\approx \hbar^2/M r_{\rm e}^2\ll \hbar^2/M b^2$ sur une résonance de Feshbach étroite, voir notre note \ref{note1_10}; même si ce n'est pas évident, l'exclusion de $n=0$ dans (\ref{eq2_16}) reste correcte dans ce cas \cite{Tignone,epls}. Cette exclusion de $n=0$ est cohérente avec l'absence d'état lié lorsque $s^2\geq 0$, voir notre note \ref{note13bis}: du côté efimovien, l'ensemble du spectre discret doit tendre vers zéro lorsque $|s|\to 0$ sachant que $E_{\rm glob}$ a une limite finie et non nulle.}
\be
\label{eq2_16}
E_n=-E_{\rm glob} \eee^{-2\pi n/|s|} \quad ,\quad n\in\mathbb{N}^*, \quad \mbox{avec} \quad E_{\rm glob}\approx\hbar^2/Mb^2
\ee
Ces états liés à $N$ corps, historiquement prédits par Efimov dans le cas $N=3$, sont dits efimoviens. Il serait très intéressant de les stabiliser dans une expérience d'atomes froids (on vérifie malheureusement que, si $s\in\ii\mathbb{R}^{+*}$ dans le problème à $(2,1)$ fermions, ce qui se produit pour $m_\uparrow/m_\downarrow>13,\!6069\ldots$ comme nous le verrons, le taux de pertes à trois corps  de l'équation (\ref{eq1_7}) ne tend pas vers zéro lorsque $b\to 0$, mais plutôt vers une quantité proportionnelle à $\hbar k_{\rm F}^2/m$ dans le gaz homogène à $T=0$, ce qui est considérable), par exemple en utilisant la note \ref{note1_10}. 

Mais revenons à notre problème mathématique: nous concluons que le gaz de $(N_\uparrow,N_\downarrow)$ fermions est stable pour une interaction $\uparrow\downarrow$ de portée nulle, et que l'hamiltonien du modèle de Wigner-Bethe-Peierls est auto-adjoint et borné inférieurement, si et seulement si les exposants d'échelle sont tous réels:
\be
\label{eq1_17}
s\in\mathbb{R}^* \quad \forall s\ \mbox{solution de}\ \Lambda_{N_\uparrow,N_\downarrow}(s)=0
\ee
Bien entendu, aucun sous-système $(n_\uparrow,n_\downarrow)$ du gaz ne doit non plus présenter d'effet Efimov, sinon (i) il faudrait introduire un paramètre à $n_\uparrow+n_\downarrow < N_\uparrow+N_\downarrow$ corps sur le modèle de l'équation (\ref{eq1_16}) et l'invariance d'échelle à l'origine de la séparabilité (\ref{eq1_14}), donc du résultat (\ref{eq1_17}), serait brisée - on perdrait la séparabilité à toutes les distances, et (ii) le sous-système pourrait s'effondrer sur lui-même dans la limite $b\to 0$ d'une interaction de portée nulle et l'énergie ne serait pas bornée inférieurement. Sans le dire, nous avons effectué un raisonnement par récurrence et l'équivalent de la condition de stabilité (\ref{eq1_17}) doit être satisfait pour tout nombre $n_\uparrow\leq N_\uparrow$ et tout nombre $n_\downarrow\leq N_\downarrow$.

\begin{rema}
On pourrait ajouter $s=0$ en troisième cas: c'est en fait précisément le seuil d'un effet Efimov.  En développant l'équation (\ref{eq1_16}) au premier ordre en $|s|$, on trouve que
\be
\label{eq2_17}
\psi\propto \ln(R/\ell) R^{-\frac{3N-5}{2}} \Phi(\Omega)
\ee
c'est-à-dire que la longueur $\ell$ règle l'amplitude relative des  solutions $R^{-(3N-5)/2}$ et $(\ln R)R^{-(3N-5)/2}$. Ceci ressemble furieusement à la définition de la longueur de diffusion de deux particules en dimension deux, voir par exemple la référence \cite{mon_cours} et la contribution de Jan Solovej \cite{Solov}. Cependant, dans le cas $\ell=O(b)$ où nous sommes, la longueur $\ell$ tend vers zéro lorsque $b\to 0$, la première solution - de coefficient $\ln(1/\ell)\to +\infty$ - l'emporte sur la deuxième et l'on garde la condition de contact à $N$ corps dans la continuité de (\ref{eq3_15}):
\be
\psi\underset{R\to 0}{\approx} R^{-(3N-5)/2}
\ee
Dans le modèle de portée nulle, il n'y a donc pas de brisure d'invariance d'échelle ni d'état lié au seuil efimovien.\footnote{Les considérations de la note \ref{note13bis} s'appliquent au seuil. En particulier, il ne faut pas croire à l'état lié d'énergie $\propto -\hbar^2/M\ell^2$ que l'équation (\ref{eq2_17}) nous inciterait à prédire: il serait d'extension spatiale $\ell$ et ne pourrait être décrit par notre modèle de portée nulle lorsque $\ell\approx b$; il n'existerait pas non plus pour $(N_\uparrow=2,N_\downarrow=1)$ dans le cas - pourtant apparemment favorable - d'une résonance de Feshbach étroite où $\ell\approx |r_{\rm e}|\gg b$, voir la note \ref{note13bis} et les références \cite{Tignone,epls}.} \footnote{Le fait que $\ln(1/\ell)$ tende lentement vers l'infini lorsque $b\to 0$ n'est pas sans conséquence pratique: si l'on veut comparer aux expériences, il vaut mieux garder la contribution en $\ln R$ dans (\ref{eq2_17}) [et le terme $(R/\ell)^{-s}$ dans (\ref{eq2_15}) pour $s>0$ assez proche de zéro] pour former la condition de contact à $N$ corps. Ainsi, on trouve que le troisième coefficient d'amas $b_{2,1}$ défini dans la section \ref{sec3.4} est en réalité une fonction régulière du rapport de masse $m_\uparrow/m_\downarrow$ donc de $s^2$, alors qu'il est de dérivée infinie en $s^2=0$ dans le modèle de portée nulle \cite{daily,virtrans} donc dans l'équation (\ref{eq1_20}).}
\end{rema}

\begin{rema}
Nous avons pris ici $a=\infty$ pour simplifier mais si le système est instable pour $a=\infty$, il le restera pour $a$ fini (toutes choses égales par ailleurs), car les $N$-mères efimoviens du gaz unitaire de taille $\ll |a|$ (il y en a autant qu'on veut pour $b\to 0$) ne font pas de différence entre une longueur de diffusion infinie et une longueur de diffusion finie $a$. 
\end{rema}

\paragraph{Complément} À la limite unitaire, la présente analyse se généralise à énergie $E$ non nulle (toujours avec le centre de masse du système au repos).  L'équation (\ref{eq1_14}) devient
\be
\label{eq1_17p}
\psi=F(R) R^{-\frac{3N-5}{2}} \Phi(\Omega)
\ee
(elle satisfait aux conditions de contact de Wigner-Bethe-Peierls en vertu de la note \ref{note1_14}) avec
\be
\label{eq2_17p}
E F(R) = -\frac{\hbar^2}{2M}\Delta_{\rm 2D} F(R) + \frac{\hbar^2 s^2}{2MR^2} F(R)
\ee
et $\Delta_{\rm 2D}$, le laplacien à 2D pour la variable $R$,  se réduit ici (en l'absence de dépendance angulaire) à $\dd^2/\dd R^2+R^{-1}\dd/\dd R$. Le cas efimovien $s^2<0$ correspond donc simplement au problème connu de \g{chute sur le centre} dans un potentiel attractif en $1/R^2$ \cite{Landaumec}. La séparabilité en coordonnées hypersphériques (\ref{eq1_17p}) s'étend même au cas piégé \cite{FWsep,livre}, il suffit d'ajouter le terme de piégeage $(1/2)M\omega^2R^2F(R)$ au second membre de l'équation (\ref{eq2_17p}),  et de  faire la substitution $E\to E-E_{\rm cdm}$,  la valeur propre de l'équation sur $F(R)$  étant l'énergie interne par opposition à celle $E_{\rm cdm}$ du centre de masse.  Pour $s^2>0$, ceci conduit au spectre
\be
\label{eq3_17p}
E-E_{\rm cdm}=(s+1+2q)\hbar\omega \quad,\quad q\in\mathbb{N}
\ee

\subsection{Ce qui est connu sur le domaine de stabilité}
\label{sec3.3}

Le problème de savoir si la condition de stabilité (\ref{eq1_17}) est satisfaite peut être attaqué par deux extrémités opposées.

\begin{figure}[t]
\includegraphics[width=12cm,clip=]{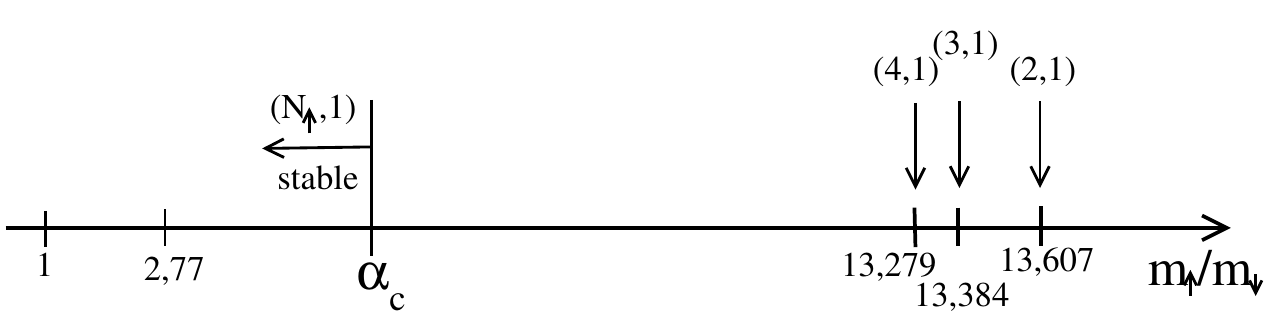}
\caption{Résultats connus sur la stabilité du système $(N_\uparrow,1)$ de $N_\uparrow$ fermions de spin $\uparrow$ et un fermion de spin $\downarrow$, pour une interaction de portée nulle, en fonction du rapport de masse $m_\uparrow/m_\downarrow$ entre une particule $\uparrow$ et la particule $\downarrow$. Flèches verticales: valeurs critiques de $m_\uparrow/m_\downarrow$ (seuils) pour l'effet Efimov à $(N_\uparrow,1)$ corps obtenues par résolution du problème correspondant, voir les références \cite{Efimov} pour $N_\uparrow=2$, \cite{MoraCastinPricoupenko} pour $N_\uparrow=3$, \cite{BazakPetrov} pour $N_\uparrow=4$;  quand $m_\uparrow/m_\downarrow$ excède ces valeurs,  l'énergie du système n'est plus bornée inférieurement.  Barre verticale avec flèche horizontale: rapport de masse critique $\alpha_{\rm c}$, dont l'existence est établie par le théorème de Moser-Seiringer \cite{MoserSeiringer}, en dessous duquel le système $(N_\uparrow,1)$ est stable $\forall N_\uparrow$; le théorème ne donne pas la valeur exacte de $\alpha_{\rm c}$ mais une minoration, $\alpha_{\rm c}>2,\!77$, plus contraignante que celle, $m_\uparrow/m_\downarrow=1$, des expériences sur les atomes froids (une supériorité cependant de ces dernières est que le gaz unitaire de fermions semble y être stable, sans effondrement ni pertes à trois corps significatives, pour toutes les valeurs de $N_\uparrow$ et $N_\downarrow$).}
\label{fig3}
\end{figure}

Par la première,  on résout le problème à $(N_\uparrow,N_\downarrow)$  fermions à $E=0$ dans le modèle de Wigner-Bethe-Peierls et on calcule les exposants d'échelle $s$ (on procède analytiquement le plus loin possible mais il y a une dernière étape numérique, en tout cas pour $N>3$). À notre connaissance, ce programme a été rempli dans le problème fermionique $(N_\uparrow>1,N_\downarrow=1)$  jusqu'à $N_\uparrow=4$, voir la figure \ref{fig3}: on trouve à chaque fois qu'un effet Efimov apparaît au-delà d'un rapport de masse critique $m_\uparrow/m_\downarrow$ (l'impureté $\downarrow$ doit être suffisamment légère), lui-même évidemment fonction décroissante de $N_\uparrow$.\footnote{Une fois qu'on a un effet  Efimov dans le problème à $(N_\uparrow,N_\downarrow)$ fermions, comme nous l'avons dit après l'équation (\ref{eq1_17}), on perd l'invariance d'échelle et on ne peut plus mettre en œuvre le raisonnement sous-tendu par les équations (\ref{eq1_14},\ref{eq3_14}) pour passer au problème à $(N_\uparrow+1,N_\downarrow)$ ou $(N_\uparrow,N_\downarrow+1)$ fermions; dans ce dernier, il n'y a plus de séparabilité en coordonnées $(R,\Omega)$ comme dans l'équation (\ref{eq1_16}), plus de spectre géométrique (\ref{eq2_16}) donc au sens strict, plus de possibilité d'effet Efimov!} On remarque que les rapports de masse critiques successifs sont de plus en plus rapprochés; mathématiquement, on ne sait pas cependant si cette séquence continue (existe-t-il un effet Efimov à (5,1) corps ? à (6,1) corps? etc.).  Le cas $(N_\uparrow>1,N_\downarrow=2)$ a été étudié pour $N_\uparrow=2$ par la référence \cite{Endopas4} qui conclut à la stabilité, tant que les sous-systèmes (2,1) et (1,2) le sont.

Par la seconde extrémité, on cherche à contraindre (plutôt qu'à calculer) les rapports de masse critiques, par une minoration du spectre de l'hamiltonien.  C'est ce qu'a fait la référence \cite{MoserSeiringer} pour le problème à $(N_\uparrow,N_\downarrow=1)$ fermions: elle démontre le magnifique 

\begin{theo*}
Il existe un rapport de masse critique $m_\uparrow/m_\downarrow=\alpha_{\rm c}$ en dessous duquel le système  fermionique $(N_\uparrow,N_\downarrow=1)$ est stable $\forall N_\uparrow$ pour une interaction de contact, et $\alpha_{\rm c}>1/0,\!36=2,\!77$.
\end{theo*}

\subsection{Développement en amas ou du viriel}
\label{sec3.4}

Certains diront que c'est le problème macroscopique à $N$ corps qui doit être en définitive l'objet de notre attention, plus que le problème à petit nombre de corps.  Ce à quoi nous répondrons  que le second peut dire quelque chose sur le premier au moyen du développement en amas, un développement de la pression  $P$ du gaz homogène à l'équilibre thermique grand-canonique en puissances des fugacités des composantes $\sigma$ (la température $T=1/k_{\rm B}\beta$ est fixée mais les potentiels chimiques $\mu_\sigma$ tendent vers $-\infty$, ce qui correspond à une limite quantiquement non dégénérée):
\be
\label{eq1_19}
\frac{P\bar{\lambda}^3}{k_{\rm B}T}=\sum_{(n_\uparrow,n_\downarrow)\in\mathbb{N}^{2*}} b_{n_\uparrow,n_\downarrow} \eee^{\beta \mu_\uparrow n_\uparrow} \eee^{\beta\mu_\downarrow n_\downarrow}
\ee
où la longueur d'onde thermique de de Broglie est prise à la masse de référence arbitraire $\bar{m}$, $\bar{\lambda}=(2\pi\hbar^2/\bar{m}k_{\rm B}T)^{1/2}$.\footnote{La référence \cite{EndoCastin2022} utilise le choix $\bar{m}^{3/2}=(m_\uparrow^{3/2}+m_\downarrow^{3/2})/2$, naturel au sens où il conduit à un premier coefficient d'amas total $b_1=(b_{1,0}+b_{0,1})/2$ égal à un.}  Le développement (\ref{eq1_19}), dans la littérature récente, est souvent confondu avec celui du viriel, qui développe en fait en puissances des densités $\rho_\sigma\lambda_\sigma^3$ dans l'espace des phases ($\lambda_\sigma=(2\pi\hbar^2/m_\sigma k_{\rm B}T)^{1/2}$).

Comment calculer les coefficients d'amas $b_{n_\uparrow,n_\downarrow}$? À la limite unitaire $a^{-1}=0$, le plus simple est d'utiliser la méthode du régulateur harmonique de la référence \cite{ComtetOuvry}, qui place d'abord le système dans les pièges harmoniques $U_\sigma(\rr)$ de l'équation (\ref{eq1_5}) puis, tous calculs faits, les ouvre pour retrouver le cas homogène par approximation d'homogénéité locale (exacte dans la limite considérée $\omega\to 0^+$).  En effet, le problème piégé est alors séparable en coordonnées hypersphériques  comme dans la section \ref{sec3.2} et, si l'on connaît les exposants d'échelle $s$ dans l'espace libre à énergie nulle pour tous les $n_\uparrow\leq n_\uparrow^{\rm cible}, n_\downarrow\leq n_\downarrow^{\rm cible}$, on connaît aussi les niveaux d'énergie du système piégé comme dans l'équation (\ref{eq3_17p}), donc toutes les fonctions de partition canoniques $Z_{n_\uparrow,n_\downarrow}$  et en définitive le coefficient $b_{n_\uparrow^{\rm cible},n_\downarrow^{\rm cible}}$.  Du coup, les coefficients  d'amas doivent être des fonctionnelles des $\Lambda_{n_\uparrow,n_\downarrow}$  de l'équation (\ref{eq1_15}). C'est bien ce que prédit la conjecture de la référence \cite{EndoCastin}, selon laquelle, à la limite unitaire,
\be
\label{eq1_20}
b_{n_\uparrow,n_\downarrow}=\frac{(n_\uparrow m_\uparrow+n_\downarrow m_\downarrow)^{3/2}}{\bar{m}^{3/2}} \left[\int_{-\infty}^{+\infty} \frac{\dd S}{4\pi} \, S \frac{\dd}{\dd S} \left(\ln\Lambda_{n_\uparrow,n_\downarrow}(\ii S)\right) + \mbox{CorrStat}_{n_\uparrow,n_\downarrow}\right]
\ee
pour tout $(n_\uparrow,n_\downarrow)\in\mathbb{N}^{*2}\setminus\{1,1\}$.\footnote{Le cas $(n_\uparrow=1,n_\downarrow=1)$ est différent et doit être exclu; il correspond, au contraire de ce que nous avons supposé, à une résonance de diffusion à $N$ corps avec $N=2$ dans l'onde $s$: à la limite unitaire, il faut garder uniquement la solution en $R^{-s}, s=1/2$, dans l'équation (\ref{eq2_15}), comme si la bonne racine de $\Lambda$ à garder était $-s$; en effet, la solution en $R^s$ correspond à la partie régulière $\propto 1/a$ de l'état de diffusion à énergie nulle (voir la note \ref{noteN2} pour plus de précisions).} Ici, le préfacteur résulte du passage du cas piégé au cas homogène et la première contribution  entre crochets est calquée sur le résultat de la référence \cite{revuecanad} obtenu pour $N=3$;  la seconde contribution entre crochets est une éventuelle correction de  statistique quantique de type gaz parfait provenant des sous-amas non monoatomiques indiscernables en lesquels se découplent les états propres  internes du système piégé $(n_\uparrow,n_\downarrow)$ à haute énergie (le centre de masse du système reste dans son état fondamental).  

\begin{figure}[t]
\includegraphics[width=8cm,clip=]{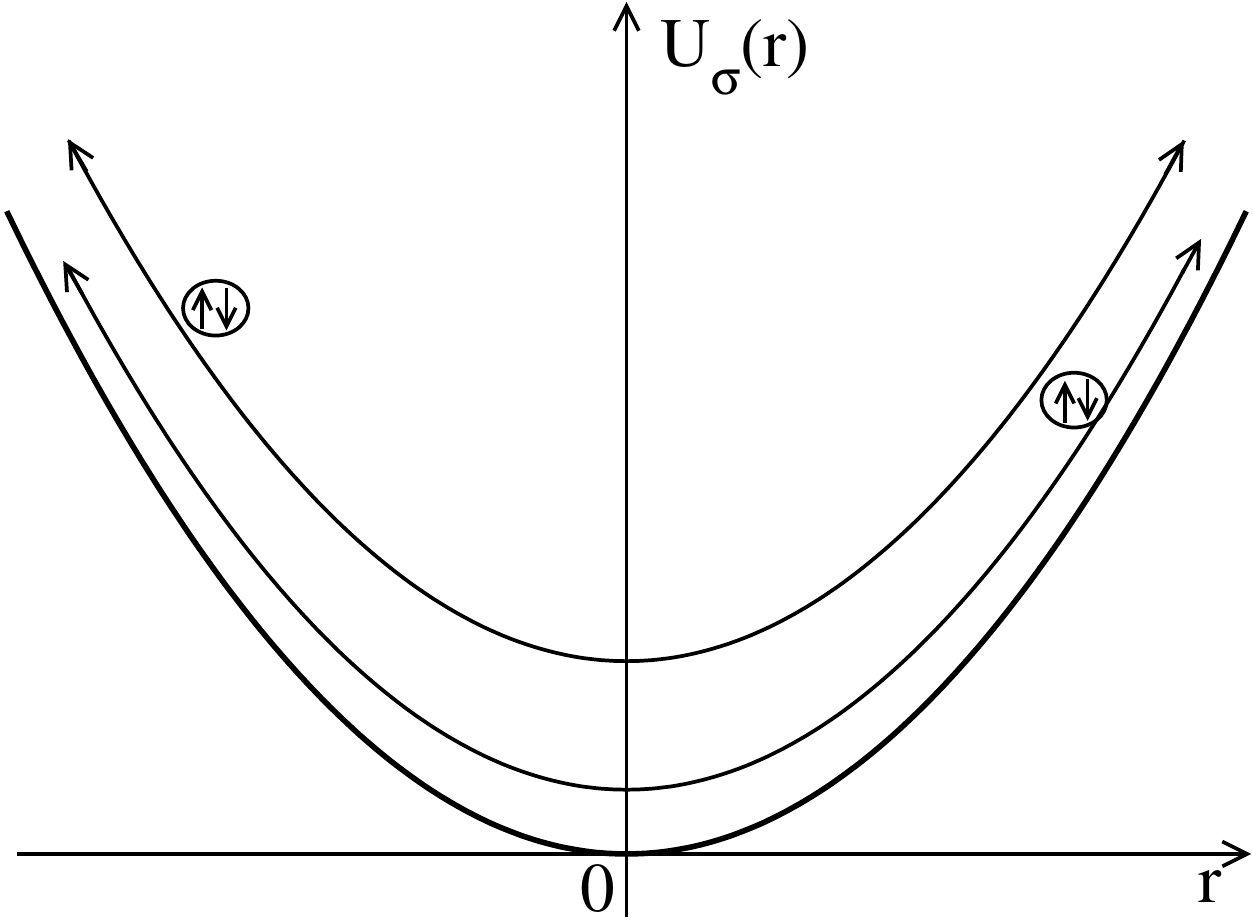}
\caption{Un comportement asymptotique possible du problème unitaire à quatre fermions $(N_\uparrow=2,N_\downarrow=2)$ dans les potentiels harmoniques $U_\sigma(\rr)$:  deux pairons oscillent furieusement (avec des mouvements de grande amplitude); les fermions $\uparrow$ et $\downarrow$ dans chaque pairon ont une énergie relative $O(1)$ (en unités de $\hbar\omega$ si l'on veut respecter la dimension) et restent fortement corrélés; les deux pairons ont une énergie relative $\to +\infty$ et sont découplés, ce qui autorise à les voir comme deux bosons identiques, dotés d'une structure interne (celle du mouvement relatif $\uparrow\downarrow$ au sein d'un pairon) mais n'interagissant pas entre eux. La correction de statistique quantique $\mbox{CorrStat}_{2,2}$ est alors non nulle dans l'équation (\ref{eq1_20}).}
\label{fig3bis}
\end{figure}

Expliquons mieux cette histoire de découplage en sous-amas sur des exemples. Si $(n_\uparrow,n_\downarrow)=(1,1)$, les états propres prennent asymptotiquement - pour des valeurs arbitrairement grandes de l'énergie - la forme de deux fermions $\uparrow$ et $\downarrow$ non corrélés dans des niveaux oscillatoires de grandes amplitudes;\footnote{Si le nombre d'onde relatif $k_{\rm rel}\to+\infty$,  l'amplitude de diffusion $f_{k_{\rm rel}}\to 0$  dans l'équation (\ref{eq2_10}) donc même à la limite unitaire, les interactions deviennent négligeables.}  les sous-amas étant monoatomiques, on a $\mbox{CorrStat}=0$. Si $(n_\uparrow,n_\downarrow)=(2,1)$, un nouveau découplage est possible, à côté de celui en trois fermions décorrélés: les particules peuvent se séparer en un atome $\uparrow$ et un pairon $\uparrow\downarrow$ de fermions fortement corrélés (le mouvement relatif au sein du pairon restant de faible amplitude), avec des niveaux oscillatoires très excités pour l'atome $\uparrow$ et pour le centre de masse du pairon $\uparrow\downarrow$; le pairon étant seul dans sa catégorie,  on a là encore $\mbox{CorrStat}=0$. La conclusion reste la même pour $(n_\uparrow,n_\downarrow)=(3,1)$, si ce n'est qu'apparaît le triplon $\uparrow\uparrow\downarrow$ comme nouveau sous-amas découplé.  En revanche, pour $(n_\uparrow,n_\downarrow)=(2,2)$, il y a découplage possible en deux pairons $\uparrow\downarrow$ de très grande énergie relative, voir la figure \ref{fig3bis}; ils n'interagissent plus mais ils sont indiscernables et conduisent, comme les bosons identiques d'un gaz parfait, à une correction de statistique quantique ignorée par l'intégrale sur $S$ dans l'équation (\ref{eq1_20}); le calcul donne $\mbox{CorrStat}=1/32$ \cite{EndoCastin}.

La conjecture (\ref{eq1_20}) est bien établie pour $N=3$, par application inverse du théorème des résidus, qui convertit la somme sur les spectres (\ref{eq3_17p}) donc sur les racines $s$ de $\Lambda_{n_\uparrow,n_\downarrow}$ en une intégrale \cite{revuecanad}.  Pour $N=4$, les propriétés analytiques de la fonction $\Lambda_{n_\uparrow,n_\downarrow}$ dans le plan complexe ne sont pas suffisamment connues  pour qu'on puisse appliquer le théorème de Cauchy;\footnote{Il faut pouvoir rabattre sur l'axe imaginaire pur le chemin d'intégration entourant les racines et les pôles de $\Lambda_{n_\uparrow,n_\downarrow}$ sur l'axe réel sans croiser de singularité - pôle ou ligne de coupure - dans les demi-plans supérieur et inférieur.} dans le cas particulier $m_\uparrow/m_\downarrow=1$, la conjecture a cependant été confirmée par un calcul très précis de Monte-Carlo quantique à petit nombre de corps \cite{HouDrut} (en revanche, les valeurs expérimentales \cite{ens,mit} ne sont pas confirmées, le problème semblant venir de l'impossibilité d'obtenir le bon polynôme de degré 4 en $z=\exp(\beta\mu)$ par ajustement de la pression $P$ ou de la densité $\rho$ mesurées sur l'intervalle de fugacité accessible expérimentalement \cite{halaug}).\footnote{Les références \cite{ens,mit} ont accès seulement au quatrième coefficient d'amas total $b_4=(b_{4,0}+b_{3,1}+b_{2,2}+b_{1,3}+b_{0,4})/2$, ce qui interdit une comparaison avec la conjecture (\ref{eq1_20}) secteur par secteur.} 

La démonstration de  l'expression (\ref{eq1_20}) dans le cas général reste donc ouverte.  Une autre question intéressante porte sur le comportement des coefficients d'amas $b_{n_\uparrow,n_\downarrow}$ aux grands ordres $n_\sigma\to +\infty$, dont la connaissance est requise si l'on veut effectuer une sommation efficace de la série (\ref{eq1_19}) après calcul de ses premiers termes, pour étendre son applicabilité au régime dégénéré $T\lesssim T_{\rm F}$ (par exemple, si le rayon de convergence est nul, on pourrait mettre en œuvre une resommation de type Borel conforme comme dans les références \cite{felixborel,zinnborel}).

\section{Questions ouvertes dans un point de vue macroscopique}
\label{sec4}

Dans cette section, le gaz de fermions en interaction, considéré dans la limite thermodynamique et à température non nulle mais arbitrairement basse, est décrit par une théorie hamiltonienne effective de basse énergie, l'hydrodynamique quantique de Landau et Khalatnikov \cite{LK}.\footnote{\label{note1_23}L'appellation \g{hydrodynamique quantique}, en physique non linéaire, est comprise par opposition à l'hydrodynamique des fluides classiques et fait référence à une équation d'Euler portant sur un champ de vitesse $\vv(\rr)$ à valeurs réelles plutôt qu'opératorielles, mais contenant un terme de pression quantique $\propto\hbar^2$, ce qui lui permet de décrire le mouvement de tourbillons quantiques - à circulation quantifiée - dans le superfluide (comme l'équation de Gross-Pitayevski sur la fonction d'onde d'un condensat de bosons réécrite en termes de la densité et du gradient de la phase).  Ici, l'appellation est à prendre au sens de la seconde quantification, le champ de vitesse étant désormais à valeur opérateur $\hat{\vv}(\rr)$.}

\subsection{Vue d'ensemble du régime superfluide considéré}
\label{sec4.1}

Le système tridimensionnel de fermions est ici spatialement homogène (dans un volume de quantification $[0,L]^3$ proche de la limite thermodynamique, cette limite étant prise à la fin des calculs), avec des particules de masses égales $m_\uparrow=m_\downarrow=m$ dans les deux états internes, non polarisé en spin (il y a le même nombre de particules dans les deux composantes, $N_\uparrow=N_\downarrow$, afin de permettre un appariement complet) et à l'équilibre thermique canonique dans une limite de basse température ($T\neq 0$ mais $T\to 0$).  

Dans ces conditions, (i) les fermions s'assemblent en paires liées $\uparrow\downarrow$ dans l'onde $s$;  en présence des mers de Fermi dans les deux états internes, c'est ce à quoi conduit l'interaction attractive entre $\uparrow$ et $\downarrow$ de la section \ref{sec3}, au travers du célèbre mécanisme de Cooper;  ceci vaut donc même pour une longueur de diffusion $a$ négative, où il n'y a pas d'état lié $\uparrow\downarrow$ dans l'espace libre, la taille d'une paire tendant toutefois vers $+\infty$ lorsque $a\to 0^-$ (dans le cas $a>0$, il existe bien un état  dimère et, sans surprise, c'est à lui que se réduit l'état de paire liée dans la limite de basse densité $\rho\to 0$);\footnote{Il n'est pas complètement évident de voir que l'interaction de contact de Wigner-Bethe-Peierls est attractive.  Une façon de faire est de l'obtenir comme la limite continue $b\to 0$ d'un modèle sur un réseau cubique $b\mathbb{Z}^3$ avec un couplage $\propto \hbar^2/mb^2$ entre sites voisins (pour représenter l'énergie cinétique)  et une interaction sur site $g_0/b^3$; à longueur de diffusion $a\neq 0$ fixée, on trouve alors que $g_0\approx -\hbar^2b/m<0$ lorsque $b\to 0$ (la constante de couplage nue $g_0$ est donc, dans le régime $b\ll|a|$ de la diffusion résonnante, fort différente de la constante de couplage effective $g=4\pi\hbar^2a/m$) \cite{varenna}.} (ii) ces paires liées, étant des sortes de bosons composites, forment un condensat dans le mode de vecteur d'onde $\KK_{\rm paire}=\zero$ de leur centre de masse (de longueur de cohérence limitée par la taille de la boîte, infinie à la limite thermodynamique) et un superfluide. 

\begin{figure}[tb]
\includegraphics[width=6cm,clip=]{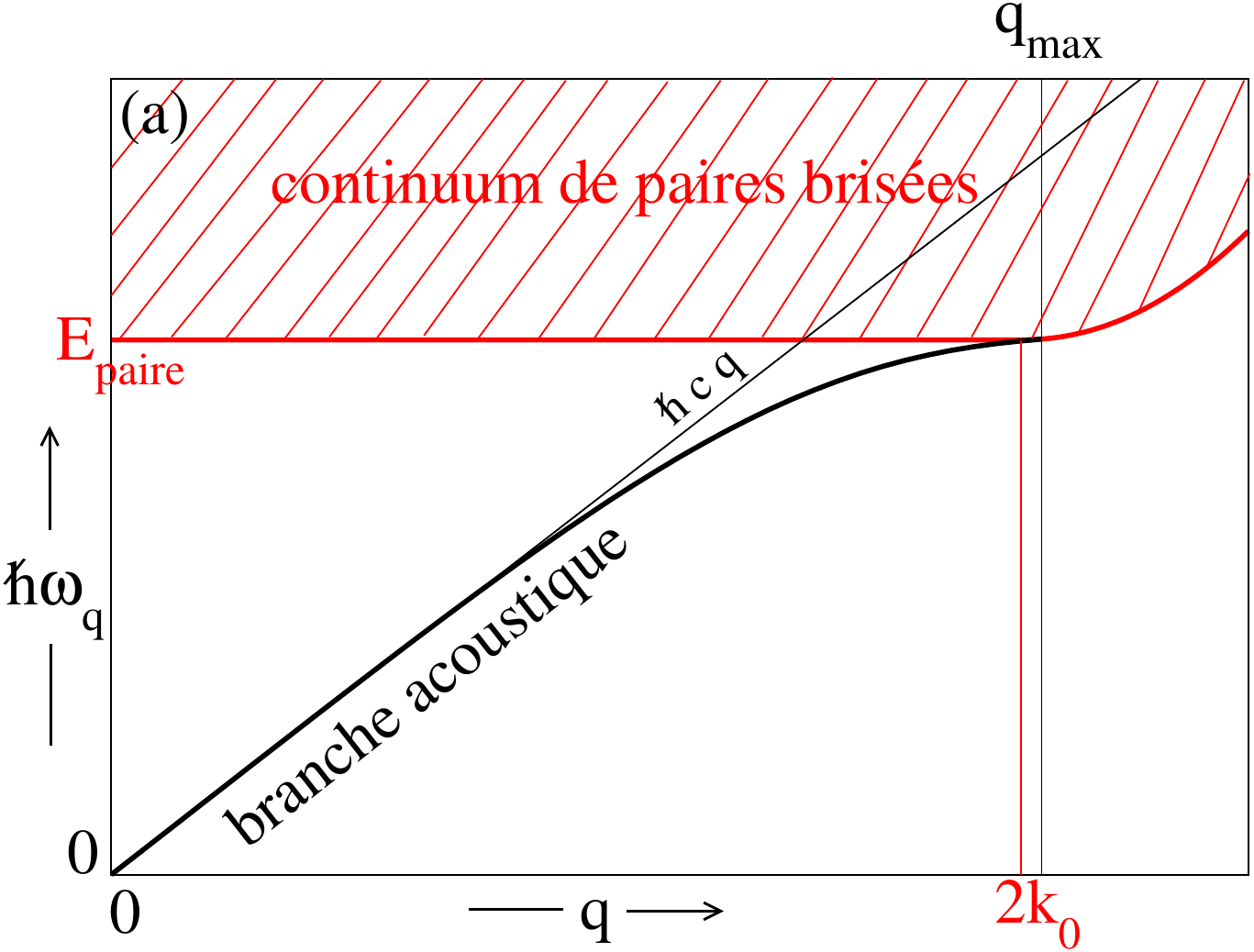}\hspace{1cm}\includegraphics[width=6cm,clip=]{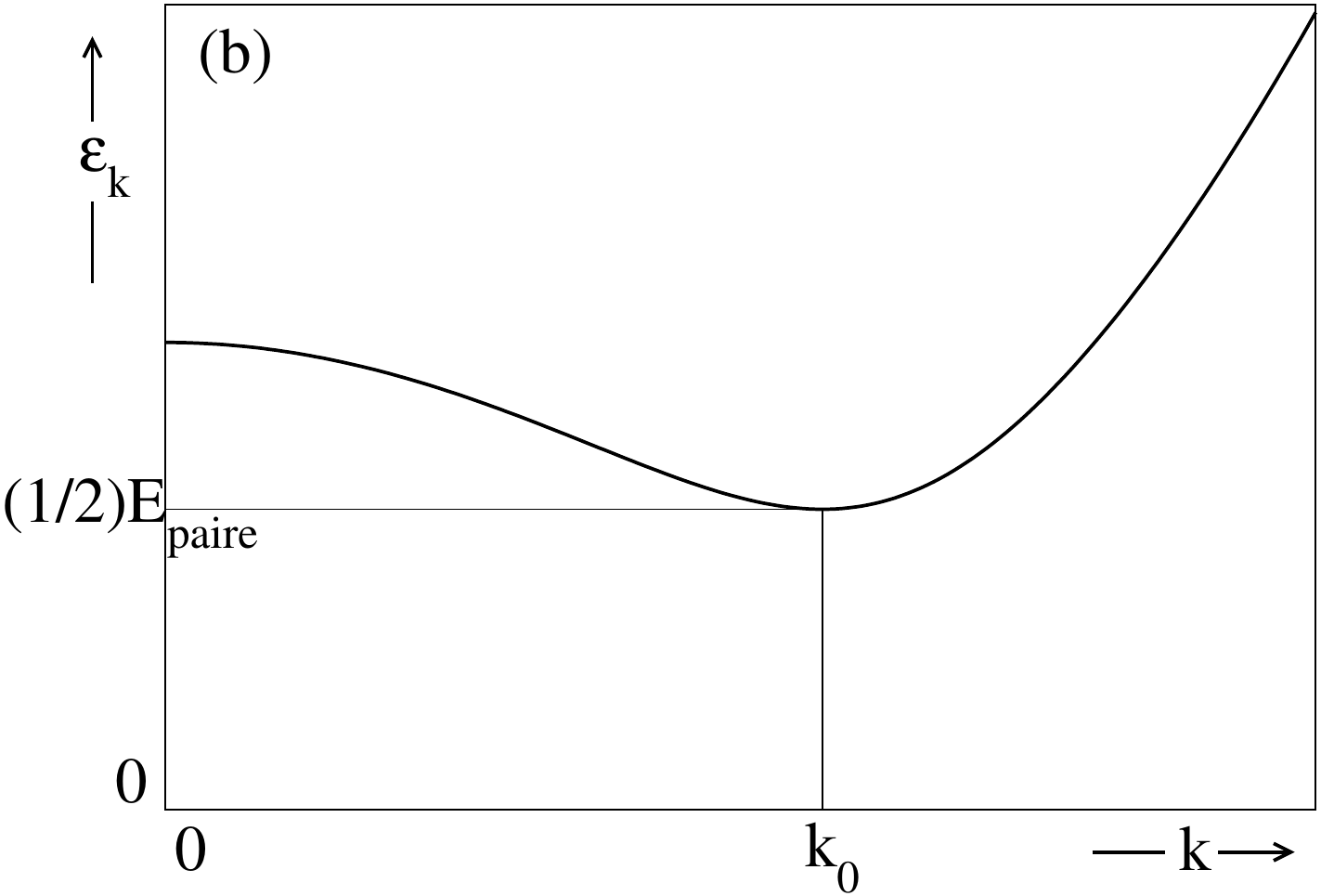}
\caption{Énergie de différents types d'excitations d'un gaz homogène non polarisé de fermions à deux composantes à température nulle, en fonction de leur nombre d'onde $q$ ou $k$. (a) Branche d'excitation acoustique $\hbar\omega_\qq$ de départ linéaire $\hbar c q$ ($c$ est la vitesse du son), limitée supérieurement par le continuum de paire brisée (zone hachurée) dont elle vient ici tangenter le bord inférieur en le point terminal de nombre d'onde $q_{\rm max}$. (b) Relation de dispersion $\veps_\kk$ d'une quasi-particule fermionique $\chi$ (voir texte). Sous l'effet d'une excitation percussionnelle de vecteur d'onde $\qq$, une paire liée $\uparrow\downarrow$ du condensat de paires, initialement au repos, se dissocie en deux quasi-particules fermionique $\chi$ de spins opposés, de vecteurs d'onde $\kk$ et $\kk'=\qq-\kk$ et d'énergies $\veps_\kk$ et $\veps_{\kk'}$; comme le vecteur $\kk$ n'est pas contraint (pas de conservation de l'énergie pour une excitation percussionnelle), il apparaît un continuum d'énergies finales $\{\veps_\kk+\veps_{\qq-\kk},\kk\in\mathbb{R}^3\}$. Pour la force des interactions choisie sur la figure ($|\Delta|/\mu=0,\!84$ soit $1/k_{\rm F}a\simeq -0,\!16$ d'après la théorie BCS, avec $\mu$ le potentiel chimique du gaz et $\Delta$ le paramètre d'ordre complexe du condensat de paires), la branche acoustique est de départ concave ($\gamma<0$ dans l'équation (\ref{eq1_24})) et la relation de dispersion $\veps_\kk$ présente un minimum $E_{\rm paire}/2$ en $k=k_0>0$; le bord inférieur du continuum vaut donc exactement $E_{\rm paire}$, du moins tant qu'on peut avoir $k=k'=k_0$ donc tant que $q=|\kk+\kk'|\leq 2 k_0$. Les relations de dispersion représentées, approchées, sont celles de la théorie BCS pour $\veps_\kk$ et de la RPA d'Anderson pour $\hbar\omega_\qq$. Selon la force des interactions, le domaine d'existence en $q$ de la branche acoustique peut également être non compact, connexe $q\in\,]0,+\infty[$ ou pas $q\in\,]0,q_{\rm max}[\bigcup]q_{\rm min},+\infty[$ ($q_{\rm min}> q_{\rm max}$) \cite{CKS}; la concavité de la branche est elle aussi variable \cite{PRAconcav}; enfin, $k_0=0$ et $E_{\rm paire}=2(\mu^2+|\Delta|^2)^{1/2}$ si $\mu<0$, $k_0=(2m\mu)^{1/2}/\hbar$ et $E_{\rm paire}=2|\Delta|$ sinon.}
\label{fig4}
\end{figure}

Du coup, on s'attend à ce que le système admette, dans son état fondamental, une branche d'excitation acoustique (par ondes sonores) de départ linéaire en le nombre d'onde $q$ avec une correction cubique, 
\be
\label{eq1_24}
\omega_\qq\underset{q\to 0}{=} cq\left[1+\frac{\gamma}{8}\left(\frac{\hbar q}{mc}\right)^2+O(q^4\ln q)\right]
\ee
et que cette branche soit limitée supérieurement en énergie par le continuum de paire brisée, de bord inférieur l'énergie de liaison d'une paire $E_{\rm paire}$, comme sur la figure \ref{fig4}a.  Ici $\omega_\qq$ est la pulsation propre au vecteur d'onde $\qq$, $c$ est la vitesse du son à température nulle et le paramètre de courbure $\gamma$ est adimensionné de façon qu'il vaille un dans la limite $k_{\rm F}a\to 0^+$ d'un gaz condensé de dimères en interaction faible  (en accord avec la théorie de  Bogolioubov).\footnote{Notre convention de signe sur $\gamma$ diffère de celle utilisée dans l'hélium 4 liquide, voir la référence \cite{Maris}.} La branche en question est souvent dite de Goldstone \cite{Goldstone}, parce qu'on l'associe à la brisure de symétrie $U(1)$ dans la condensation de paires; son pendant de Higgs est discuté dans la section \ref{sec5}.

Notre régime de basse température satisfait dans la suite aux deux conditions
\be
\label{eq2_24}
0< k_{\rm B}T\ll mc^2 \quad \mbox{et}\quad 0<k_{\rm B}T\ll E_{\rm paire}
\ee
 la première assurant que soit peuplée thermiquement seulement la partie linéaire de la branche acoustique, la seconde qu'il y ait une densité négligeable de paires brisées (d'après la loi de Boltzmann, cette densité comporte un facteur d'activation $\exp(-E_{\rm paire}/2 k_{\rm B}T)$, les fragments issus de la dissociation d'une paire liée - les quasi-particules fermioniques $\chi$ de la figure \ref{fig4}b - ayant individuellement une énergie minimale $E_{\rm paire}/2$).  Notre système se réduit alors à un gaz thermique de phonons, si l'on convient d'appeler comme tels les quanta de la branche acoustique (eu égard à son départ linéaire).\footnote{L'équation (\ref{eq1_24}) ne vaut que pour une interaction $V(\rr_{ij})$ à courte portée, décroissant assez vite lorsque $r_{ij}\to +\infty$. Dans le cas d'une interaction dipolaire, comme dans les gaz d'atomes froids magnétiques, la vitesse du son est anisotrope \cite{Axel}, voir aussi la contribution de Wilhelm Zwerger à ce dossier thématique \cite{Zwergerdt} et le cours 2023-2024 de Jean Dalibard au Collège de France \cite{jd}.  Dans le cas d'une interaction coulombienne, comme dans les gaz d'électrons supraconducteurs, la branche acoustique fait place à une branche de plasmons à bande interdite (la pulsation propre $\omega_\qq$ a une limite $>0$ en $q=0$) \cite{Anderson}.  Ici, nos atomes sont neutres et de moment dipolaire négligeable.}

On peut alors se poser trois types de questions, partiellement ouvertes:
\begin{enumerate}
\item nous allons le voir, les phonons (abrégés en $\phi$) interagissent entre eux, le superfluide de fermions sous-jacent constituant un milieu non linéaire pour le son.  Quels sont les effets de ces interactions sur les phonons de vecteur d'onde $\qq$? On s'attend en particulier à ce qu'ils s'amortissent  avec un taux $\Gamma_\qq(T)$ et qu'ils subissent un déplacement de pulsation thermique $\Delta_\qq(T)$ (on ne compte pas le déplacement à température nulle, qui donne naissance par définition au spectre (\ref{eq1_24}) - le terme négligé en $q^5\ln q$ provient précisément de l'effet croisé des interactions et des fluctuations quantiques du champ de phonons \cite{espagnols,insuffisance}).\footnote{Dans le cas convexe $\gamma>0$,  il s'accompagne - toujours à $T=0$ - d'une partie imaginaire non nulle $\approx q^5$, correspondant au mécanisme d'amortissement de Belyaev $\qq\to\kk,\kk'$,  voir plus loin.}
\item quelles sont les conséquences de la dynamique collisionnelle des phonons sur l'évolution d'une variable macroscopique  du gaz particulièrement intéressante, l'opérateur phase $\hat{\phi}_0(t)$ du condensat de paires?

\item on enrichit le problème en considérant le cas partiellement polarisé $N_\uparrow\neq N_\downarrow$.  Dans le cas faiblement polarisé, par exemple $N_\uparrow-N_\downarrow=O(1)$, les fermions non appariés (car surnuméraires) de la composante de spin majoritaire forment, dans le gaz en interaction, des quasi-particules fermioniques (abrégées en $\chi$) d'une relation de dispersion $\veps_\kk$ différente de celles des fermions libres: elle présente  notamment un minimum non nul, donc une bande d'énergie interdite, donné par la demi-énergie de liaison d'une paire $\uparrow\downarrow$, au voisinage duquel elle varie quadratiquement en nombre d'onde $k$ (voir la figure \ref{fig4}b).  Se pose alors la question de leurs interactions $\phi-\chi$ avec les phonons et $\chi-\chi$ entre elles; en particulier, un désaccord persiste sur l'expression de l'amplitude de diffusion $\phi-\chi$ à basse énergie (les références \cite{cras} et \cite{italiens} diffèrent).\footnote{Ce problème intéresse également l'hélium 4 superfluide, dont la branche d'excitation admet elle aussi un minimum quadratique, le minimum de roton; notre problème de diffusion $\phi-\chi$ est donc formellement équivalent à la diffusion roton-phonon étudiée déjà dans la référence \cite{LK}, à la différence près que les rotons sont des bosons. Les prédictions de \cite{LK} sont cependant incomplètes et en désaccord avec \cite{cras,italiens}.} Dans le cas fortement polarisé, où $N_\uparrow-N_\downarrow$ est extensif comme $N$,  on s'attend à ce que la condensation des paires liées à $T=0$ puisse se faire dans une superposition d'ondes planes de leur centre de masse (plutôt que dans $\KK_{\rm paire}=\zero$ comme supposé ici), voir les références \cite{ff,lo}, ce qui donne naissance à un superfluide modulé spatialement (un supersolide suivant la terminologie à la mode), sans que son domaine d'existence dans l'espace des paramètres $((N_\uparrow-N_\downarrow)/N,1/k_{\rm F}a,T/T_{\rm F})$ soit parfaitement bien connu théoriquement, le problème étant complexifié par sa grande sensibilité aux fluctuations thermiques \cite{Leo} (il n'y a pas encore de résultats expérimentaux dans les gaz d'atomes froids tridimensionnels \cite{Hulet}). 
\end{enumerate}

Le cas le plus intéressant pour les points 1 et 2 est celui d'une branche acoustique de départ concave, $\gamma<0$ dans l'équation (\ref{eq1_24}), très différent du gaz de bosons en interaction faible assez bien connu (où $\gamma\simeq 1>0$ comme nous l'avons dit).\footnote{Qualitativement, ce cas $\gamma<0$ s'obtient lorsque l'énergie de liaison $E_{\rm paire}$ est assez faible: si l'on réduit $E_{\rm paire}$, le continuum de paire brisée sur la figure \ref{fig4}a s'abaisse, pousse sur la branche acoustique et finit par la faire se courber vers le bas.  C'est certainement le cas dans la limite BCS $k_{\rm F}a\to 0^-$ où $E_{\rm paire}/mc^2=O(\exp(-\pi/2k_{\rm F}|a|))$ tend rapidement vers zéro; ce n'est plus le cas dans la limite CBE $k_{\rm F}a\to 0^+$ où $E_{\rm paire}\sim E_{\rm dim}=\hbar^2/ma^2\gg mc^2$. On ne sait pas avec certitude de quel côté de la limite unitaire $1/k_{\rm F}a=0$ (c'est-à-dire pour quel signe de la longueur de diffusion $a$) se produit le changement de signe du paramètre de courbure $\gamma$, voir la section \ref{sec5}. Vu la forme du bord inférieur du continuum sur la figure \ref{fig4}a - et ceci indépendamment du signe de $\gamma$, l'effet de répulsion sur la branche acoustique est le plus fort à grand $q$ mais le plus faible à petit $q$ (là où la différence d'énergie entre le bord et la branche est le plus grand). On s'attend donc à avoir un intervalle de valeurs de $1/k_{\rm F}a$ sur lequel la branche, convexe à faible $q$, est concave à grand $q$ \cite{PRAconcav}.} En particulier, l'amortissement des phonons pour $\gamma<0$ peut se produire seulement à $T\neq 0$, puisque la désintégration d'un phonon en un nombre quelconque $n>1$ de phonons est interdite par la conservation de l'énergie-impulsion pour une branche acoustique concave; à l'ordre dominant en température, il résulte pour la même raison, non pas comme pour $\gamma>0$ de processus à trois phonons de type Belyaev $\phi\to\phi\phi$ ou Landau $\phi\phi\to\phi$ \cite{Belyaev,Martin}, mais des processus à quatre phonons $\phi\phi\to\phi\phi$ de Landau et Khalatnikov \cite{LK}. À notre connaissance, cet amortissement à quatre phonons n'a encore été observé expérimentalement dans aucun système.  Il pourrait l'être dans un gaz d'atomes froids fermioniques dans une boîte de potentiel \cite{epl}.  Il pourrait l'être aussi dans l'hélium 4 superfluide (un liquide de bosons) si l'on augmente suffisamment la pression pour rendre $\gamma<0$ (le minimum de roton s'abaisse ce qui finit par rendre concave le départ de la branche acoustique) et si l'on abaisse suffisamment la température pour réduire la densité de rotons (au travers du facteur d'activation $\exp(-E_{\rm roton}/k_{\rm B}T)$) et rendre négligeable l'amortissement parasite des phonons par les rotons \cite{prlrevi}.\footnote{Dans l'hélium 4 liquide, les processus de diffusion à quatre phonons entre des faisceaux de phonons produits intentionnellement (non thermiques) ont déjà fait l'objet d'études théoriques et expérimentales \cite{Adamenko}.}

\subsection{Quelle théorie macroscopique utiliser?}
\label{sec4.2}

Une théorie effective de basse énergie renonce à décrire le système en dessous d'une certaine échelle de longueur $\ell$; on s'attend en revanche à ce que la théorie soit exacte aux grandes longueurs d'onde, ici à l'ordre dominant en température. 

\begin{figure}[t]
\includegraphics[width=4cm,clip=]{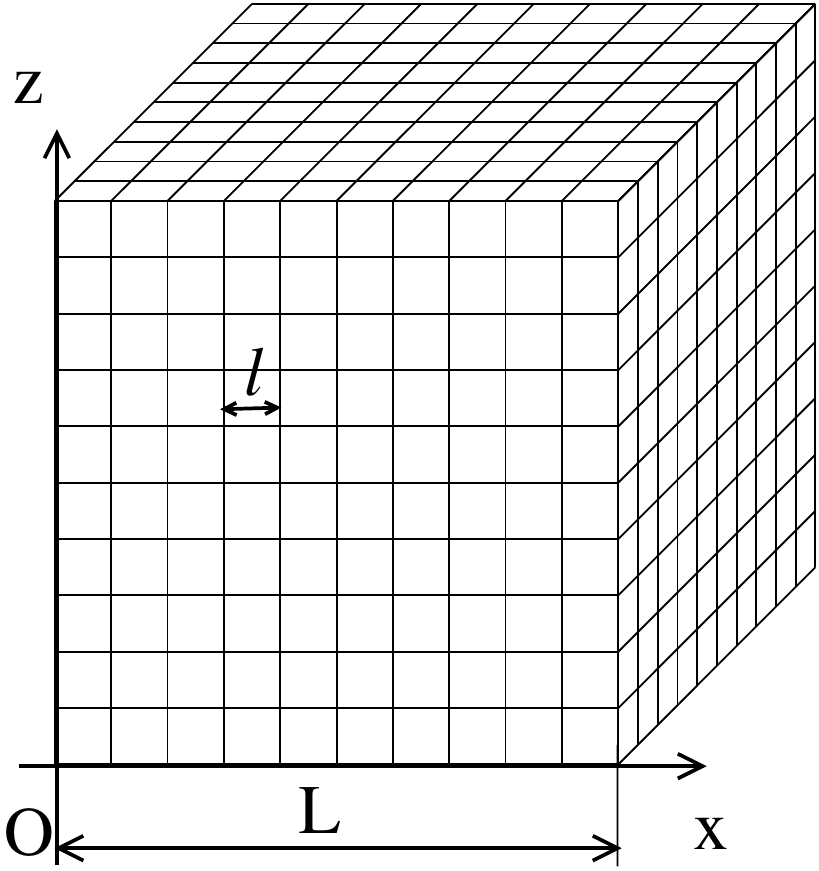}
\caption{Découpage du gaz en portions cubiques mésoscopiques de côté $\ell$, dans l'hydrodynamique quantique de Landau et Khalatnikov (cette théorie effective ne décrit pas les échelles de longueur $<\ell$).  Voir texte pour le choix de $\ell$. }
\label{fig5}
\end{figure}

 Dans ces conditions, il est légitime de découper le gaz en portions de taille $\ell$, par exemple en boîtes cubiques de côté $\ell$ centrées sur le réseau cubique $\ell\mathbb{Z}^3$, voir la figure \ref{fig5}.  Énonçons quelques contraintes sur le choix de $\ell$:
\begin{enumerate}
\item  on doit avoir $\ell\gg\xi$ (ici $\xi=\hbar/mc$ est la longueur dite de relaxation ou de corrélation du superfluide) et $\rho\ell^3\gg 1$ (il y a un grand nombre de fermions par site)  de façon que (i) chaque portion cubique puisse être considérée comme mésoscopique et relever de la notion d'équation d'état liant de manière univoque la pression ou le potentiel chimique à la densité (comme c'est le cas à la limite thermodynamique), et (ii) le pas $\ell$ du réseau fournisse une coupure en nombre d'onde $\pi/\ell\ll mc/\hbar$ aux excitations phononiques du gaz, les restreignant ainsi à la partie quasi linéaire de la branche (\ref{eq1_24}), partie universelle car décrite par deux paramètres, $c$ et $\gamma$.
\item on doit avoir $k_{\rm B}T\ll \hbar c \pi/\ell$ (c'est l'énergie du mode de phonons fondamental dans une portion) de façon qu'on puisse considérer que (i)  chaque portion est à température nulle, et (ii) chaque portion est spatialement homogène à l'échelle de la longueur d'onde typique $q_{\rm th}^{-1}=\hbar c/k_{\rm B}T$ des ondes sonores thermiques.
\item il faut aussi que $\ell\ll\ell_{\rm coh}$ où $\ell_{\rm coh}$ est la longueur de cohérence des paires de fermions, de façon que la notion de phase globale $\hat{\phi}$ ait un sens dans chaque portion (comme pour un condensat).  Cette contrainte est inopérante ici puisque $\ell_{\rm coh}\approx L$ (les paires liées sont condensées à 3D). 
\end{enumerate}

On admet alors qu'on peut représenter le système par deux opérateurs de champ, le champ de densité $\hat{\rho}(\rr)$ et le champ de phase $\hat{\phi}(\rr')$, avec $\rr,\rr'\in\ell\mathbb{Z}^3$; ces variables sont canoniquement conjuguées au sens où
\be
\label{eq1_28}
[\hat{\rho}(\rr)\ell^3,\hat{\phi}(\rr')]=\ii\,\delta_{\rr,\rr'}
\ee
comme si $\hat{\phi}$ était un opérateur impulsion et $\hat{\rho}\ell^3$ un opérateur position en mécanique quantique ordinaire \cite{LK,tome9landau}.\footnote{Ces références historiques se placent pour simplifier dans un espace continu.  La nécessité de discrétiser l'espace pour éviter les infinis et rendre la théorie renormalisable est souligné dans la publication \cite{brouilfer}.  Ici, nous mettons ces difficultés sous le tapis; par exemple, nous ne distinguons pas dans (\ref{eq1_29}) entre la notion d'équation d'état nue $e_{0,0}(\rho)$ - celle qui entre dans l'hamiltonien -  et l'équation d'état vraie ou effective $e_0(\rho)$ -  celle qu'on observe dans une expérience. }  Le champ de phase donne accès au champ de vitesse par simple différentiation (il s'agit ici d'un gradient discret):\footnote{\label{note_passuper}Le fait que l'opérateur champ de vitesse soit un vecteur gradient ne signifie nullement que l'écoulement soit entièrement superfluide (ceci serait d'ailleurs physiquement faux même à l'équilibre thermique, à température non nulle). Expliquons-le en deux remarques. (i) On ne confondra pas l'opérateur $\hat{\vv}(\rr)$ de l'hydrodynamique quantique (qui contient toutes les fluctuations quantiques et thermiques possibles) avec le champ de vitesse moyen $\vv(\rr)$ de l'hydrodynamique ordinaire; en particulier, le caractère superfluide ou pas de l'écoulement dépend du caractère irrotationnel ou pas de $\vv(\rr)$ (sans chapeau). (ii) En général, on a $\vv(\rr)\neq\langle\hat{\vv}(\rr)\rangle$ où la moyenne est prise dans l'état quantique du système, car $\vv(\rr)$ est défini en termes de la densité moyenne de courant de matière, $\vv(\rr)=\langle\hat{\mathbf{j}}(\rr)\rangle/\langle\hat{\rho}(\rr)\rangle$ avec ici $\mathbf{\hat{j}}(\rr)=[\hat{\rho}(\rr)\hat{\vv}(\rr)+\hat{\vv}(\rr)\hat{\rho}(\rr)]/2$ (par définition, l'équation d'évolution de $\hat{\rho}$ en point de vue de Heisenberg s'écrit $\partial_t\hat{\rho}+\mathrm{div}\,\hat{\mathbf{j}}=0$ et celle de la densité moyenne $\rho(\rr)=\langle\hat{\rho}(\rr)\rangle$ s'écrit $\partial_t\rho+\mathrm{div}\,(\rho \vv)=0$); il serait donc faux de croire que $\vv(\rr)=(\hbar/m)\mathbf{grad}\langle[\hat{\phi}(\rr)-\hat{\phi}(\rr_0)]\rangle$ (où $\rr_0$ est une position de référence arbitraire) et d'en déduire que $\vv(\rr)$ est forcément un vecteur gradient.}
\be
\label{eq2_28}
\hat{\vv}(\rr)=\frac{\hbar}{m}\mathbf{grad}\,\hat{\phi}(\rr)
\ee
L' hamiltonien s'obtient en sommant l'énergie interne et l'énergie cinétique associée à la vitesse locale d'écoulement du fluide dans chaque portion:
\be
\label{eq1_29}
H=\sum_\rr \frac{1}{2} m\hat{\vv}(\rr)\cdot \hat{\rho}(\rr)\ell^3 \hat{\vv}(\rr)+\ell^3 e_0(\hat{\rho}(\rr))
\ee
En effet, $e_0(\rho)$ est ici l'énergie volumique à température nulle du gaz homogène de fermions de densité $\rho$, et $m\hat{\rho}(\rr)\ell^3$  est la quantité de matière (la masse) dans la portion centrée en $\rr$.  Les équations du mouvement sur $\hat{\rho}$ et $\hat{\vv}$ en point de vue de Heisenberg dérivant de l'hamiltonien $H$ prennent la forme d'une équation de continuité et d'une équation d'Euler (sans terme de viscosité) à valeur opérateur,\footnote{Voir la note \ref{note_passuper} pour l'équation sur $\hat{\rho}(\rr)$. L'équation sur $\hat{\vv}(\rr)$ s'obtient en prenant le gradient de celle sur $\hat{\phi}(\rr)$, $\hbar\partial_t\hat{\phi}=-\mu_0(\hat{\rho})-m\hat{\vv}^2/2$ où $\mu_0(\rho)$ est la fonction potentiel chimique à température nulle comme dans l'équation (\ref{eq2_30}).} d'où le nom d'hydrodynamique quantique donné à la théorie (avec le risque de confusion signalé dans la note \ref{note1_23}). 

On l'aura compris, la grande force de cette théorie effective est qu'elle ne dépend pas de la nature des particules bosoniques ou fermioniques constituant le superfluide sous-jacent, ni de leurs interactions  (fortes ou faibles, en phase liquide ou gazeuse) pourvu qu'elles restent à courte portée, si ce n'est au travers de l'équation d'état $e_0(\rho)$ et du paramètre de courbure $\gamma$ à température nulle. Elle s'applique donc également bien aux gaz de bosons en interaction faible, aux gaz de fermions en interaction forte et à l'hélium 4 liquide (système pourtant extrêmement dense, défiant toute théorie microscopique). 

Il ne faut cependant croire au formalisme que dans une limite de basse température, $T\to 0$, où les fluctuations spatiales de densité sont faibles et les gradients de phase aussi; on doit donc développer l'équation (\ref{eq1_29}) jusqu'à l'ordre pertinent (ici l'ordre quatre) en puissances de
\be
\label{eq2_29}
\delta\hat{\rho}(\rr)\equiv\hat{\rho}(\rr)-\hat{\rho}_0 \quad \mbox{et}\quad \delta\hat{\phi}(\rr)\equiv\hat{\phi}(\rr)-\hat{\phi}_0
\ee
où $\hat{\rho}_0$ et $\hat{\phi}_0$ sont les composantes de Fourier des champs $\hat{\rho}(\rr)$ et $\hat{\phi}(\rr)$ de vecteur d'onde nul (physiquement, $\hat{\rho}_0=\hat{N}/L^3$ où $\hat{N}$ est l'opérateur nombre total de fermions, $L^3$ est le volume de la boîte de quantification $[0,L]^3$ et $\hat{\phi}_0$  est l'opérateur phase du condensat de paires liées \cite{brouilfer}). L'hamiltonien développé s'écrit formellement
\be
\label{eq3_29}
H=H_0+H_2+H_3+H_4+\ldots
\ee
où $H_n$ est la contribution de degré total $n$ en $\delta\hat{\rho}$ et $\delta\hat{\phi}$.

La contribution d'ordre 0 $H_0$ est une constante sans grand intérêt;  celle d'ordre 1 est exactement nulle (car $\sum_\rr\ell^3\delta\hat{\rho}(\rr)=0$ par construction) et a été directement omise dans l'équation (\ref{eq3_29}).  La contribution quadratique $H_2$ se diagonalise par transformation de Bogolioubov:\footnote{\label{note32} Ladite transformation correspond aux développements modaux $\delta\hat{\rho}(\rr)=L^{-3/2}\sum_{\kk\neq\zero} \rho_\kk (\hat{b}_\kk+\hat{b}_{-\kk}^\dagger) \exp(\ii\kk\cdot\rr)$ et $\delta\hat{\phi}(\rr)=L^{-3/2}\sum_{\kk\neq\zero}\phi_\kk (\hat{b}_\kk-\hat{b}_{-\kk}^\dagger)\exp(\ii\kk\cdot\rr)$ où $\rho_{\kk}=(\hbar\rho k/2 m c)^{1/2}$ et $\phi_\kk=(-\ii)(mc/2\hbar\rho k)^{1/2}$ sont les amplitudes des fluctuations quantiques de densité et de phase dans le mode de phonons de vecteur d'onde $\kk$. On notera la relation $-\ii\omega_\kk\delta\rho_\kk-\rho(\hbar/m)k^2\phi_\kk=0$ imposée par l'équation de continuité linéarisée.}
\be
\label{eq4_29}
H_2=\mbox{cte}+\sum_{\kk\neq\zero} \hbar\omega_\kk \hat{b}_\kk^\dagger\hat{b}_\kk
\ee
où les opérateurs de création $\hat{b}_\kk^\dagger$ et d'annihilation $\hat{b}_\kk$ d'une excitation élémentaire (un phonon) de vecteur d'onde $\kk$  obéissent aux habituelles relations de commutation bosoniques
\be
\label{eq1_30}
[\hat{b}_\kk,\hat{b}_{\kk'}]=0 \enspace\mbox{et}\enspace[\hat{b}_\kk,\hat{b}_{\kk'}^\dagger]=\delta_{\kk,\kk'}
\ee
Le spectre obtenu est ici exactement linéaire, $\omega_\kk=c k$, avec la vitesse du son donnée par
\be
\label{eq2_30}
mc^2=\rho\frac{\dd^2}{\dd\rho^2} e_0(\rho) = \rho \frac{\dd}{\dd\rho}\mu_0(\rho)
\ee
où $\mu_0(\rho)$ est le potentiel chimique à température nulle du gaz de fermions de densité $\rho$.  La relation (\ref{eq2_30}) est exacte (c'est celle bien connue de l'hydrodynamique des superfluides \cite{tome9landau}) mais l'absence  systématique de courbure dans le spectre n'est pas réaliste physiquement: cette pathologie vient du fait que nous avons omis dans l'hamiltonien $H$ des corrections dites de gradient \cite{SonWingate}; pour simplifier, comme l'ont fait d'illustres prédécesseurs \cite{LK}, nous remplaçons ici $\omega_\kk$ dans $H_2$  à la main par son approximation cubique (\ref{eq1_24}), ce que justifie d'ailleurs la référence \cite{Annalen}. 

L'approximation $H_2$ correspond à un gaz parfait de phonons, et ne peut décrire l'atténuation du son.  L'interaction entre phonons à l'origine de leur amortissement provient des contributions cubique $H_3$ et quartique $H_4$.  Pour alléger, nous donnons ici seulement l'expression de la partie la plus utile de $H_3$, et encore en la simplifiant, pour bien faire comprendre la physique:
\be
\label{eq3_30}
H_3|_{\rm simpl} = \frac{\mathcal{A}}{2L^{3/2}}\sum_{\kk,\kk',\qq} (k k' q)^{1/2} \hat{b}_\kk^\dagger\hat{b}_{\kk'}^\dagger\hat{b}_\qq \delta_{\kk+\kk',\qq} + (k' k q)^{1/2} \hat{b}_{\kk'}^\dagger\hat{b}_\kk \hat{b}_\qq \delta_{\kk',\kk+\qq}+\ldots
\ee
avec l'amplitude constante (indépendante des nombres d'onde) mise en facteur,
\be
\label{eq4_30}
\mathcal{A}=(\xi/2)^{3/2}\rho^{-1/2}\left[3 mc^2+\rho^2 \frac{\dd^3}{\dd\rho^3}e_0(\rho)\right]
\ee
et $\xi=\hbar/mc$  comme précédemment.\footnote{La véritable amplitude de couplage dépend des angles entre les trois vecteurs d'onde $\kk_i$ mis en jeu;  comme l'amortissement est en réalité dominé à basse température par des processus aux petits angles entre  les vecteurs d'onde, à cause de l'effet petit dénominateur décrit plus bas, nous l'avons écrite directement à angles nuls $\kk_i\cdot\kk_j/k_ik_j=1$.} L'ellipse dans l'équation (\ref{eq3_30}) contient des termes en $\hat{b}^\dagger\hat{b}^\dagger\hat{b}^\dagger$ et $\hat{b}\hat{b}\hat{b}$ sans grande importance car ils ne conservent pas l'énergie $H_2$.  Les contributions en $\hat{b}^\dagger\hat{b}^\dagger\hat{b}$ et en $\hat{b}^\dagger\hat{b}\hat{b}$ sont en revanche centrales dans l'amortissement: elles correspondent respectivement  aux processus de Belyaev (le phonon $\qq$ se désintègre en deux phonons $\kk$ et $\kk'$) et de Landau (le phonon $\qq$ fusionne avec un phonon $\kk$ en un seul phonon $\kk'$), dont on peut donner la représentation diagrammatique suivante en faisant jouer un rôle privilégié au phonon dont on  étudie l'amortissement:
\be
\label{eq1_31}
\begin{tabular}{c}
\includegraphics[width=5cm,clip=]{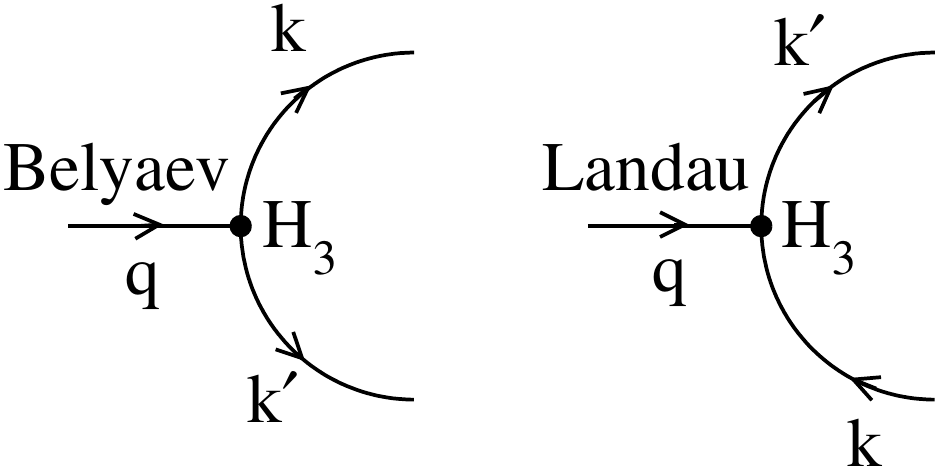}
\end{tabular}
\ee
 Comme $H_3$ est cubique, chaque sommet du diagramme représentant son action est le point de rencontre de trois lignes de phonons. On pourrait procéder de même avec $H_4$ (les sommets seraient alors à quatre lignes) mais nous ne le ferons pas car les processus quartiques jouent en général un rôle sous-dominant dans l'amortissement par rapport aux processus cubiques (par exemple, dans l'amortissement à quatre phonons $\phi\phi\to\phi\phi$ pour $\gamma<0$, l'amplitude du processus direct $\kk,\qq\to\kk',\kk''$ induit par $H_4$ au premier ordre de la théorie des perturbations est en pratique négligeable devant celle du processus indirect de même état initial et final $\kk,\qq\to\kk+\qq\to\kk',\kk''$ induit par $H_3$ traité au second ordre, à cause de l'apparition dans cette dernière d'un dénominateur d'énergie très petit aux petits angles entre $\kk$ et $\qq$).\footnote{Le dénominateur d'énergie en question $\hbar\omega_\qq+\hbar\omega_\kk-\hbar\omega_{\kk+\qq}$ serait même exactement nul à angle nul sans les termes de courbure dans la relation de dispersion (\ref{eq1_24}), ce qui montre d'ailleurs toute la singularité d'une théorie à $\gamma=0$ (ce sur quoi nous reviendrons dans la section \ref{sec4.3}).}

\subsection{Comment calculer l'amortissement des phonons?}
\label{sec4.3}

Imaginons qu'on applique sur le gaz, initialement à l'équilibre thermique, une brève excitation de type Bragg induisant un petit déplacement cohérent de Glauber d'amplitude $\alpha\in\mathbb{C}^*$ dans le mode de phonons de vecteur d'onde $\qq$ sans toucher aux autres modes, ce qui correspond donc à l'opérateur d'évolution unitaire $U_{\rm exc}=\exp(\alpha\hat{b}_\qq^\dagger-\alpha^*\hat{b}_\qq)$.\footnote{Dans une expérience d'atomes froids, l'excitation de Bragg est induite par la superposition de deux faisceaux laser loin de résonance de vecteurs d'onde $\kk_1$ et $\kk_2$ avec $\kk_1-\kk_2=\qq$; même si les modes acoustiques $\pm \qq$ sont initialement vides ($\hbar\omega_\qq\gg k_{\rm B} T$)  et peuvent seulement recevoir des phonons, les processus Raman (à deux photons) absorption d'un photon dans un faisceau laser-émission stimulée dans l'autre induisent les changements d'impulsion  $\pm\hbar(\kk_1-\kk_2)=\pm\hbar\qq$ dans le gaz de fermions et excitent en général les deux modes en question; on peut cependant jouer sur la durée de l'excitation de Bragg pour que le mode $-\qq$ sorte intact de la procédure d'excitation \cite{Cartago}.} On a alors, juste après l'excitation, une moyenne non nulle pour l'opérateur d'annihilation correspondant:
\be
\label{eq1_32}
\langle\hat{b}_\qq(0^+)\rangle=\alpha\neq 0
\ee
Ceci conduit à une modulation observable de la densité moyenne du gaz aux vecteurs d'onde $\pm\qq$ puisque $\langle\delta\hat{\rho}(\rr,0^+)\rangle$, combinaison linéaire des $\langle\hat{b}_\kk(0^+)\rangle$ et des $\langle\hat{b}_\kk^\dagger(0^+)\rangle$ comme dans la note \ref{note32}, est alors $\neq 0$. 

\begin{figure}[t]
\includegraphics[width=10cm,clip=]{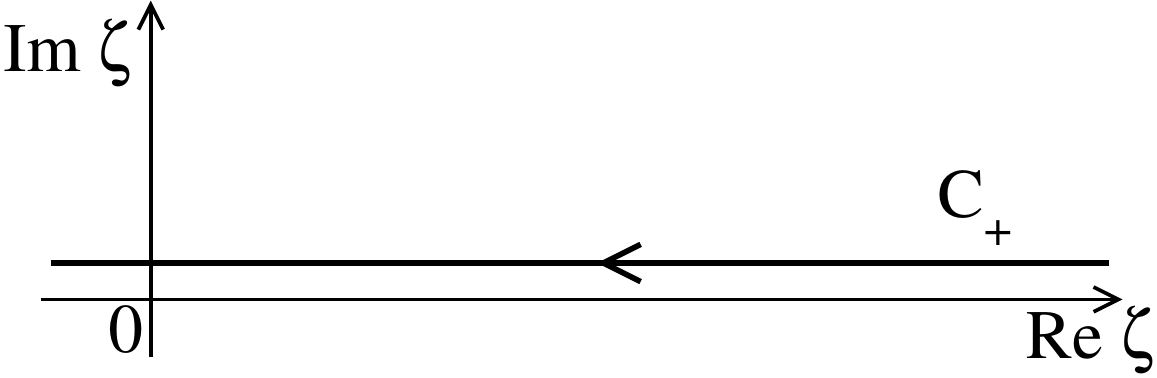}
\caption{Chemin d'intégration dans le plan complexe suivi par l'équation (\ref{eq2_32}).}
\label{fig6}
\end{figure}

Dans la limite $\alpha\to 0$, c'est-à-dire dans le régime de réponse linéaire, le formalisme des fonctions de Green à $N$ corps appliqué à la théorie effective de basse énergie, donc à l'hamiltonien de phonons (\ref{eq3_29}) [plutôt qu'à une description microscopique du gaz de fermions avec potentiel d'interaction $V(\rr_{ij})$ comme le font les références \cite{tome9landau,FetterW} par exemple], conduit à l'expression exacte
\be
\label{eq2_32}
\langle\hat{b}_\qq(t)\rangle \stackrel{t>0}{=} \alpha \eee^{-\ii\omega_\qq t} \int_{C_+} \frac{\dd\zeta}{2\ii\pi} \frac{\eee^{-\ii\zeta t/\hbar}}{\zeta-\Sigma_\qq(\zeta)}
\ee
Dans cette expression, $C_+$ est le chemin d'intégration parallèle à l'axe réel dans le demi-plan complexe supérieur, décrit de droite à gauche (de $\re\zeta=+\infty$ à $\re\zeta=-\infty$), voir la figure \ref{fig6}, et  $\Sigma_\qq(\zeta)$ est la fonction énergie propre au vecteur d'onde $\qq$ et à l'énergie complexe $\zeta$,\footnote{\label{note1_33p}Par rapport à la variable énergie $z$ habituelle (de la référence \cite{FetterW} par exemple), la variable énergie utilisée ici est translatée de l'énergie non perturbée du mode $\qq$, $\zeta=z-\hbar\omega_\qq$.  Ceci explique pourquoi  on a pu sortir le facteur de phase de l'évolution non perturbée dans (\ref{eq2_32}) et pourquoi $\Sigma_\qq(\zeta)$ est prise en $\zeta=\ii 0^+$  dans l'approximation (\ref{eq1_34}) à venir (cela correspond effectivement à $z =\hbar\omega_\qq+\ii 0^+$).} dont on ne connaît pas l'expression explicite mais qu'on définit par son développement perturbatif à tous les ordres en l'interaction phonon-phonon.  Ici, nous nous limitons à l'interaction cubique $H_3$ (voir la section \ref{sec4.2}) et le développement prend la forme diagrammatique suivante
\be
\label{eq1_33}
\begin{tabular}{c}
\includegraphics[width=8cm,clip=]{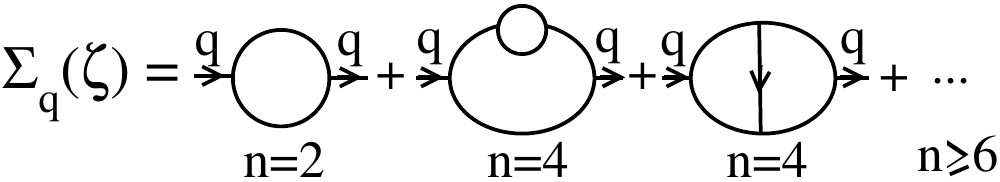}
\end{tabular}
\ee
où l'entier $n$ donne l'ordre en $H_3$ \cite{insuffisance}.  Nous n'indiquons que la topologie, il reste à sommer sur toutes les orientations possibles mais non redondantes des lignes internes, voir l'exemple de la portion de diagramme (\ref{eq1_31});\footnote{Dans le premier diagramme de (\ref{eq1_33}) (à une boucle), (i) orienter la ligne du haut vers la droite et la ligne du bas vers la gauche et (ii) orienter la ligne du haut vers la gauche et la ligne du bas vers la droite correspondent à la même contribution, par invariance du diagramme par rotation d'angle $\pi$ autour de son axe horizontal. Il en va de même pour la boucle interne du deuxième diagramme de (\ref{eq1_33}), par invariance par rotation locale de cette dernière. Pour la même raison d'invariance par rotation (cette fois globale) des diagrammes d'ordre $n=4$, on décide, pour éviter un double comptage, de mettre la boucle interne dans la branche du haut et, dans le troisième diagramme de (\ref{eq1_33}), d'orienter le pont vers le bas.}  une valeur précise  peut alors être attribuée à chaque diagramme, mettant en jeu une somme sur les vecteurs d'onde et les fréquences de Matsubara des lignes internes \cite{FetterW}. 

\begin{figure}[tb]
\includegraphics[width=8cm,clip=]{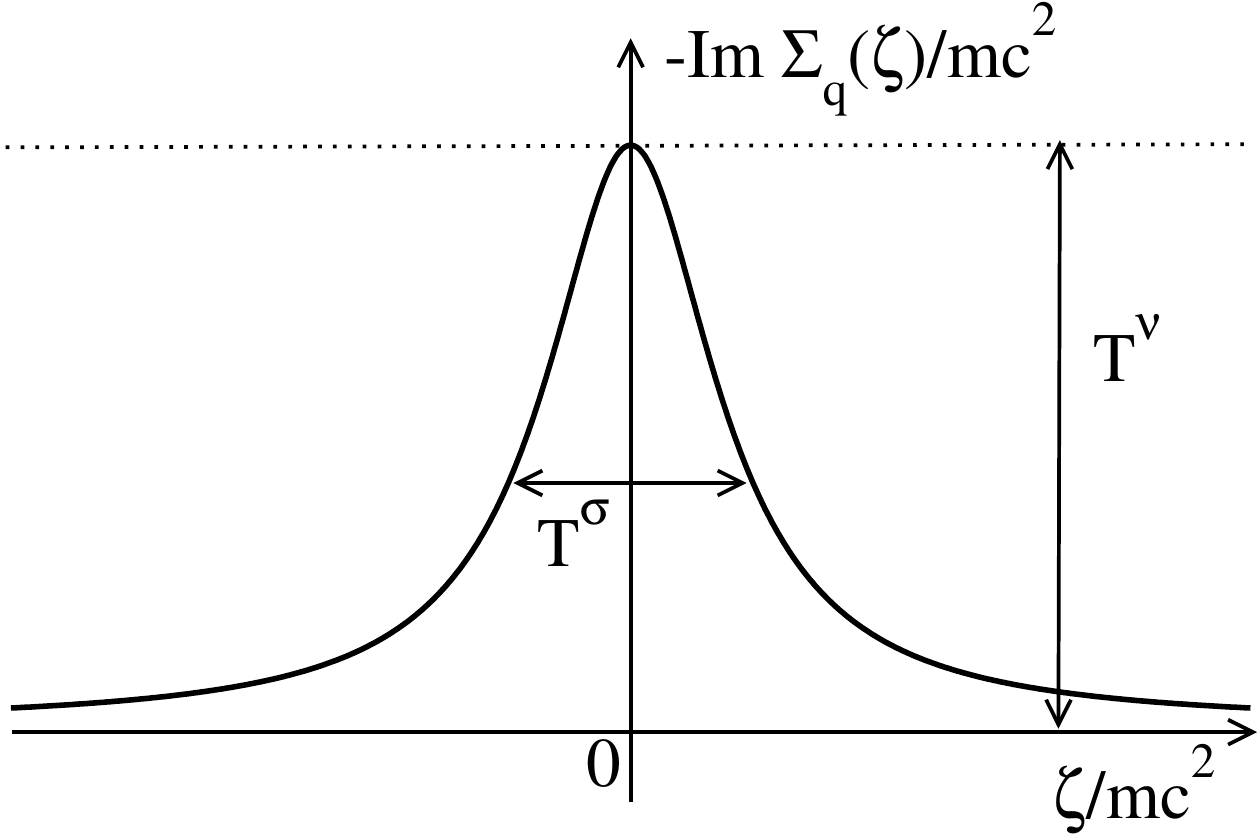}
\caption{Dans la limite $T\to 0$, on admet que l'ordre de grandeur et la largeur typique de la fonction $\im\Sigma_\qq(\zeta)$ près de $\zeta=\ii 0^+$ sont caractérisés par deux lois de puissance en température, d'exposants $\nu$ et $\sigma$ (en tenant compte d'une possible dépendance en température du paramètre de courbure $\gamma$).}
\label{fig7}
\end{figure}

Afin d'obtenir des résultats explicites sur l'amortissement, on effectue traditionnellement les deux approximations suivantes:\footnote{\label{note2_33}Notre théorie effective étant exacte à l'ordre dominant en température, il vaut mieux ne l'utiliser que dans la limite $T\to 0$, qui plus est en fixant le rapport $\bar{q}=\hbar c q/k_{\rm B}T$  afin que le mode $\qq$ soit lui aussi décrit exactement.  On a alors $\omega_\qq/\Gamma_{\rm th}\to +\infty$ où $\Gamma_{\rm th}=\Gamma_{q=k_{\rm B} T/\hbar c}$ est le taux de thermalisation du gaz de phonons et $\Gamma_q$ est la fonction (\ref{eq3_34}), puisque l'exposant $\nu$ introduit dans la figure \ref{fig7} page \pageref{fig7} est toujours $>1$, voir le tableau \ref{table1} page \pageref{table1}: le taux de thermalisation $\Gamma_{\rm th}$ tend vers zéro plus vite que la pulsation propre $\omega_\qq$ et le mode entre par définition dans le régime faiblement collisionnel.  Dans le régime opposé $\omega_\qq\ll\Gamma_{\rm th}$ dit hydrodynamique, le gaz de phonons a le temps d'atteindre un équilibre thermique local en chaque point d'oscillation de l'onde sonore $\qq$ et l'amortissement se décrit au moyen de coefficients de type viscosité dans les équations hydrodynamiques classiques d'un modèle à deux fluides \cite{Khalatlivre}.}
\begin{enumerate}
\item l'approximation de Markov (le gaz de phonons vu par le mode $\qq$ constitue un réservoir sans mémoire c'est-à-dire avec un temps de corrélation négligeable): on ignore la dépendance en énergie de la fonction énergie propre comme suit (voir la note \ref{note1_33p}),
\be
\label{eq1_34}
\Sigma_\qq(\zeta)\simeq \Sigma_\qq(\ii 0^+)
\ee
L'intégrale dans l'équation (\ref{eq2_32}) se calcule alors par le théorème des résidus (en refermant le contour par un demi-cercle infini dans le demi-plan  complexe inférieur),
\be
\label{eq2_34}
\langle\hat{b}_\qq(t)\rangle|_{\rm Markov} \stackrel{t\geq 0}{=} \alpha \eee^{-\ii\omega_\qq t} \eee^{-\ii \Sigma_\qq(\ii 0^+)t/\hbar}
\ee
La décroissance du signal est dans ce cas exponentielle, avec un taux correspondant à la partie imaginaire de $\Sigma_\qq(\ii 0^+)$ (la partie réelle donne le changement de pulsation propre du mode).
\item  l'approximation de Born: on calcule $\im\Sigma_\qq(\ii 0^+)$ perturbativement au premier ordre non nul $n$ en $H_3$.  Le taux d'amortissement des phonons $\qq$ vaut alors\footnote{Pour donner un autre éclairage à l'approximation (\ref{eq3_34}), signalons pour $n=2$ qu'on la retrouve exactement par la méthode de l'équation pilote bien connue en optique quantique (on obtient une équation d'évolution fermée sur l'opérateur densité $\hat{\rho}_S(t)$ d'un petit système $S$ -  ici le mode de phonons $\qq$ -  couplé à un gros réservoir $R$ - ici les autres modes de phonons $\kk\neq\qq$, en recourant justement à l'approximation de Born-Markov) \cite{CCTbordeaux,mon_cours} ou, plus simplement, au moyen de la règle d'or de Fermi (on calcule $\dd\langle\hat{n}_\qq\rangle/\dd t$ où $\langle\hat{n}_\qq\rangle=\langle\hat{b}_\qq^\dagger\hat{b}_\qq\rangle$ est le nombre moyen de phonons dans le mode  $\qq$ hors d'équilibre, en sommant les flux entrants -  processus de population $\kk,\kk'\to \qq$ et $\kk'\to \kk,\qq$ -  et sortants -  processus inverses de dépopulation $\qq\to\kk,\kk'$ et $\qq,\kk\to\kk'$ -  puis en  linéarisant en l'écart à l'équilibre thermique $\delta n_\qq(t)\equiv \langle n_\qq\rangle(t)-\bar{n}_\qq$ juste après l'excitation de Bragg, sous la forme $(\dd/\dd t)\delta n_\qq(t=0^+)=-\Gamma_\qq \delta n_\qq(t=0^+)$).  Pour $n=4$, on retrouve le même résultat (\ref{eq3_34}) en étendant la règle d'or de Fermi aux ordres supérieurs \cite{Landaumec}.}\ \footnote{\label{note_sub} Il y a ici une petite subtilité mathématique: si le premier ordre non nul $n$ est $\geq 4$, l'approximation de Markov ne doit plus se contenter de remplacer $\Sigma_\qq(\zeta)$ par $\Sigma_\qq(\ii 0^+)$ mais doit l'approximer par un développement limité autour de $\zeta=\ii 0^+$. Par exemple, pour $n=4$, on prend $\Sigma_\qq(\zeta)\simeq \Sigma_\qq(\ii 0^+)+\zeta\frac{\dd}{\dd\zeta}\Sigma_\qq(\ii 0^+)$ si bien qu'à l'ordre 4 en $H_3$, $\zeta-\Sigma_\qq(\zeta)\simeq[1-\frac{\dd}{\dd\zeta}\Sigma_\qq(\ii 0^+)][\zeta-\Sigma_\qq^{(2)}(\ii 0^+)-\Sigma_\qq^{(4){\rm eff}}(\ii 0^+)]$ avec $\Sigma_\qq^{(4){\rm eff}}(\ii 0^+)=\Sigma_\qq^{(4)}(\ii 0^+)+\Sigma_\qq^{(2)}(\ii 0^+)\frac{\dd}{\dd\zeta}\Sigma_\qq^{(2)}(\ii 0^+)$. Il faut alors remplacer $\Sigma_\qq^{(4)}(\ii 0^+)$ par $\Sigma_\qq^{(4){\rm eff}}(\ii 0^+)$ dans l'expression (\ref{eq3_34}) du taux d'amortissement. Dans le cas tridimensionnel concave, ceci ne change rien à $\Gamma_\qq|_{\rm Born-Markov}$ car $\frac{\dd}{\dd\zeta}\Sigma_\qq^{(2)}(\ii 0^+)$ est une quantité réelle, tout comme $\Sigma_\qq^{(2)}(\ii 0^+)$; dans le cas bidimensionnel concave, la conclusion est moins évidente mais reste la même, voir la note \ref{note_scott}. En revanche, ce remplacement doit être effectué dans le calcul du déplacement de pulsation thermique $\Delta_\qq$ du mode, $\hbar\Delta_\qq|_{\rm Born-Markov}=\re[\Sigma_\qq^{(2)}(\ii 0^+)+\Sigma_\qq^{(4){\rm eff}}(\ii 0^+)]-\mbox{idem à }T=0$.}
\be
\label{eq3_34}
\Gamma_\qq|_{\rm Born-Markov} = -\frac{2}{\hbar} \im\Sigma_\qq^{(n)}(\ii 0^+)
\ee
 où l'exposant donne l'ordre en $H_3$.  Dans le cas d'une branche acoustique convexe ($\gamma>0$) à 3D, il suffit d'aller à l'ordre $n=2$: c'est l'amortissement à trois phonons de Belyaev-Landau, très étudié théoriquement et observé dans l'hélium 4 liquide \cite{Roach,hel4} et, dans une moindre mesure, dans les gaz d'atomes froids bosoniques, seul l'amortissement de Belyaev y ayant été vu \cite{NirDav}.  Dans le cas concave ($\gamma<0$), il faut aller à l'ordre $n=4$ (la première contribution dans l'équation (\ref{eq1_33}) est purement réelle pour $\zeta=\ii 0^+$ car ses dénominateurs d'énergie, de la forme $\hbar\omega_\qq+\hbar\omega_\kk-\hbar\omega_{\kk+\qq}$ et $\hbar\omega_\qq-(\hbar\omega_\kk+\hbar\omega_{\qq-\kk})$, ne peuvent s'annuler, mais les deux suivantes ne le sont pas dans les processus $\phi\phi\to\phi\phi$); ce cas a été peu étudié théoriquement (la référence \cite{Annalen} a d'ailleurs relevé et corrigé une erreur dans le calcul originel \cite{LK}, et la référence \cite{diffpferm} a obtenu une expression beaucoup plus explicite du résultat, le généralisant même à un potentiel chimique de phonons $\mu_\phi$ non nul\footnote{Si l'on se limite à l'ordre dominant en température aux processus collisionnels $\phi\phi\rightarrow\phi\phi$, le nombre de phonons devient une quantité conservée, ce qui autorise à prendre $\mu_\phi<0$.}) et n'a, à notre connaissance, jamais été observé expérimentalement (aucune mesure précise de $\Gamma_\qq^{\gamma <0}$ n'a été faite dans aucun système). 
\end{enumerate}

 Déterminons la validité de l'approximation de Born au moyen de l'estimation de l'ordre $n \in 2\mathbb{N}^*$ donnée dans la référence \cite{insuffisance}:
\be
\label{eq1_35}
\Sigma_{\qq}^{(n)}(\ii 0^+)\approx \int\left(\prod_{i=1}^{n/2}\dd^dk_i\right) \frac{\langle\ \ |L^{d/2}H_3|\ \ \rangle^n}{(\Delta E)^{n-1}}\approx |\gamma|T^3 \left(\epsilon_{d{\rm D}}=\frac{T^{2d-4}}{|\gamma|^{(5-d)/2}}\right)^{n/2}
\ee
 où $d\geq 2$ est la dimension de l'espace.  L'écriture au second membre représente symboliquement le produit de $n$ éléments de matrice de l'interaction cubique entre phonons au numérateur et le produit de $n-1$ dénominateurs d'énergie (associés à $n-1$ états intermédiaires) au dénominateur, et l'intégrale est prise sur les vecteurs d'onde de phonons indépendants $\kk_i$.  L'ordre de grandeur au troisième membre est obtenu comme dans la référence \cite{insuffisance} en se limitant aux petits angles entre $\kk_i$ et $\qq$, en $O(|\gamma|^{1/2}T)$,  ce qui est légitime lorsque $T\to 0$ \cite{LK}; nous omettons ici la dépendance en la densité $\rho$, en $\xi=\hbar/mc$ et  en la constante de couplage $\mathcal{A}$ de l'équation (\ref{eq4_30}), au contraire de la référence \cite{insuffisance}, mais nous gardons celle en le paramètre de courbure $\gamma$, car nous lierons bientôt $\gamma$ et $T$. En résumé, le développement de Born est légitime en dimension $d = 3$ si son petit paramètre tend vers zéro à basse température:
\be
\label{eq2_35}
\epsilon_{{\rm 3D}}=\frac{T^2}{|\gamma|}\underset{T\to 0}{\to} 0
\ee
 Pour discuter la validité de l'approximation de Markov, admettons que le comportement de la fonction énergie propre au voisinage de $\zeta=\ii 0^+$ soit caractérisé par deux exposants, $\nu$ et $\sigma$, celui donnant les valeurs typiques $\propto T^\nu$ de sa partie imaginaire et celui donnant son échelle de variation typique $\propto T^\sigma$, comme sur la figure \ref{fig7}.\footnote{Les exposants introduits ici diffèrent d'une unité de ceux de la référence \cite{insuffisance} par un choix de convention différent.} \footnote{Pour la clarté de l'exposition, nous avons supposé sur la figure \ref{fig7} que la fonction représentée admet un maximum en l'origine. Ceci n'est pas nécessairement vrai (le cas bidimensionnel convexe de la référence \cite{insuffisance} en fournit un contre-exemple, voir son équation (114)). La véritable définition des exposants $\nu$ et $\sigma$ est que la fonction mise à l'échelle $\im \Sigma_\qq\left(\zeta=\bar{\zeta}mc^2(k_{\rm B}T/mc^2)^\sigma\right)/\left[mc^2(k_{\rm B}T/mc^2)^\nu\right]$ admet une limite finie et non nulle lorsque $T\to 0$ à énergie complexe réduite $\bar{\zeta}$ fixée ($\im\bar{\zeta}>0$).}
D'après l'équation (\ref{eq3_34}), on a donc
\be
\label{eq3_35}
\Gamma_\qq \underset{T\to 0}{\approx} T^\nu
\ee
La fonction $\Sigma_\qq(\zeta)$ admet alors une variation en énergie lente (négligeable) à l'échelle du taux d'amortissement (qui est bien l'inverse du temps caractéristique dans l'équation (\ref{eq2_32})) si elle est plus large que haute, ce qui impose
\be
\label{eq1_36}
\nu>\sigma
\ee
L'exposant $\nu$ s'obtient par un calcul explicite du second membre de l'équation (\ref{eq3_34}) dans la limite $T\to 0$, comme il a été fait dans la référence \cite{Annalen} pour le cas tridimensionnel\footnote{La référence historique \cite{LK} pour $\gamma<0$ fixé trouve elle aussi $\nu = 7$ mais la dépendance en $q$ de $\Gamma_\qq$ est différente, par exemple $\Gamma_\qq\approx q T^6$ dans \cite{LK} au lieu de $q^3T^4$ dans \cite{Annalen} lorsque $q\to 0$.} et dans la référence \cite{insuffisance} pour le cas convexe bidimensionnel; on peut aussi, plus simplement, utiliser l'estimation (\ref{eq1_35}) avec $n=2$ si $\gamma>0$ et $d\in\{2,3\}$, $n=4$ si $\gamma<0$ et $d=3$.\footnote{\label{note_scott} Pour $\gamma<0$ fixé et $d=2$, Alice Sinatra a obtenu en 2021, dans la formulation des références \cite{epl,Annalen}, le résultat non publié que $\im\Sigma_\qq^{(n=4)}(\ii 0^+)=0$ à l'ordre $T^3$ (l'ordre dominant attendu en température). Pour le voir, il est en fait plus simple d'utiliser les expressions (84) et (85) de la référence \cite{insuffisance}: (i) dans (84), on peut ignorer les processus $\phi\leftrightarrow\phi\phi\phi$ et se limiter au processus $\phi\phi\to\phi\phi$ (deuxième contribution), seul capable de conserver l'énergie-impulsion; (ii) dans l'intégrande de (85), on a le droit de remplacer $\zeta$ par $0$ au numérateur de la grande fraction pour la raison similaire que les processus $\phi\leftrightarrow\phi\phi$ ne conservent pas l'énergie-impulsion - ceci rend le numérateur réel ; (iii) on vérifie alors que, si l'écart d'énergie mis à l'échelle $\Delta E/\left[k_{\rm B}T (k_{\rm B}T/mc^2)^2\right]$ à côté de $\ii 0^+$ s'annule au dénominateur de la grande fraction, comme l'impose le Dirac $\delta(\Delta E)$ de la règle d'or de Fermi généralisée \cite{Landaumec}, le numérateur s'annule aussi (on le montre en remplaçant formellement $\gamma$ au numérateur par son expression annulant $\Delta E$, une fonction rationnelle des modules et des angles des vecteurs d'onde de phonons).  
En d'autres termes, la limite en $\zeta=0$ (mais aussi en $\zeta=-\Delta E$) de l'amplitude de transition au numérateur de la grande fraction, considérée comme une fraction rationnelle des angles, peut s'écrire $\Delta E\times P/Q$, où les polynômes $\Delta E$, $P$ et $Q$ sont deux à deux premiers entre eux.  

Ce raisonnement néglige cependant de possibles effets de bord dans l'intégrale sur les nombres d'onde de phonons, au sens de la référence \cite{fermi1D}, où l'un des nombres d'onde tend vers zéro, ce qui fait tendre vers zéro l'un des dénominateurs d'énergie des processus $\phi\leftrightarrow\phi\phi$. En incluant ces effets de bords, nous trouvons que les processus $\phi\to\phi\phi\phi$, $\phi\phi\to\phi\phi$ et $\phi\phi\phi\to\phi$, abrégés en $1\to 3$, $2\to 2$ et $3\to 1$ dans la référence \cite{insuffisance}, apportent chacun une contribution non nulle à $\im\Sigma_\qq^{(n=4)}(\ii 0^+)$ à l'ordre $T^3$, mais que la somme de ces contributions donne exactement zéro (la contribution du bord $q_1'+q_2'=q$ dans $2\to 2$, à savoir $C\int_0^{\bar{q}}\dd\bar{k}\,(\bar{n}_k^{\rm lin}+\bar{n}_{q-k}^{\rm lin}+1)\bar{k}(\bar{q}-\bar{k})/\bar{q}$, est compensée exactement par $1\to 3$, et celle des bords $q_1'=0$ et $q_2'=0$ dans leur ensemble, à savoir $2C\int_{\bar{q}}^{+\infty}\dd\bar{k}\, (\bar{n}_{k-q}^{\rm lin}-\bar{n}_k^{\rm lin})\bar{k}(\bar{k}-\bar{q})/\bar{q}$, est compensée exactement par $3\to 1$; ici, $\bar{k}=\hbar c k/k_{\rm B}T$, $\bar{n}_k^{\rm lin}=1/(\exp\bar{k}-1)$, $\Lambda=\rho^2\left(\frac{\dd^3}{\dd\rho^3}e_0(\rho)\right)/(3 mc^2)$ et $C=k_{\rm B}T (k_{\rm B}T/mc^2)^2\left[9(1+\Lambda)^2/8\rho\xi^2\right]^2/\left[\pi(3\gamma)^2\right]$). 

La conclusion n'est pas changée par la correction $\Sigma_\qq^{(2)}(\ii 0^+)\frac{\dd}{\dd\zeta}\Sigma_\qq^{(2)}(\ii 0^+)$ de notre note \ref{note_sub} car on trouve que les facteurs $\Sigma_\qq^{(2)}(\ii 0^+)$ et $\frac{\dd}{\dd\zeta}\Sigma_\qq^{(2)}(\ii 0^+)$ sont tous les deux réels. C'était évident pour le premier facteur (les effets de bord qu'il présente à 1D \cite{fermi1D} sont supprimés à 2D par abaissement de la densité d'états des phonons à faible nombre d'onde). Ça ne l'était pas pour le second facteur: à cause des effets de bord dans l'intégration sur $\kk$ (voir l'équation (39) de la référence \cite{insuffisance}), les processus de Belyaev et de Landau donnent chacun une contribution non nulle à $\im\frac{\dd}{\dd\zeta}\Sigma_\qq^{(2)}(\ii 0^+)$ (elle vaut $2[9(1+\Lambda)^2/8\rho\xi^2]/[\bar{q}(3|\gamma|)^{3/2}]$ pour Landau à l'ordre dominant en température) mais ces contributions sont exactement opposées, en particulier parce que les dénominateurs d'énergie $\veps_\qq-(\veps_\kk+\veps_{\qq-\kk})$ et $\veps_\qq+\veps_\kk-\veps_{\qq+\kk}$ sont opposés $\sim\mp\hbar c k[1-(v_q/c)\cos\theta]$ à l'ordre dominant en $k$ ($v_q=\dd\veps_\qq/\hbar\dd q$ est la vitesse de groupe et $\theta$ est l'angle entre $\kk$ et $\qq$) et entrent dans la dérivée de la distribution de Dirac $\delta'(\veps)$, qui est une fonction impaire de son argument.

Le cas bidimensionnel concave est donc spécial: la quantité $\im\Sigma_\qq^{(4){\rm eff}}(\ii 0^+)$ de la note \ref{note_sub} - considérée à tous les ordres en température - ne donne pas la bonne loi d'échelle en température $\propto T^3$ de la fonction énergie propre d'ordre 4 en $H_3$ sur un voisinage en $O(T^3)$ de $\zeta=\ii 0^+$; elle ne donne pas non plus, d'ailleurs, le taux d'amortissement $\Gamma_\qq$ puisqu'il y a échec des approximations de Born et de Markov comme dans le cas bidimensionnel convexe, voir la dernière ligne de notre tableau \ref{table1}. 
Notons cependant, toujours pour $\gamma<0$, que le cas limite $\rho\xi^2\to +\infty$ d'une interaction très faible dans le superfluide sous-jacent doit être mis à part car on y dispose du petit paramètre supplémentaire $1/\rho\xi^2$ aidant à la validité de Markov (comme dans la section 3.2 de la référence \cite{insuffisance}) et de Born (comme dans l'équation (17) de cette même référence); ce cas limite est inaccessible dans un gaz de fermions de spin $1/2$ en interaction de contact - on y a $\rho\xi^2=O(1)$ lorsque $\gamma<0$ \cite{csdm} - mais il l'est dans un gaz de bosons avec une interaction de portée $\gtrsim\xi$ comme l'envisage la référence \cite{Annalen}.} L'exposant $\sigma$ s'obtient en généralisant ladite estimation au cas $\zeta\neq 0$, c'est-à-dire en ajoutant $\zeta$ à $\Delta E$ dans (\ref{eq1_35}); or, dans un développement aux petits angles entre $\kk_i$ et $\qq$, la partie de $\Delta E$ linéaire en les nombres d'onde s'annule et il ne reste que les contributions cubiques $\approx\gamma T^3$ si bien que, indépendamment de la dimensionnalité $d$,
\be
\label{eq2_36}
\frac{1}{\zeta+\Delta E} \approx \frac{1}{\zeta+\gamma T^3} \quad\mbox{et donc}\quad T^\sigma\approx|\gamma|T^3
\ee
en tenant compte de la dépendance en le paramètre $\gamma$, ce dernier - nous l'avons dit - pouvant varier en température. La condition de validité $T^\nu=o(\gamma T^3)$ qui en résulte dans (\ref{eq1_36}) admet une interprétation simple: dans la limite $T\to 0$ prise avec la loi d'échelle $q\approx T$, il faut que le taux d'amortissement $\Gamma_\qq$ tende vers zéro plus vite que le terme cubique en $q$ dans $\omega_\qq$, 
\be
\Gamma_q\stackrel{\hbar c q/k_{\rm B}T\,\mbox{\scriptsize fixé}}{\underset{T\to 0}{=}}o(\omega_\qq^{(3)})\quad\mbox{avec}\quad \omega_\qq^{(3)}=\gamma mc^2(q\xi)^3/8\hbar
\ee
ce qui est donc la vraie marque de la nature markovienne de l'amortissement (plutôt que la condition perturbative irréfléchie $\Gamma_\qq=o(\omega_\qq)$).

La situation est résumée sur le tableau \ref{table1} ci-dessous.\footnote{Dans la troisième ligne du tableau, on omet de possibles facteurs logarithmiques $\ln(1/T)$ pour simplifier. Ces facteurs proviennent du fait que, pour les lois d'échelle $\gamma\propto T^2$ et $q\propto T$, les termes en $q^3$ et en $q^5\ln q$ sont du même ordre de grandeur dans la relation de dispersion (\ref{eq1_24}): dans ce régime de faible courbure, la contribution logarithmico-quintique à $\omega_\qq$ n'est donc plus une petite correction et doit être gardée.} L'approximation de Born-Markov est donc utilisable en dimension 3, sauf sur un étroit intervalle de valeurs de $\gamma$, de largeur $\approx (k_{\rm B}T/mc^2)^2$ autour de $\gamma = 0$; en $\gamma = 0$, la relation de dispersion des phonons (\ref{eq1_24}) a d'ailleurs un premier écart quintique à la loi linéaire $cq$, ce qui, on le sent bien, est un cas spécial.  

\begin{table}[h]
\begin{tabular}{|l||c|c|c||c|c|}
\hline
& $\nu$ & $\sigma$ & Markov & $\epsilon_{d{\rm D}}$ &  Born \\
\hline
$d=3,\gamma>0$ \mbox{fixé}& 5 & 3 & oui & $\approx T^2\to 0$ & oui \\
\hline
$d=3,\gamma<0$ \mbox{fixé}& 7 & 3 & oui & $\approx T^2\to 0$ & oui \\
\hline
$d=3,\gamma=O(T^2)$ & 5 & 5 & non & $\approx T^0\not\to 0$ & non \\
\hline
$d=2,\gamma>0$ \mbox{fixé}& 3 & 3 & non & $\approx T^0\not\to 0$ & non \\
\hline
\end{tabular}
\medskip
\caption{Dans l'étude de l'amortissement des phonons d'un superfluide, validité de l'approximation de Born-Markov dans la limite de basse température $T\to 0$ selon la dimension de l'espace $d$ et le paramètre de courbure $\gamma$ de la branche acoustique (plus précisément son signe et sa variation en température, la troisième ligne valant quel que soit le signe de $\gamma$). Les exposants $\nu$ et $\sigma$ entrant dans la définition (\ref{eq1_36}) du régime markovien sont ceux de la figure \ref{fig7}, et le petit paramètre du développement de Born $\epsilon_{d{\rm D}}$ est donné dans l'équation (\ref{eq1_35}).}
\label{table1}
\end{table}

Le calcul précis du taux d'amortissement $\Gamma_\qq$ (ou ce qui en tient lieu pour une décroissance non exponentielle, comme l'inverse de la largeur de $|\langle\hat{b}_\qq(t)\rangle|^2$ à la hauteur relative $1/\eee$) pour ces faibles valeurs de la courbure constitue, à notre connaissance, une question ouverte; elle est d'une grande pertinence expérimentale, la force des interactions annulant $\gamma$ étant semble-t-il proche de la limite unitaire (voir la référence \cite{PRAconcav} et notre section \ref{sec5.1}), point de prédilection des expériences sur les atomes froids car dans un régime de valeurs assez élevées de $T_{\rm c}/T_{\rm F}$ et de propriétés collisionnelles propices au refroidissement par évaporation \cite{Shlyap}.

Pour faire bonne mesure, nous avons aussi considéré le cas bidimensionnel convexe dans le tableau \ref{table1}: l'approximation de Born-Markov y échoue, et la référence \cite{insuffisance} a dû faire appel à une approximation heuristique non perturbative sur la fonction énergie propre $\Sigma_\qq(\zeta)$ pour arriver à un bon accord avec des simulations de champ classique (opérateurs $\hat{b}_\qq,\hat{b}_\qq^\dagger$ de l'hydrodynamique quantique remplacés par des nombres complexes $b_\qq,b_\qq^*$), ceci dans le régime d'interaction faible $\rho\xi^2\gg 1$ du superfluide bosonique sous-jacent, où l'on pensait pourtant disposer d'un petit paramètre assurant le succès de la règle d'or de Fermi même dans la limite $k_{\rm B}T/mc^2\to 0$ (cette attente raisonnable, confirmée dans la section 3.2 de la référence \cite{insuffisance} à l'ordre deux en $H_3$, est infirmée dans la section 4.3 de cette même référence par un calcul à l'ordre quatre).\footnote{Nous n'avons pas parlé ici du cas très particulier de la dimension $d=1$, où deux vecteurs d'onde font un angle très petit (nul!) dès qu'ils sont de même sens. Disons simplement que le petit paramètre de Born reste donné par l'équation (\ref{eq1_35}), obtenue pourtant sous l'hypothèse $d\geq 2$. En rétablissant la dépendance en densité comme dans \cite{insuffisance}, nous trouvons plus précisément $\epsilon_{1\rm D}=1/[\gamma^2\rho\xi(k_{\rm B}T/mc^2)^2]$, le préfacteur dans (\ref{eq1_35}) s'écrivant $\gamma (k_{\rm B}T)^3/(mc^2)^2$. Dans la limite de basse température $k_{\rm B}T/mc^2\to 0$ à $\rho\xi$ fixé considérée ici, $\epsilon_{1\rm D}\to +\infty$ et il faut immédiatement faire appel à des approximations non perturbatives sur $\Sigma_\qq(\zeta)$ et $\Gamma_\qq$, comme le calcul autocohérent des références \cite{auto1,auto2}. Dans la limite opposée d'interaction faible $\rho\xi\to+\infty$ à $k_{\rm B}T/mc^2$ fixé, $\epsilon_{1\rm D}\to 0$ et l'on peut utiliser la règle d'or de Fermi comme dans la référence \cite{fermi1D}; plus précisément, on s'attend à ce que la condition de validité de la règle d'or s'écrive $\rho\xi(k_{\rm B}T/mc^2)^2\gg \phi_{1\rm D}(\bar{q})$ où $\phi_{1\rm D}$ est une certaine fonction de $\bar{q}=\hbar c q/k_{\rm B}T$, en oubliant la dépendance en $\gamma$ pour simplifier (à $\bar{q}$ fixé, l'approximation de Born impose cette condition, mais l'approximation de Markov est alors satisfaite aussi car on a $\hbar\Gamma_\qq|^{\scriptsize\mbox{règle}}_{\scriptsize\mbox{d'or}}\approx\gamma[(k_{\rm B}T)^3/(mc^2)^2]\epsilon_{1\rm D}\approx k_{\rm B}T/(\gamma\rho\xi)\ll\Delta E\approx\gamma(k_{\rm B}T)^3/(mc^2)^2$ où $\Delta E$, dénominateur d'énergie de Belyaev-Landau typique, donne la largeur en $\zeta$ de la fonction énergie propre comme dans l'équation (\ref{eq2_36})). À 2D, comme le montre la référence \cite{insuffisance}, le développement perturbatif en $H_3$ de la fonction énergie propre est soumis à une condition de validité similaire, $\rho\xi^2(k_{\rm B}T/mc^2)^2\gg \phi_{2\rm D}(\bar{q})$, voir son équation (96), mais qui, contrairement au cas 1D, ne s'obtient pas par simple comptage de puissances.}

\subsection{Diffusion de phase du condensat de paires}
\label{sec4.4}

Une question d'ordre à la fois pratique et fondamental porte sur le temps de cohérence du condensat de paires à l'équilibre thermique dans un gaz de fermions parfaitement isolé de son environnement.

Pour un système de fermions $\uparrow$ et $\downarrow$ infini non polarisé, le temps de cohérence est infini, comme l'affirme le phénomène de brisure de symétrie $U(1)$: dans l'ensemble grand canonique (terme $-\mu\hat{N}$ ajouté à l'hamiltonien du gaz de fermions où $\mu$ est le potentiel chimique et $\hat{N}$ l'opérateur nombre total de particules), le paramètre d'ordre complexe $\Delta(\rr,t)$ est uniforme et constant; dans l'ensemble canonique, il évolue donc avec le facteur de phase non amorti $\exp(-2\ii \mu t/\hbar)$,\footnote{Il y a un facteur 2 sous l'exponentielle car $\Delta$ est un paramètre d'ordre de paires alors que $\mu$ est le potentiel chimique des fermions. Il n'y a pas de facteur 2 dans l'équation (\ref{eq1_37}) car l'opérateur phase $\hat{\phi}_0$ est conjugué à la densité de fermions.} avançant à la pulsation immuable $2\mu/\hbar$; en tout cas, le temps de cohérence est infini. 

Qu'en est-il dans un système de taille finie (boîte de quantification $[0,L]^3$, nombre total fixé $N$ de fermions)? Pour le savoir, suivons la référence \cite{brouilfer} et écrivons l'équation d'évolution de l'opérateur phase du condensat, que nous notons $\hat{\phi}_0$ comme dans l'équation (\ref{eq2_29}), dans le régime de l'hydrodynamique quantique:\footnote{Pour obtenir cette équation, il a fallu éliminer par lissage temporel des termes en $\hat{b}_\qq\hat{b}_{-\qq}$ et $\hat{b}^\dagger_\qq\hat{b}^\dagger_{-\qq}$; c'est sans conséquence car ils oscillent avec une période $\approx\hbar/k_{\rm B}T$ bien plus courte que les échelles de temps collisionnelles qui nous intéressent ici (voir plus loin) et se moyennent donc automatiquement à zéro.}
\be
\label{eq1_37}
-\hbar\frac{\dd}{\dd t}\hat{\phi}_0 = \mu_0(\rho) + \sum_{\qq\neq\zero} \hat{b}_\qq^\dagger\hat{b}_\qq \frac{\dd}{\dd N}(\hbar\omega_\qq) \equiv \hat{\mu}
\ee
Au second membre, $\mu_0(\rho)=\dd E_0/\dd N$ est le potentiel chimique des $N$ fermions dans l'état fondamental d'énergie $E_0$ à la densité $\rho$ et la somme sur $\qq$ peut être interprétée comme la dérivée adiabatique (comprendre aux opérateurs nombres d'occupation $\hat{b}_\qq^\dagger\hat{b}_\qq$ des modes de phonons fixés) par rapport à $N$ de la somme correspondante dans l'hamiltonien de phonons $H_2$ (\ref{eq4_29}). Le second membre dans son ensemble est donc la dérivée isentropique de l'hamiltonien par rapport au nombre total de particules. En ce sens, il constitue un opérateur potentiel chimique des fermions, d'où la notation $\hat{\mu}$ au troisième membre, et l'équation (\ref{eq1_37}) n'est autre qu'une version quantique de la fameuse seconde relation de Josephson,  reliant la dérivée temporelle de la phase (classique) du paramètre d'ordre au potentiel chimique à l'équilibre $\mu$. 

Dans une réalisation donnée de l'expérience, que nous supposons correspondre à un état propre à $N$ corps $|\psi_\lambda\rangle$ d'énergie $E_\lambda$ échantillonnant l'ensemble canonique, les nombres d'occupation $\hat{b}_\qq^\dagger\hat{b}_\qq$ fluctuent et se décorrèlent sous l'effet des collisions incessantes entre phonons dues en particulier à $H_3$,  voir l'équation (\ref{eq3_30}). Aux temps assez longs pour qu'un grand nombre de collisions aient eu lieu, on s'attend donc à un étalement diffusif de la phase du condensat, avec un déphasage aléatoire de variance croissant linéairement en temps:
\be
\label{eq1_38}
\mbox{Var}_\lambda[\hat{\phi}_0(t)-\hat{\phi}_0(0)] \underset{\Gamma_{\rm coll}^{\phi}t\gg 1}{\sim} 2 D_\lambda t
\ee
et un coefficient de diffusion $D_\lambda$ sous-intensif, c'est-à-dire $\approx 1/N$ à la limite thermodynamique. Ici $\Gamma_{\rm coll}^{\phi}=\Gamma_{q=k_{\rm B}T/\hbar c}$ est le taux de collision typique entre phonons thermiques (la fonction $\Gamma_q$ étant celle de l'équation (\ref{eq3_34})). L'étalement (\ref{eq1_38}) induit une perte de cohérence temporelle exponentielle de taux $D_\lambda$, en vertu de la relation de Wick (on s'attend à ce que la statistique du déphasage dans $|\psi_\lambda\rangle$ soit approximativement gaussienne \cite{vraiediff}),
\begin{multline}
\label{eq2_38}
\left\langle\exp\left\{-\ii\left[\hat{\phi}_0(t)-\hat{\phi}_0(0)\right]\right\}\right\rangle_\lambda \simeq \exp\left[-\ii\langle\hat{\phi}_0(t)-\hat{\phi}_0(0)\rangle_\lambda\right] \exp\left\{-\frac{1}{2}\mathrm{Var}_\lambda[\hat{\phi}_0(t)-\hat{\phi}_0(0)]\right\} \\
\underset{\Gamma_{\rm coll}^{\phi}t\gg 1}{\simeq} \exp\left[-\ii\langle\hat{\mu}\rangle_\lambda t/\hbar\right] \exp(-D_\lambda t)
\end{multline}
ce que confirme d'ailleurs l'analyse par résolvante de la référence \cite{brouilfer}.\footnote{\label{note1_38} Si l'on admet que le gaz de phonons en interaction est un système quantique ergodique \cite{eth1,eth2}, la moyenne $\langle\hat{\mu}\rangle_\lambda$ dans l'état stationnaire $|\psi_\lambda\rangle$ ne dépend que des deux quantités conservées, l'énergie $E$ et le nombre de particules $N$,  et coïncide pour un grand système avec le potentiel chimique microcanonique $\mu_{\rm mc}(E=E_\lambda,N)$. Si l'énergie $E$ fluctue d'une réalisation à l'autre de l'expérience autour de la moyenne $\bar{E}$, comme dans l'ensemble canonique, le facteur de phase au troisième membre de l'équation (\ref{eq2_38}) fluctue et conduit à un brouillage gaussien en temps: la linéarisation de $\mu_{\rm mc}(E,N)$ autour de $\bar{E}$ donne $\mbox{Var}\,[\hat{\phi}_0(t)-\hat{\phi}_0(0)]\sim[\partial_E \mu_{\rm mc}(\bar{E},N)]^2(\mbox{Var}\, E)t^2/\hbar^2$, effet parasite $\approx t^2/N$ masquant rapidement la diffusion de phase (\ref{eq1_38}) $\approx t/N$ \cite{superdiff}.}

Dans le cas $\gamma>0$ d'une branche acoustique convexe, la situation ressemble à celle des condensats de bosons en interaction faible bien étudiée dans la référence \cite{praphase}: les collisions dominantes sont celles à trois phonons $\phi\leftrightarrow\phi\phi$ de Belyaev et de Landau, et $D_\lambda$ a été calculé pour le condensat de paires de fermions à basse température dans la référence \cite{diffpferm}; nous en donnons ici l'expression simplifiée suivante, ne gardant que les lois d'échelle en $N$, $T$ et $\gamma$ (sous l'hypothèse $\gamma=O(1)$):
\be
\label{eq3_38}
D_\lambda^{\gamma>0} \approx N^{-1} T^4 \gamma^0
\ee
Il n'y a eu encore aucune vérification expérimentale dans les gaz d'atomes froids (seuls fluides quantiques suffisamment bien isolés pour que la perte de cohérence du condensat soit intrinsèque), même pour les bosons. 

En revanche, dans le cas concave $\gamma<0$, la question reste largement ouverte.  La tentative de calcul de $D_\lambda$ de la référence \cite{diffpferm}, tenant compte seulement des processus de collision Landau-Khalatnikov à quatre phonons $\phi\phi\to\phi\phi$ aux petits angles, de taux typique $\Gamma_{\rm coll}^{\phi}\propto (k_{\rm B}T/mc^2)^7mc^2/\hbar|\gamma|\approx T^7$, a conduit à un coefficient de diffusion infini,
\be
\label{eq1_39}
D_\lambda^{\gamma<0} = +\infty
\ee
plus précisément à une loi d'étalement superdiffusive (simplifiée comme dans (\ref{eq3_38}))\footnote{Dans tous les cas, voir les équations (\ref{eq3_38}) et (\ref{eq2_39}), on trouve bien qu'il n'y a pas d'étalement de phase à la limite thermodynamique $N\to +\infty$: dans un gaz isolé, le temps de cohérence limité du condensat est un effet de taille finie.}
\be
\label{eq2_39}
\mbox{Var}_\lambda^{\gamma<0} [\hat{\phi}_0(t)-\hat{\phi}_0(0)] \approx N^{-1} T^{20/3} |\gamma|^{1/3} t^{5/3}
\ee
en particulier parce que les collisions $\phi\phi\to\phi\phi$ conservent le nombre total de phonons $N_\phi$ (au contraire de $\phi\leftrightarrow\phi\phi$).\footnote{$N_\phi$ doit alors être ajouté à la liste des constantes du mouvement, à côté de $E$ et $N$, dans la note \ref{note1_38}.}\ \footnote{Le fait que, pour $\gamma<0$, le taux d'amortissement des phonons $\Gamma_\qq$ tende vers zéro comme $q^3$ (au lieu de $q$ pour $\gamma>0$) joue aussi un rôle; cependant, sans la conservation de $N_\phi$, il conduirait à une loi d'étalement en $t\,\ln t$ marginalement superdiffusive (voir l'équation (C.20) de la référence \cite{diffpferm} et la moralité énoncée après son équation (72)).} Aller au-delà et obtenir la vraie valeur (a priori finie) de $D_\lambda$ reste une question ouverte: il faudrait tenir compte des processus sous-dominants à cinq phonons $\phi\phi\leftrightarrow\phi\phi\phi$ qui changent $N_\phi$ et se produisent à un taux $\approx T^9$ \cite{Khalat}, de même ordre de grandeur que celui des processus $\phi\phi\to\phi\phi$ aux grands angles \cite{LK,Annalen}, ce qui n'est pas  aisé.\footnote{La publication \cite{diffpferm}, comprenant mal la référence \cite{Khalat}, y avait vu un taux d'amortissement à cinq phonons en $T^{11}$.  Erreur corrigée ici.  En effet, la référence \cite{Khalat}, considérant un quasi-équilibre thermique avec un petit potentiel chimique de phonons non nul $\mu_\phi\to 0^-$, obtient l'équation d'évolution $L^{-3}\dd N_\phi/\dd t=-\Gamma_\phi\mu_\phi$ sur le nombre moyen de phonons, où $\Gamma_\phi\approx T^{11}$ n'est pas le taux cherché malgré les apparences;  comme $L^{-3}\dd N_\phi/\dd t\approx T^2\dd\mu_\phi/\dd t$ pour la loi de Bose $\bar{n}_\qq=1/\{\exp[(\hbar c q -\mu_\phi)/k_{\rm B}T]-1\}$, on a en fait $-\dd\mu_\phi/\dd t\propto (\Gamma_\phi/T^2)\mu_\phi$, de taux $\approx T^9$.}

\section{Questions ouvertes requérant une théorie microscopique du problème à $N$ corps}
\label{sec5}

L'hydrodynamique quantique de la section \ref{sec4} n'est qu'une théorie effective de basse énergie.  Elle présente donc des limitations de deux types, que nous passons ici brièvement en revue, et qui empêchent de faire l'économie d'un calcul microscopique à $N$ corps. 

\subsection{Déterminer les ingrédients de l'hydrodynamique quantique}
\label{sec5.1}

L'hydrodynamique quantique fait intervenir deux quantités qui lui sont extérieures, l'équation d'état du gaz de fermions non polarisé à température nulle (au travers de l'énergie volumique $e_0(\rho)$ ou du potentiel chimique $\mu_0(\rho)$ - sa dérivée - à la densité $\rho$) et le paramètre de courbure $\gamma$ de la branche acoustique (\ref{eq1_24}). 

Dans le présent cas de masses égales $m_\uparrow=m_\downarrow=m$, l'équation d'état a été mesurée expérimentalement \cite{Salomon,mit} et différentes méthodes de calcul approchées donnent des résultats satisfaisants, comme le Monte-Carlo quantique diffusif à surface nodale fixée \cite{Giorgini,Gezerlis} ou l'approximation des fluctuations gaussiennes dans une formulation de champ par intégrale de chemin \cite{Randeria,Drummond}. 

La situation est beaucoup plus ouverte pour le paramètre de courbure $\gamma$.  L'approximation de la phase aléatoire d'Anderson (RPA) \cite{Anderson}, équivalente pour ce problème au calcul des pulsations propres des équations BCS dépendant du temps linéarisées ou même à l'approximation des fluctuations gaussiennes pourtant plus performante \cite{Strinati,Randeria},\footnote{Ces différentes approches conduisent exactement à la même équation implicite liant la pulsation propre $\omega_\qq$, le potentiel chimique $\mu$ et le paramètre d'ordre $\Delta$, et exactement à la même équation liant $\mu, \Delta$ et la longueur de diffusion $a$  dans l'onde $s$ \cite{PRAconcav}; elles diffèrent seulement par l'équation d'état $\mu=\mu_0(\rho)$ reliant $\mu$ à $\rho$ dans l'état fondamental, celle des fluctuations gaussiennes étant la plus précise.  Par exemple, à la limite unitaire $a^{-1}=0$, les approches donnent toutes $mc^2/\mu=2/3$ (c'est exact par invariance d'échelle, $\mu_0(\rho)\propto \rho^{2/3}$  dans l'équation (\ref{eq2_30})), $|\Delta|/\mu\simeq 1,\!16$ (proche de la valeur expérimentale $0,\!44 E_{\rm F}/0,\!376 E_{\rm F}\simeq 1,\!17$ sachant que $|\Delta|=E_{\rm paire}/2$ dans ces théories et que $E_{\rm paire}/2E_{\rm F}\simeq 0,\!44$ dans l'expérience \cite{Ketterle}) mais le rapport $\mu/E_{\rm F}\simeq 0,\!376$ dans l'expérience \cite{mit}, très mal reproduit $\simeq 0,\!59$ par la RPA et BCS, est bien meilleur $\simeq 0,\!40$ dans les fluctuations gaussiennes.}  conduit à une expression analytique assez simple de $\gamma$  en termes de $\mu/|\Delta|$ et $(\partial\mu/\partial|\Delta|)_a$, exacte $\gamma\to 1$ dans la limite $k_{\rm F}a\to 0^+$ d'un condensat de dimères, raisonnable dans la limite BCS $k_{\rm F}a\to 0^-$ ($\gamma\to -\infty$ exponentiellement en $1/k_{\rm F}|a|$ sous l'effet de l'écrasement de la branche acoustique par le continuum de paire brisée) et présentant une annulation avec changement de signe pour $|\Delta|/\mu\simeq 0,\!87$, soit $1/k_{\rm F}a\simeq -0,\!14$ pour l'équation d'état assez approximative de la théorie BCS, un changement de signe proche de la limite unitaire en tout cas \cite{PRAconcav}.

En particulier, $\gamma$ a même valeur positive à la limite unitaire dans ces trois approches (branche acoustique de départ convexe):
\be
\label{eq1_41}
\gamma_{a^{-1}=0}^{\rm RPA}\simeq 0,084
\ee
L'erreur commise est cependant non contrôlée, et l'on n'est même pas sûr du signe. 

Une méthode complètement différente procède par extension du problème à une dimension spatiale $d$ quelconque puis développement autour de la dimension quatre, en puissances donc du petit paramètre $\epsilon=4-d$.  À la limite unitaire, elle conduit elle aussi à un départ convexe \cite{Rupak}:\footnote{Nous avons obtenu l'expression (\ref{eq2_41}) en reportant directement l'équation (50) de la référence \cite{Rupak} dans la relation de dispersion (48) de cette même référence et en utilisant la propriété $mc^2=2\mu/3$ exacte par invariance d'échelle. En procédant différemment, c'est-à-dire en passant par son équation (52) et son résultat $c_2/c_1=O(\epsilon^2)\simeq 0$ avec $d=3$ dans son équation (48), on trouve la valeur assez proche $\gamma_{a^{-1}=0}^{\rm dimension}=8/45\simeq 0,\!18$.}
\be
\label{eq2_41}
\gamma_{a^{-1}=0}^{\rm dimension} = \frac{1}{3}\left[1-\frac{1}{4}\epsilon+O(\epsilon^2)\right]\underset{d=3}{\stackrel{\epsilon=1}{\simeq}}\frac{1}{4}>0
\ee


Expérimentalement, une mesure récente de la branche acoustique par excitation de Bragg dans un gaz d'atomes froids  fermioniques conduit au contraire à un départ concave à la limite unitaire \cite{Moritz}:
\be
\label{eq1_42}
\gamma_{a^{-1}=0}^{\rm exp}=\frac{8\mu}{3E_{\rm F}}\zeta\simeq \zeta\quad\mbox{avec}\quad\zeta=-0,\!085(8)<0
\ee
sachant que le rapport $\mu/E_{\rm F}$ vaut $\simeq 3/8$ pour le gaz unitaire dans l'état fondamental \cite{mit} et que $\zeta$ est le paramètre de courbure de la branche acoustique pour l'adimensionnement de $q$ par $k_{\rm F}$, $\omega_\qq=cq(1+\zeta q^2/k_{\rm F}^2+\ldots)$.  Mais le résultat (\ref{eq1_42}) souffre de deux  limitations \cite{comment}: (i) un ajustement cubique de la branche sur un intervalle de valeurs de $q$ assez élevées, $q/k_{\rm F}\in[0,\!29\,;1,\!63]$, plutôt que sur un étroit voisinage de $q=0$ (pour la relation de dispersion de la RPA, par exemple, qui présente un point d'inflexion en $q\simeq 0,\!5 k_{\rm F}$, un tel ajustement, mélangeant aveuglément des parties convexe et concave, ne donnerait pas le bon signe de $\gamma_{a^{-1}=0}^{\rm RPA}$), et (ii) une température relativement élevée, $T=0,\!128(8)T_{\rm F}\simeq 0,\!8\!T_{\rm c}$: même si l'on part de la branche de la RPA de paramètre $\gamma>0$ dans l'état fondamental, l'hydrodynamique quantique prédit un changement thermique $\delta\gamma_{\rm th}$ de la courbure (par interaction du mode $\qq$ avec les phonons thermiques) assez négatif pour en changer le signe:
\be
\label{eq2_42}
\delta\gamma_{\rm th}\sim -\frac{8\pi^2}{9(3\mu/E_{\rm F})^{1/2}}\left(\frac{T}{T_{\rm F}}\right)^2\simeq -0,\!14<-\gamma_{a^{-1}=0}^{\rm RPA}
\ee

La question du signe de $\gamma$ à la limite unitaire, qui détermine crucialement la nature à trois phonons ($\gamma>0$) ou à quatre phonons ($\gamma<0$) des mécanismes d'amortissement du son dans le régime faiblement collisionnel à basse température, reste donc largement ouverte.\footnote{L'amortissement étudié expérimentalement dans la référence \cite{Zwierlein_visco} est dans le régime hydrodynamique, au sens de la note \ref{note2_33}. Cette référence ne permet donc pas de trancher.}

\subsection{Décrire les modes de haute fréquence}
\label{sec5.2}

L'hydrodynamique quantique, avec sa branche acoustique presque linéaire en le nombre d'onde $q$, ne peut décrire de manière fiable les ondes sonores de pulsation $\omega_\qq>mc^2/\hbar$ du superfluide de fermions.  Ignorante de la nature composite des paires liées $\uparrow\downarrow$, elle est totalement inapplicable aux pulsations $\omega\approx E_{\rm paire}/\hbar$, où $E_{\rm paire}$ est l'énergie de liaison d'une paire: à ces pulsations, les paires peuvent se briser en deux excitations fermioniques $\chi$ (la conservation de l'énergie ne l'interdit plus), voir la figure \ref{fig4}a. 

Il faut alors avoir recours à une description microscopique du gaz de fermions.  À température nulle, la principale méthode disponible est celle de la théorie variationnelle BCS dépendant du temps \cite{BlaizotRipka}.  Sa spécialisation au régime de réponse linéaire donne l'équation aux valeurs propres suivante sur l'énergie $z$ des modes de vecteur d'onde $\qq$:
\be
\label{eq1_43}
\mathrm{det}\, M(\qq,z)=0\quad\mbox{avec}\quad M(\qq,z)=\begin{pmatrix} M_{|\Delta||\Delta|}(\qq,z) & M_{|\Delta|\theta}(\qq,z)\\ M_{\theta|\Delta|}(\qq,z) & M_{\theta\theta}(\qq,z)\end{pmatrix}
\ee
où les coefficients de la matrice $2\times 2$ correspondent à une réponse en le module $|\Delta|$ ou en la phase $\theta$ du paramètre d'ordre complexe $\Delta(\rr,t)$.  Dans la limite BCS d'interaction faible $k_{\rm F}a\to 0^-$, les éléments non diagonaux sont habituellement négligés (à juste titre) et la dynamique se découple en mode de module et mode de phase; dans le cas général, cette distinction ne vaut plus. 

L'exploration des solutions de l'équation (\ref{eq1_43}) a commencé. À nombre d'onde $q$ fixé, on trouve sous le bord $\veps_\qq^{\rm bord}$ du continuum de paire brisée  au plus une racine,  celle $\hbar\omega_\qq$ de la branche acoustique.  Sur l'intervalle $z\in\,]\veps_\qq^{\rm bord},+\infty[$, la fonction $\mathrm{det}\, M(\qq,z)$ admet une ligne de coupure,\footnote{Les éléments de matrice de $M(\qq,z)$ comportent une intégrale sur le vecteur d'onde $\kk$ d'un des fragments de dissociation d'une paire liée du condensat, avec dans l'intégrande le dénominateur d'énergie correspondant $z-(\veps_\kk+\veps_{\qq-\kk})$; par définition, le dénominateur peut donc s'annuler quand $z$ appartient au continuum de paire brisée, voir la légende de la figure \ref{fig4}.} il faut mettre un décalage infinitésimal $\ii 0^+$ dans $z$ pour lui donner un sens; elle acquiert alors une partie imaginaire, qu'on n'arrive pas à annuler simultanément avec la partie réelle, et l'équation (\ref{eq1_43}) n'admet pas de solution.  En revanche, on peut en trouver une, complexe $z_\qq$ de partie imaginaire non infinitésimale $<0$, en prolongeant analytiquement la fonction $z\mapsto\mathrm{det}\, M(\qq,z)$ du demi-plan complexe supérieur au demi-plan inférieur à travers sa ligne de coupure (ce qu'indique la flèche $\downarrow$ en indice):
\be
\label{eq1_44}
\mathrm{det}M_{\downarrow}(\qq,z_\qq)=0\quad\mbox{avec}\quad \im z_\qq<0
\ee
Il existe donc un mode collectif dans le continuum, qui s'amortit exponentiellement en temps par émission de paires brisées.  Le calcul a été fait d'abord dans la limite BCS $k_{\rm F}a\to 0^-$, aussi bien pour des fermions neutres que pour les électrons d'un supraconducteur, dans la référence \cite{AndrianovPopov}.  Il a été ensuite généralisé aux gaz d'atomes froids fermioniques pour une valeur quelconque de $k_{\rm F}a$, sans qu'on puisse plus négliger les éléments non diagonaux $M_{|\Delta|\theta}$ et $M_{\theta|\Delta|}$ \cite{prlhiggs,crashiggs}. La branche d'Andrianov-Popov subsiste jusqu'à $1/k_{\rm F}a=0,\!55$ (point d'annulation $\mu=0$ du potentiel chimique dans la théorie BCS) et présente toujours un départ à $2|\Delta|$ quadratique en $q$ avec un coefficient complexe:
\be
\label{eq1_45}
z_\qq\stackrel{\mu>0}{\underset{q\to 0}{=}}2|\Delta|+\zeta\frac{\hbar^2q^2}{4 m_*}+O(q^3) \quad (\im\zeta<0)
\ee
où $m_*$ est la masse effective d'une quasi-particule fermionique $\chi$ à l'endroit $k=k_0$ de son minimum d'énergie.\footnote{On l'aura compris, la masse effective est telle que $\veps_\kk-E_{\rm paire}/2\sim\hbar^2(k-k_0)^2/2m_*$ quand $k\to k_0$. La mise à l'échelle par $m_*$ dans l'équation (\ref{eq1_45}) assure que $\zeta$ a une limite finie et non nulle lorsque $k_{\rm F}a\to 0^-$ \cite{AndrianovPopov}. Dans ce même régime, la référence pourtant connue \cite{LV} prédit un comportement fantaisiste de $z_\qq$ à faible $q$,  avec une partie imaginaire tendant vers zéro linéairement en $q$, voir son équation (2.38). La quantité $\zeta$ n'a ici rien à voir avec celle de l'équation (\ref{eq1_42}), il y a une coïncidence malheureuse de notations.} Nous avons écrit ici $2|\Delta|$ plutôt que $E_{\rm paire}$, où $\Delta$ est le paramètre d'ordre à l'équilibre, même si la théorie BCS est incapable de distinguer (on a exactement $E_{\rm paire}=2|\Delta|$ pour tout $\mu>0$ dans cette théorie), afin d'évoquer le mécanisme de Higgs \cite{Higgs} dont on pense que le mode collectif du continuum relève \cite{Varma};\footnote{Comme ce mécanisme résulte de la brisure de symétrie $U(1)$, ici par condensation des paires liées, il doit être caractérisé par l'échelle d'énergie associée au paramètre d'ordre, c'est-à-dire $|\Delta|$ à un facteur près; c'est bien ce que trouve la référence \cite{Higgs}, voir son équation (2b). En revanche, l'échelle d'énergie $E_{\rm paire}$ est reliée à la brisure de paires, pas à leur condensation, donc n'a a priori aucun rapport avec la branche de Higgs. Le fait d'avoir $E_{\rm paire}=2|\Delta|$ est source de confusion et empêche de découpler les deux phénomènes. Il serait par ailleurs intéressant de voir si la propriété $E_{\rm paire}=2|\Delta|$ reste rigoureusement vraie à température nulle dans une théorie plus élaborée que BCS ou dans les expériences.} d'ailleurs, dans la limite opposée $k_{\rm F}a\to 0^+$ d'un condensat de dimères bosoniques, où $2|\Delta|\ll E_{\rm paire}\sim 2|\mu|\sim\hbar^2/ma^2$ (on a cette fois $\mu<0$), on trouve bien une branche d'excitation collective commençant quadratiquement à $2|\Delta|$ et non pas à $E_{\rm paire}$ \cite{crashiggs}. L'extension de l'équation (\ref{eq1_43}) à température non nulle (au-delà d'une simple généralisation de type BCS en champ moyen, peut-être insuffisante\footnote{Cette généralisation n'est pas une panacée, comme on le voit sur la partie imaginaire de la branche acoustique. Pour $\gamma>0$ (mais pas pour $\gamma<0$), ce reproche peut être fait à la RPA déjà à température nulle, puisqu'elle prédit à tort une pulsation propre $\omega_\qq$ purement réelle. Cependant, ça ne semble pas être très grave car la partie imaginaire $(-1/2)\Gamma_\qq(T=0)\approx q^5$ obtenue par l'hydrodynamique quantique (amortissement de Belyaev) vient se perdre dans les termes sous-sous-dominants négligés dans l'équation (\ref{eq1_24}). Ce problème est plus visible à faible $q$ à température non nulle, où $\Gamma_\qq(T>0)$ commence linéairement en $q$ avec un coefficient en $T^4$ (l'exposant $\nu$ vaut 5 dans le tableau \ref{table1} pour la loi d'échelle $q\propto T$ de la note \ref{note2_33}), ce dont la théorie de type BCS en champ moyen ne peut rendre compte (elle prédit un coefficient en $O[-\exp(E_{\rm paire}/2k_{\rm B}T)]$ \cite{Kulik,Tempere} puisque les seuls nombres d'occupation thermiques qu'elle fait apparaître sont ceux $\bar{n}_\kk=1/[\exp(\veps_\kk/k_{\rm B}T)+1]$ des quasi-particules fermioniques $\chi$). En d'autres termes, la linéarisation des équations BCS dépendant du temps ou, ce qui revient au même, l'approximation des fluctuations gaussiennes prend en considération le couplage $\phi-\chi$ mais pas le couplage $\phi-\phi$.}) reste à notre connaissance une question ouverte.  

D'un point de vue expérimental, dans les atomes froids ou les supraconducteurs, l'excitation à des pulsations $\omega>E_{\rm paire}/\hbar$ a été effectuée seulement à nombre d'onde nul, où il n'y a d'après les théories à température nulle pas de mode collectif du continuum, le poids spectral du mode tendant vers 0 lorsque $q\to 0$ \cite{prlhiggs}; on observe simplement aux temps longs des oscillations du paramètre d'ordre à la pulsation $E_{\rm paire}/\hbar$ (c'est l'effet du bord non nul du continuum) qui s'amortissent en loi de puissance $t^{-\alpha}$ \cite{Shimano,ValePRL} par le même mécanisme que l'étalement du paquet d'ondes gaussien d'une particule libre en mécanique quantique ordinaire (l'excitation percussionnelle crée un \g{paquet d'ondes} de paires brisées $(\kk,-\kk)$ dans le continuum, dont l'évolution gouvernée par la relation de dispersion $2\veps_\kk$ est de manière effective unidimensionnelle pour $k_0>0$ ($\mu>0$), auquel cas $\alpha=1/2$ \cite{Volkov}, et tridimensionnelle pour $k_0=0$ ($\mu<0$), auquel cas $\alpha=3/2$ \cite{Gurarie}).\footnote{Les observations de la référence \cite{ValePRL} dans un gaz unitaire de fermions à $T\neq 0$ soulèvent cependant plusieurs questions : (i) l'exposant $\alpha\simeq 1\pm 0,\!15$ mesuré est fort différent de la valeur prédite théoriquement ($\alpha=1/2$ à la limite unitaire), (ii) il ne peut être exclu que la décroissance de l'amplitude des oscillations soit en réalité exponentielle, (iii) au contraire de l'amplitude, la pulsation des oscillations ne présente aucune réduction observable lorsque $T$ se rapproche de la température de transition $T_{\rm c}$ (où il n'y a plus de brisure de symétrie $U(1)$ et $|\Delta|$ tend vers zéro) ce qui semble incompatible avec la qualification d'oscillations de Higgs utilisée dans cette référence (la pulsation mesurée n'est pas proportionnelle à $|\Delta|/\hbar$), mais suggère aussi une constance de $E_{\rm paire}$ en température assez troublante (la pulsation mesurée devrait être donnée par $E_{\rm paire}/\hbar$ puisque l'excitation est faite à $q=0$). Rappelons par ailleurs que la température n'est jamais très faible dans l'expérience, $T\gtrsim 0,\!1T_{\rm F}$, voir notre section \ref{sec2}, ce qui rend la théorie à $T=0$ stricto sensu inapplicable.} L'observation du mode du continuum (à $q>0$) et la mesure précise de sa relation de dispersion $z_\qq$ restent donc à faire (des pistes sont données dans les références \cite{prlhiggs,Bragg}). 


\selectlanguage{english}
\section*{English version (la version fran\c{c}aise commence \`a la page \pageref{debf})}
\label{debe}

\raz

\textit{To obtain the English version, we translated the French version with \texttrademark DeepL Pro and a self-made glossary of technical terms. The result was improved with \texttrademark DeepL Write.}

\section{Introduction and general presentation}
\label{sece1}

This text is essentially the transcript of our 90-minute talk at the prospective symposium “Open questions in the quantum many-body problem” held at the Institut Henri Poincaré in Paris, from July 8 to 12, 2024, hence its style and level of precision differ from that of a usual research article. It is more complete than the presentation on Section \ref{sece4} (treated briefly at the oral presentation) and on Section \ref{sece5} (omitted at the oral presentation due to lack of time). Footnotes can be ignored on first reading. The presentation was recorded and is available online on the IHP Carmin channel (click \href{https://www.carmin.tv/fr/collections/symposium-open-questions-in-the-quantum-many-body-problem}{here}).

The system considered is inspired by experiments on cold atoms: it is a three-dimensional two-component Fermi gas (meaning two internal states $\uparrow$ and $\downarrow$) in an immaterial trap - made of light, at very low temperatures in the microkelvin range. This is the worthy descendant of laser-cooled atomic gases (in the famous “optical molasses”, see the 1997 Nobel Prize in Physics awarded to Steven Chu, William Phillips and Claude Cohen-Tannoudji) and then of evaporation-cooled gaseous atomic Bose-Einstein condensates (see the 2001 Nobel Prize in Physics awarded to Eric Cornell, Carl Wieman and Wolfgang Ketterle). 

Compared to their illustrious predecessors, cold fermionic atomic gases have the advantage (i) of being composed of fermions, thus covering both possible statistics (the gas can always be “bosonized” by forming strongly bound pairs $\uparrow\downarrow$) and making a direct link with the electron systems (fermions!) of solid-state physics, (ii) of remaining collisionally stable (weak three-body losses by recombination to deep molecular states) even in the strongly interacting regime as in the famous “unitary limit” described below (unlike, for the moment, bosonic cold-atom gases), and (iii) of constituting beautiful, simple and universal model systems in this regime, thanks to the negligible range of van der Waals interactions between $\uparrow$ and $\downarrow$ (more precisely, the associated van der Waals length is negligible); as we shall see, this allows the interaction to be replaced by contact conditions on the $N$-body wave function, depending solely on the $s$-wave scattering length $a$, which experimentalists can adjust at will by means of a Feshbach resonance, simply by applying a well-chosen uniform magnetic field.

Our system is not unrelated to those of other presentations at the symposium. The link is obvious with Tilman Enss's contribution on the viscosity of strongly interacting Fermi gases \cite{Enss}, which complements our own. If we place our fermions in an optical lattice, with about one particle per site (close to half-filling), we recover the problems of strongly correlated fermions and superconductivity at high critical temperature discussed by Antoine Georges. In a regime of strong on-site interaction $U_{\uparrow\downarrow}$ compared to tunnel coupling $t$ between neighboring sites, $U_{\uparrow\downarrow}\gg t$, the system is described by a Heisenberg-type spin model Hamiltonian, with magnetic coupling $J\propto t^2/U_{\uparrow\downarrow}$, which links up with Sylvain Capponi's talk \cite{Cappo}. If we return to a uniform system (without lattice), but apply an artificial gauge field (a fictitious magnetic field) to our fermionic yet neutral cold atoms - which is what experimentalists know how to do, see Sylvain Nascimbène's talk \cite{Nascim} - we come across problems related to the talks by Thierry Jolicœur \cite{Jolic} (on the 2D fractional quantum Hall effect) and Carlos Sá de Melo \cite{SadeM} (spin-orbit coupling in one spatial dimension). All these bridges to solid-state physics are not so easy to build, however, because of non-conservative parasitic effects, finite sample sizes and difficulty in getting down to sufficiently low temperatures (in units of the Fermi temperature $T_{\rm F}$ or the magnetic coupling temperature $J/k_{\rm B}$), see the presentations by Wolfgang Ketterle, Sylvain Nascimbène and Antoine Georges. 

Let's finish with the outline of our contribution. In Section \ref{sece2}, we start from reality, outlining the progress of experiments on cold atoms since the 1980s and the situation reached in the case of fermions. In Section \ref{sece3}, we adopt a microscopic point of view, of interactions replaced by contact conditions, and review some open questions in the few-fermion problem. In Section \ref{sece4}, on the other hand, we adopt a macroscopic point of view, that of a low-energy effective theory (quantum hydrodynamics), and review some open questions related to the interaction between phonons (the quanta of sound waves) in the superfluid phase. Finally, in the short Section \ref{sece5}, we cross points of view and list some open questions that require a microscopic theoretical treatment of the full many-body problem.

\section{A fairly recent physical system}
\label{sece2}

Let's start by putting our Fermi gases into context, with a brief history of cold atoms. 

The adventure began in the early 1980s with the laser cooling of alkalis. The low temperatures reached are spectacular when expressed in kelvins, $T\approx 1\,\mu{\rm K}$, but the spatial densities are unfortunately very low, $\rho\lesssim 10^{10}{\rm at/cm}^3$, so that the gases have very low quantum degeneracy, i.e.\ very low density in phase space, $\rho\lambda^3\ll 1$, where $\lambda$ is the thermal de Broglie wavelength of atoms with mass $m$: 
\begin{equation}\lambda=\left(\frac{2\pi\hbar^2}{m k_{\rm B} T}\right)^{1/2}
\end{equation} The effects of quantum statistics (bosonic or fermionic) are imperceptible. 

This all changed in 1995, when Eric Cornell and Carl Wieman at JILA \cite{CBEJila}, shortly followed by Wolfgang Ketterle at MIT \cite{CBEMit}, achieved Bose-Einstein condensation (BEC), obviously on bosonic isotopes, thanks to evaporative cooling in non-dissipative trapping potentials with harmonic bottoms.\footnote{Reference \cite{Schreck} later succeeded, using clever tricks, in obtaining a Bose-Einstein condensate without evaporation, by laser cooling alone (see also reference \cite{Vuletic}); in particular, this involved (i) using a narrow atomic line with low saturation to keep the temperature limit of laser cooling \cite{josa} as low as possible, and (ii) preventing spontaneously emitted photons, which carry away part of the energy of atomic motion, from depositing it back into the gas by reabsorption.} Transition temperatures remain in the range of laser cooling, $T_{\rm c}^{\rm BEC}\simeq 0.1\ \mbox{to}\ 1\,\mu{\rm K}$, but spatial densities are considerably higher, $\rho=10^{12}\ \mbox{to}\ 10^{15}\mbox{at/cm}^3$, enabling quantum degeneracy $\rho\lambda^3\gtrsim 1$ to be achieved. 

Finally, in 2004, evaporative cooling was successfully extended to fermionic isotopes down to the transition temperature \cite{Jin,Zwier}; gases with two internal states $\uparrow$ and $\downarrow$ no longer form Bose-Einstein condensates, but condense in $\uparrow\downarrow$ pairs by the BCS mechanism \cite{revuefer}: the attractive van der Waals interactions between $\uparrow$ and $\downarrow$, in the presence of a Fermi sea in each internal state, lead to the formation of bound pairs, the famous Cooper pairs, “composite bosons”, which can form a condensate at sufficiently low temperatures, $T<T_{\rm c}^{\rm BCS}$. The lowest experimentally accessible temperatures are of the order of $0.1 T_{\rm F}$, where the Fermi temperature $T_{\rm F}$ remains of the order of a microkelvin; this is nevertheless sufficient to go below $T_{\rm c}^{\rm BCS}$ because the interactions between $\uparrow$ and $\downarrow$ are made very strong by means of a two-body scattering resonance (magnetic Feshbach resonance): the transition temperature $T_{\rm c}^{\rm BCS}$ is then a fraction of $T_{\rm F}$ and we avoid the extreme situation of BCS superconductors, for which $T_{\rm c}^{\rm BCS}\lll T_{\rm F}$ by several orders of magnitude. 

\begin{figure}[tb]
\includegraphics[width=8cm,clip=]{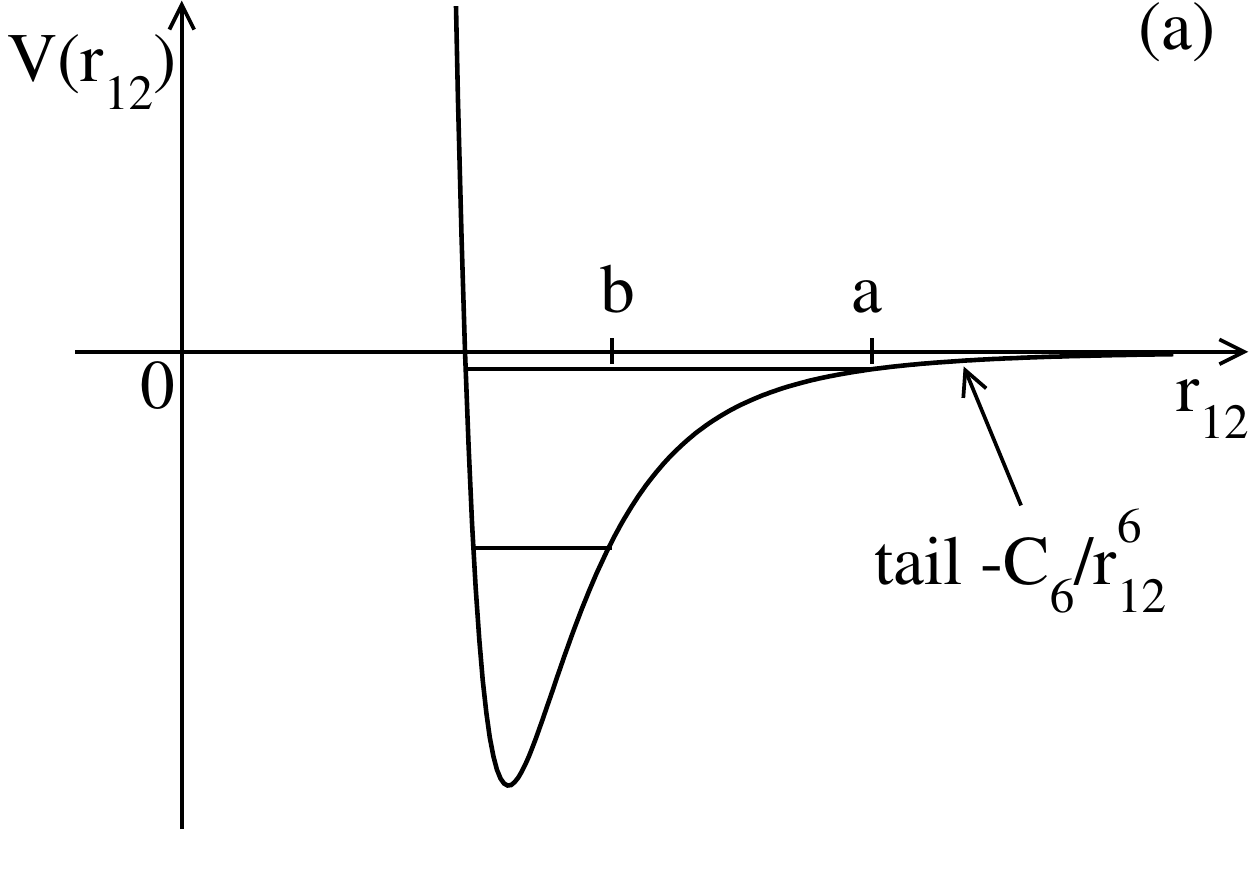}
\includegraphics[width=8cm,clip=]{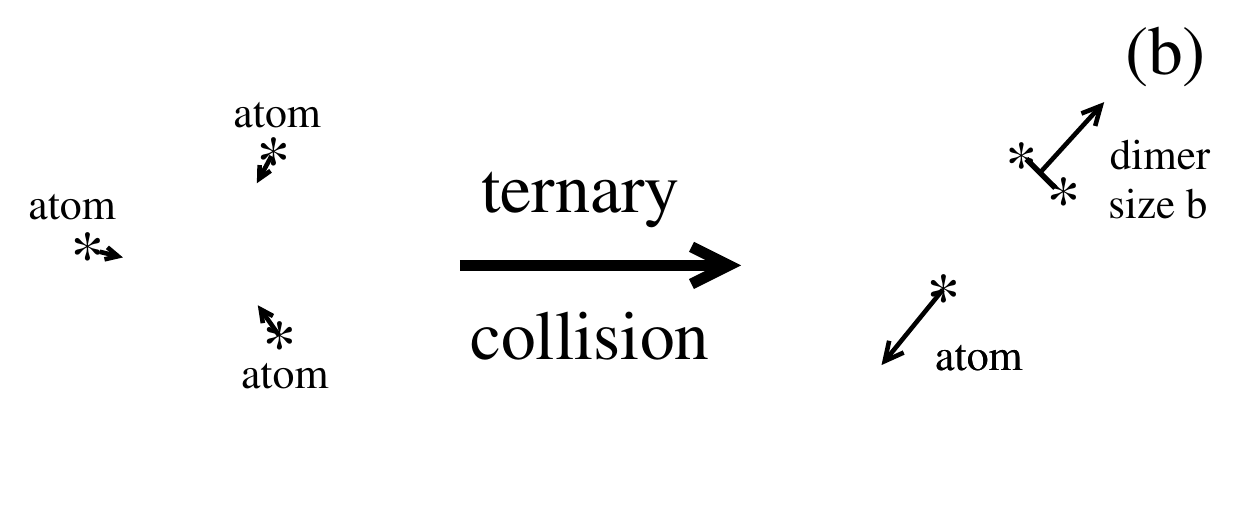}
\caption{(a) Schematic representation of the resonant van der Waals (more precisely, Lennard-Jones) interaction ($|a|\gg b$) between $\uparrow$ and $\downarrow$ fermions as a function of their relative distance.  The potential supports at least one strongly bound state of extension of the order of the van der Waals length $b\simeq (m C_6/\hbar^2)^{1/4}$ and therefore of binding energy $\approx \hbar^2/mb^2$ and, in the case of a scattering length $a>0$ as in the figure, a last energy level of extension $a$ (of binding energy $\hbar^2/ma^2$) on “the verge of disappearing” (here $m_\uparrow=m_\downarrow=m$ as in the experiments); if $a$ were large but negative ($|a|\gg b, a<0$), this weakly bound state would be on “the verge of appearing”. (b) The strongly bound dimer state can be populated by three-body collisions, which causes particle losses in the Fermi gas, known as three-body losses (the collision products carry away the considerable binding energy $\approx \hbar^2/m b^2$ in the form of kinetic energy and leave the trap). The arrows represent the linear momenta before and after the collision.}
\label{fige1}
\end{figure}

Let's now describe our system of fermionic cold atoms in broad outline, in an attempt to idealize experimental reality. (i) The fermions have two internal states $\uparrow$ and $\downarrow$; as we are not considering here a Rabi coupling between $\uparrow$ and $\downarrow$, our considerations also apply to the case of a mixture of two formally spinless fermionic chemical species; for this reason, we do not assume that the masses $m_\sigma$ of the particles are equal in the two internal states\footnote{In the case where $\uparrow$ and $\downarrow$ are two spin states of the same chemical species, we naturally have $m_\uparrow=m_\downarrow$ in the experiment. However, by applying an optical lattice that couples differently to the two internal states (within a low filling factor limit), we could produce different  effective masses $m_\sigma$. This remains to be done.} and we consider the ratio $m_{\uparrow}/m_{\downarrow}$ as a free parameter. (ii) The fermions are trapped either in isotropic harmonic potentials of the same trapping angular frequency $\omega$ for the two components $\sigma$, 
\begin{equation}\label{eqe1_5}
U_\sigma(\rr)=\frac{1}{2} m_\sigma\omega^2 r^2
\end{equation} where $\rr$ is the 3D position vector, or in the cubic quantization box $[0,L]^3$ common to both components, with the usual periodic boundary conditions.\footnote{\label{notee6} Experimentally, flat-bottom potentials can be produced using Laguerre-Gauss or Bessel-Gauss beams and laser light sheets, after gravity compensation (levitation of atoms) by a magnetic field gradient \cite{Zoran,buxida,revbox}.} (iii) The van der Waals interaction between the two internal states $\uparrow$ and $\downarrow$, shown schematically in Figure \ref{fige1}a, is effectively made very strong (resonant) in $s$ wave (relative orbital angular momentum $l=0$) by application of a suitable magnetic field\footnote{\label{notee2_5}Without going into too much detail, it should be pointed out that, to understand this resonance, it is necessary to take into account the internal structure of the atoms and describe their binary interaction a minima by a two-channel model, with an open channel $V_{\rm o}(r_{12})$ interaction potential and a closed channel $V_{\rm f}(r_{12})$ interaction potential - think of the singlet and triplet interaction potentials of two spin-1/2 fermions. In a collision, atoms $\uparrow$ and $\downarrow$ enter through the open channel and, by conservation of energy, also exit through the open channel, as their relative incoming kinetic energy is less than the difference of dissociation limits $V_{\rm f}(+\infty)-V_{\rm o}(+\infty)>0$. As there is a coupling between the two channels, however, the atoms virtually populate the closed channel during the collision. The applied magnetic field $B$ induces a different Zeeman shift in the two channels. A clever choice of $ B$ is then all that's needed to ensure that the energy of a bound state in $V_{\rm f}(r_{12})$ - not the bare energy, but the energy shifted by the coupling - almost coincides with the dissociation limit $V_{\rm o}(+\infty)$, inducing a two-body scattering (or collision) resonance in the open channel and making the scattering length $a$ diverge.} so that the scattering length $a$ between two atoms $\uparrow$ and $\downarrow$ (defined mathematically in Section \ref{sece3.1}) is sufficiently large in absolute value (it can be positive or negative) so that 
\begin{equation}\label{eqe1_6}
\rho^{1/3}|a|\gtrsim 1
\end{equation} We recall that the theory of weakly interacting Bose gases relies on the small parameter $(\rho a^3)^{1/2}\ll 1$, see Jan Solovej's contribution \cite{Solov} to the symposium proceedings; condition (\ref{eqe1_6}) is therefore, on the contrary, the mark of a strongly interacting gas. 
The scattering length $a$ is also much greater in absolute value than the interaction range $b$, defined in Figure \ref{fige1}a, 
\begin{equation}|a|\gg b
\end{equation} which is indeed the hallmark of a two-body scattering resonance. As $b$ is of the order of a few nanometers in the experiments, we also have 
\begin{equation}b\ll \rho^{-1/3},\lambda
\end{equation} which gives us the idea of constructing a model system, by taking the limit of a zero-range interaction $b\to 0$ at fixed $a$, characterized only by the algebraic length $a$. This idea will be implemented in Section \ref{sece3.1}. (iv) On the other hand, the interaction is not resonant in $p$ wave (relative orbital angular momentum $l=1$), so the $\uparrow\uparrow$ and $\downarrow\downarrow$ interactions, which occur predominantly in this wave at low energy (due to fermionic antisymmetry), are negligible.

As we shall see in Section \ref{sece3}, the existence of a well-defined model (of energy bounded from below when $b\to 0$) is a mathematically non-trivial problem. We can already propose a necessary condition, inspired by experimental reality. Since the van der Waals interaction supports (at least) a two-body bound state of size $\approx b$, as shown in Figure \ref{fige1}a, the gas phase considered so far and seen in experiments is only a metastable phase, temporarily escaping the solidification predicted by the laws of equilibrium physics, a solidification of which three-body losses are the precursors (see Figure \ref{fige1}b). These losses occur at a rate estimated as follows in reference \cite{Shlyap} for equal masses: 
\begin{equation}\Gamma_{\rm losses}^{\scriptsize\mbox{3-body}}\propto \frac{\hbar}{mb^2} \ \mbox{Proba}(\mbox{3 fermions $\uparrow\uparrow\downarrow$ or $\downarrow\downarrow\uparrow$ in a same ball of radius $b$})
\label{eqe1_7}
\end{equation} 
The first factor represents the relevant energy scale of this recombination process: this is the binding energy of the strongly bound dimer formed, and the length scale $|a|\gg b$ cannot come into play. The second factor takes into account the fact that the three-fermion process cannot occur if one of the fermions is separated from the other two by a distance $\gg b$, by quasi-locality in position space: the interaction range and the size of the strongly bound dimer are both of the order of $b$. The proportionality factor in Equation (\ref{eqe1_7}) depends on the details of the microscopic physics. This leads to an experimental stability condition for the Fermi gas in the $b\to 0$ limit of a contact interaction: 
\begin{equation}\Gamma_{\rm losses}^{\scriptsize\mbox{3-body}}\underset{b\to 0}{\to} 0
\end{equation} The study of this system, though gaseous, is made non-trivial by the interaction strength. For example, since $k_{\rm F}|a|\approx 1$, where $k_{\rm F}=(3\pi^2\rho)^{1/3}$ is the Fermi wave number, the superfluid transition temperature is a priori of the order of the Fermi temperature $T_{\rm F}=E_{\rm F}/k_{\rm B}$ (there is no other scale available than the Fermi energy $E_{\rm F}=\hbar^2k_{\rm F}^2/2m$) and will be difficult to calculate accurately: BCS theory will be qualitative at best, and quantum Monte Carlo methods are difficult to apply to fermions; the challenge has, however, been met by reference \cite{SvisunovTc}, in the symmetric case of equal masses and chemical potentials in the two components, where Monte Carlo methods free of the famous “sign problem” exist.

\section{Open questions from a microscopic point of view}
\label{sece3}

In this section, fermion interactions are replaced in a zero-range limit by contact conditions on the $N$-body wave function, the Hamiltonian operator then reducing to that of the ideal gas (Wigner-Bethe-Peierls model \cite{Wigner,BethePeierls}). 

\subsection{Defining the Wigner-Bethe-Peierls model}
\label{sece3.1}

\begin{figure}[t]
\includegraphics[width=12cm,clip=]{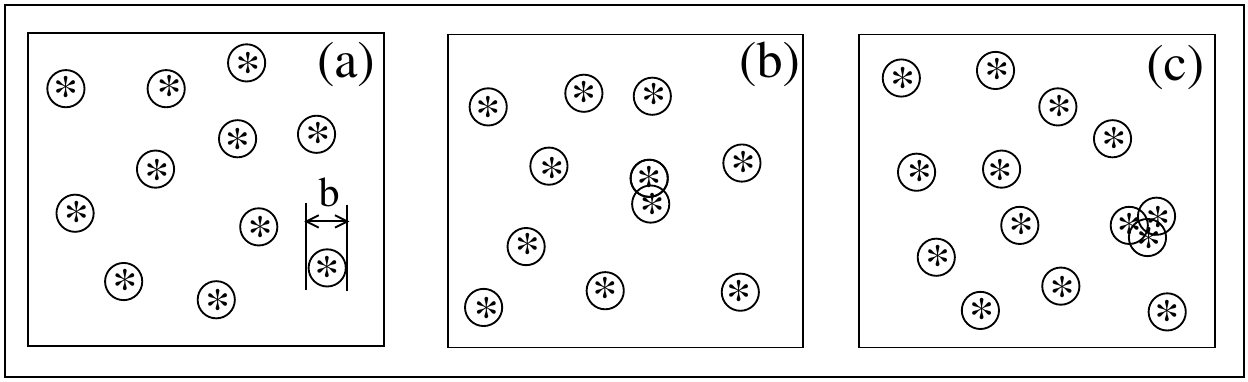}
\caption{Photograph of the Fermi gas showing the positions (stars) of the $N$ fermions, as would be obtained with a quantum gas microscope, in the limit of an interaction range $b\to 0$ (for illustrative purposes, we have surrounded each star with a circle of diameter $b$). (a) Typical case: the particles are separated by a distance $\gg b$ and do not interact. This sets the Hamiltonian operator (\ref{eqe2_9}) of the Wigner-Bethe-Peierls model. (b) Case where two particles $\uparrow$ and $\downarrow$, well separated from the others, undergo a binary collision. This sets the contact conditions (\ref{eqe1_12}) of the model. (c) Case of an isolated ternary collision. This raises the question of the need for three-body contact conditions.}
\label{fige2}
\end{figure}

To build the model, let's start from the simple perception that a photograph of the gas would give us, i.e.\ a measurement of the positions of the $N$ fermions, as quantum gas microscopes have recently been able to do in the homogeneous case \cite{Tarik}. In the limit where the interaction range $b$ tends to zero, the typical photograph looks like Figure \ref{fige2}a: the fermions are separated two by two by a distance $\gg b$ and the interaction potential $V(\rr_i-\rr_j)$ is negligible. The $N$-body wave function obeys in this case the stationary Schrödinger equation 
\begin{equation}\label{eqe1_9}
E\psi = H_{\rm ideal\, gas}\psi
\end{equation} with the Hamiltonian operator of the ideal gas, the sum of the kinetic energy $\pp^2/2m_\sigma$ and trapping $U_\sigma(\rr)$ terms in each internal state $\sigma$:
\begin{equation}\label{eqe2_9}
H_{\rm ideal\, gas}=\sum_{i=1}^{N_\uparrow}\left(\frac{\pp_i^2}{2m_\uparrow} + U_\uparrow(\rr_i)\right) + \sum_{j=N_\uparrow+1}^{N}\left(\frac{\pp_j^2}{2m_\downarrow} + U_\downarrow(\rr_j)\right)
\end{equation} It is convenient here to number the particles so that the first $N_\uparrow$ are in internal state $\uparrow$ and the last $N_\downarrow$ are in internal state $\downarrow$; the wave function $\psi(\rr_1,\ldots,\rr_N)$ is then an antisymmetric function of the first $N_\uparrow$ positions and an antisymmetric function of the last $N_\downarrow$ positions. 
 Some pictures, however, will resemble Figure \ref{fige2}b: two fermions $i$ and $j$, of different internal states, respectively $\uparrow$ and $\downarrow$, are separated from the others by a distance $\gg b$ but are separated from each other by a distance $\approx b$ and are therefore affected by the interaction potential $V(\rr_i-\rr_j)$. The correct way of looking at this is to say that $i$ and $j$ are undergoing isolated two-body scattering in the gas, which has two consequences, one qualitative, the other quantitative.

Qualitatively, we understand that in our very low density gas (in the sense of $\rho b^3\ll 1$), it is better to characterize the interaction between $\uparrow$ and $\downarrow$ by its two-body scattering amplitude, more generally by a transmission operator known as the $T$ matrix, than by the function $V(\rr)$ itself; as the interaction occurs in the $s$ wave, the scattering amplitude $f$ is isotropic and depends only on the relative wave number $k_{\rm rel}$ of the two particles; in the limit $b\to 0$ at a fixed scattering length $a$, we then have the low-energy expansion\footnote{If $V(\rr)$ decays faster than $1/r^7$ at infinity, we can put an $O(k_{\rm rel}^4b^3)$ in the denominator.} 
\begin{equation}
\label{eqefk}
f_{k_{\rm rel}} = \frac{-1}{a^{-1}+\ii k_{\rm rel}-(1/2) k_{\rm rel}^2 r_{\rm e} + O(k_{\rm rel}^3b^2)}
\end{equation} We assume in the following that the effective range $r_{\rm e}$ of the interaction is $O(b)$ and therefore becomes negligible as $b\to 0$.\footnote{\label{notee1_10} Experimentally, however, there are so-called narrow magnetic Feshbach resonances, for which negative $r_{\rm e}$ is gigantic on the atomic scale and can be of the order of $1/k_{\rm F}$, due to an unusually weak coupling between the open and closed channels of our note \ref{notee2_5}, see reference \cite{Gur}. These resonances are difficult to use, as they require very good control of the magnetic field. The existence of a non-zero effective range as $b\to 0$, however, has the advantage of stabilizing the gas in the unstable regime of Section \ref{sece3.2} (the spectrum remains bounded from below and the three-body loss rate (\ref{eqe1_7}) tends to zero), and should enable the preparation and observation of long-lived Efimovian bound states, provided the mass ratio $m_\uparrow/m_\downarrow$ is large enough. The experiment remains to be done.} So the scattering amplitude reduces to the universal form for a contact interaction 
\begin{equation}\label{eqe2_10}
f_{k_{\rm rel}}^{\rm contact}=\frac{-1}{a^{-1}+\ii k_{\rm rel}}\end{equation}

Quantitatively, we expect the two nearby $\uparrow$ and $\downarrow$ fermions to decouple from the $N-2$ others in the $N$-body wave function, in the sense that 
\begin{equation}\label{eqe3_10}
\psi(\rr_1,\ldots,\rr_N) \underset{r_{ij}=O(b)}{\simeq} \phi(\rr_i-\rr_j) A_{ij}(\RR_{ij}; (\rr_k)_{k\neq i,j})
\end{equation} where $\RR_{ij}=(m_\uparrow\rr_i+m_\downarrow \rr_j)/(m_\uparrow+m_\downarrow)$ is the center of mass position of the particles $i$ and $j$, $\rr_{ij}=\rr_i-\rr_j$ is their relative position, $(\rr_k)_{k\neq i,j}$ is the $(N-2)$-uplet of the positions of the other particles, the function $A_{ij}$ is generally not known, and $\phi(\rr)$ is a two-body scattering state, solution of Schrödinger equation 
\begin{equation}\label{eqe1_11}
\veps\phi(\rr)= -\frac{\hbar^2}{2 m_{\rm rel}} \Delta \phi(\rr) + V(\rr)\phi(\rr)
\end{equation} for the relative motion of mass $m_{\rm rel}=m_\uparrow m_\downarrow/(m_\uparrow+m_\downarrow)$ at an energy $\veps$ given in reference \cite{Gen} (see its Equation (85)) but which we will retain only as $\approx \hbar^2 k_{\rm typ}^2/2 m_{\rm rel}$, where the typical wave number $k_{\rm typ}$ in the gas is of the order of $k_{\rm F}$ for $T=O(T_{\rm F})$. In the limit $b\to 0$, it is in fact sufficient to analyze Equation (\ref{eqe1_11}) in the interval
\begin{equation}\label{eqe2_11}
b\ll r \ll k_{\rm typ}^{-1}
\end{equation}
with the case $r\leq b$ providing only non-universal details of the interaction and the case $r>k_{\rm typ}^{-1}$ invalidating the factorization (\ref{eqe3_10}) (the pair $ij$ is no longer well isolated as in the photo in Figure \ref{fige2}b). The first inequality in Equation (\ref{eqe2_11}) allows $V(\rr)$ to be set to zero in the right-hand side, and the second allows energy $\veps$ to be set to zero in the left-hand side of Equation (\ref{eqe1_11}), hence the simplified Schrödinger equation 
\begin{equation}0=-\frac{\hbar^2}{2 m_{\rm rel}} \Delta \phi(\rr)
\end{equation} Its general solution in the $s$ wave (rotationally symmetric) is a linear combination of the constant solution $1$ (the incoming wave of zero energy) and the Coulomb solution (the scattered wave) with a relative amplitude fixed by $V(\rr)$ at short distances: 
\begin{equation}\label{eqe4_11}
\phi(\rr)=\mathcal{N}\left(1-\frac{a}{r}\right) = \frac{1}{a}-\frac{1}{r}
\end{equation} By definition, see Jan Solovej's contribution \cite{Solov}, the quantity $a$ is the scattering length of the potential. In the third expression, we have chosen the convenient normalization (factor $\mathcal{N}$ taken equal to $1/a$ in the second expression) to have a finite result at the $a^{-1}=0$ scattering resonance. 

We thus arrive naturally at the definition of the Wigner-Bethe-Peierls model for our three-dimensional system of $(N_\uparrow,N_\downarrow)$ two-component fermions with zero-range interaction and $s$-wave scattering length $a\neq 0$: 
\begin{enumerate}\item the Hamiltonian operator is the same as that of the ideal gas, as in Equations (\ref{eqe1_9},\ref{eqe2_9}) \item there is fermionic antisymmetry of the state vector $\psi$ for the first $N_\uparrow$ and for the last $N_\downarrow$ positions
\item the interaction is described not by a potential $V$ but by the following contact conditions on $\psi$: for any index $i\in\{1,\ldots,N_\uparrow\}$ and any index $j\in\{N_\uparrow+1,\ldots,N=N_\uparrow+N_\downarrow\}$, there exists a function $A_{ij}$ such that\footnote{The functions $A_{ij}$ are not independent. Fermionic antisymmetry dictates that $A_{ij}(\RR_{ij},(\rr_k)_{k\neq i,j})=(-1)^{i-1}(-1)^{j-(N_\uparrow+1)} A_{1,N_\uparrow+1}(\RR_{ij}, (\rr_k)_{k\neq i,j})$ (to bring them to the first position in their respective internal states and thus reveal the function $A_{1, N_\uparrow+1}$, we had to pass $\rr_i$ through $i-1$ position vectors of $\uparrow$ fermions and $\rr_j$ through $j-(N_\uparrow+1)$ position vectors of $\downarrow$ fermions, hence the signs).}
\begin{equation}\label{eqe1_12}
\psi(\rr_1,\ldots,\rr_N)\underset{r_{ij}\to 0}{=}A_{ij}(\RR_{ij},(\rr_k)_{k\neq i,j})\left(\frac{1}{r_{ij}}-\frac{1}{a}\right)+O(r_{ij})
\end{equation} where the distance $r_{ij}$ between particles $i$ and $j$ tends to zero at fixed positions of their center of mass $\RR_{ij}$ and other particles $\rr_k$, with the constraints $\RR_{ij}\neq\rr_k \forall k\neq i,j$ and the $\rr_k$ two-by-two distinct (as in Figure \ref{fige2}b). 
\end{enumerate}
Mathematically, point 3 means that the domain of the Hamiltonian operator is not the same as that of the ideal gas: in the absence of interaction ($a=0$), we rightly eliminate solutions that diverge as $1/r_{ij}$, as stated in any good book on quantum mechanics. That's the only difference, but it's a big one!\footnote{A key point is that the scattering state $\phi(\rr)=1/r-1/a$ is square-summable on a neighborhood of the origin, $\int_{r< r_{\rm max}}\dd^3r\, |\phi(\rr)|^2<\infty$: there is therefore no cutoff to be put at short distance and $a$ is the only length associated with the interaction. The situation is different in waves of angular momentum $l>0$: $\phi(\rr)=Y_l^{m_l}(\theta,\varphi)\left(r^l+a_{\rm gen}^{2l+1}/r^{l+1}\right)$ (where the parameter $a_{\rm gen}\neq 0$ generalizing $a$ is a length and $Y_l^{m_l}$ is a spherical harmonic) is then no longer square-summable, and a cutoff and thus a second length must be introduced to characterize the interaction \cite{Ludoondel}.}

Figure \ref{fige2}c, which shows a trio of atoms in close proximity, well separated from the others and undergoing three-body scattering, raises a legitimate question: should we complete the model with three-body contact conditions? four-body? etc. Answer in the next section. 

\subsection{Questions of existence}
\label{sece3.2}

It is not obvious that the Wigner-Bethe-Peierls model, as defined on page \pageref{eqe1_12}, leads to a self-adjoint Hamiltonian (without additional contact conditions) and, above all, to an energy spectrum bounded from below. In fact, we have boldly stretched an energy scale to $-\infty$, the $-\hbar^2/m_{\rm rel} b^2$ scale associated with the interaction range, by taking the $b\to 0$ limit at a fixed scattering length $a$, i.e.\ without letting the interaction strength tend to zero, which could cause the system to collapse, as in the well-known Thomas effect in nuclear physics \cite{Thom}!

The discussion becomes clearer in the special case $a^{-1}=0$, known as the unitary limit (the scattering amplitude (\ref{eqe2_10}) of the model reaches the maximum modulus $k_{\rm rel}^{-1}$ allowed in the $s$ wave by the unitarity of the scattering matrix $S$), as the contact conditions (\ref{eqe1_12}) become scale invariant (this is also the most interesting and open regime, because it corresponds to a maximally interacting gas phase). For simplicity's sake, let's restrict ourselves to $E=0$ energy eigenstates in free space, with the center of mass of the $N$ fermions at rest. As there is no external energy or potential to introduce a length scale, we expect the eigenstate $\psi$ to be scale invariant, i.e.\ a homogeneous function of the coordinates (invariant up to a factor by the homothety $\rr_i\to \lambda \rr_i$ of ratio $\lambda$ applied to the $N$ positions), of the form \cite{FWsep,livre}
\begin{equation}\label{eqe1_14}
\psi(\rr_1,\ldots,\rr_N)=R^{s-\frac{3N-5}{2}}\Phi(\Omega)
\end{equation} where (i) $R$ is the internal hyperradius, the root-mean-square deviation of the mass-weighted positions of the $N$ particles relative to their center of mass $\CC$, 
\begin{equation}\label{eqe2_14}
MR^2 = \sum_{i=1}^{N} m_i (\rr_i-\CC)^2
\end{equation} with $M=\sum_{i=1}^{N} m_i$ the total mass and $M\CC=\sum_{i=1}^{N} m_i \rr_i$; (ii) the scaling exponent (the degree of homogeneity) is conveniently defined by the quantity $s$ after translation of $(3N-5)/2$ - to reveal an $s\leftrightarrow -s$ symmetry; (iii) $\Phi$ is an unknown function of the $3N-4$ hyperangles completing $R$ in the parameterization of $\rr_i-\CC$ in hyperspherical coordinates. Carrying over the ansatz (\ref{eqe1_14}) into the Schrödinger equation (\ref{eqe1_9}) (with $E=0$ and $U_\sigma\equiv 0$ as mentioned) gives an eigenvalue equation for $\Phi$: 
\begin{equation}\label{eqe3_14}
\left[-\Delta_\Omega +\left(\frac{3N-5}{2}\right)^2\right]\Phi(\Omega) = s^2 \Phi(\Omega)
\end{equation} whose eigenvalues are precisely $s^2$! Since the Laplacian $\Delta_\Omega$ is taken on a compact, the unit hypersphere $S_{3N-4}$, the possible values of $s^2$ form a discrete set, belonging to $\mathbb{R}$ if one assumes that the Hamiltonian is Hermitian; we generally don't know how to calculate them, because of the difficult contact conditions (\ref{eqe1_12}) on $\Phi(\Omega)$.\footnote{\label{notee1_14}On the other hand, contact conditions (\ref{eqe1_12}) do not constrain the dependence of $\psi$ on the hyperradius. This is because, if $\psi$ obeys the contact conditions, $f(R)\psi$ also obeys them, provided that the factor $f(R)$ is a regular function of $R$. Indeed, in the limit $r_{ij}\to 0$ with $\RR_{ij}$ and $(\rr_k)_{k\neq i,j}$ fixed, we have $MR^2=m_i(\rr_i-\CC)^2+m_j(\rr_j-\CC)^2+\mbox{const}=m_i\rr_i^2+m_j\rr_j^2+\mbox{const}=(m_i+m_j)\RR_{ij}^2+m_{\rm rel} r_{ij}^2+\mbox{const}=O(r_{ij}^2)+\mbox{const}$. Then $(1/r_{ij}-1/a)O(r_{ij}^2)$ is $O(r_{ij})$ and negligible.}

In the following, we'll simply write formally that $s$ is the root of an even transcendental function, the so-called Efimov function, 
\begin{equation}\label{eqe1_15}
\Lambda_{N_\uparrow,N_\downarrow}(s)=0
\end{equation} without specifying this function (the most direct way to obtain it is to impose the contact conditions (\ref{eqe1_12}) on a Faddeev ansatz written in reciprocal space,\footnote{Let's briefly recall the construction of the ansatz. First, we write the Schrödinger equation for zero energy in the sense of distributions, $H_{\rm ideal\, gas}\psi= \sum_{i=1}^{N_\uparrow} \sum_{j=N_\uparrow+1}^{N} (2\pi\hbar^2/m_{\rm rel})\delta(\rr_{ij}) A_{ij}(\RR_{ij},(\rr_k)_{k\neq i,j})$ where Dirac distributions arise from the action of kinetic energy operators on singularities $1/r_{ij}$, by virtue of the Poisson equation $\Delta_\rr(1/r)=-4\pi\delta(\rr)$, and $m_{\rm rel}$ is the reduced mass of two opposite-spin fermions as we said. We then take its Fourier transform ($\psi\to\tilde{\psi}$, $\Delta_\rr\to -k^2$). Taking advantage of fermionic antisymmetry as in note \ref{notedep} and of translational invariance (the center of mass is at rest), we reduce to $\tilde{\psi}(\kk_1,\ldots,\kk_N)=\frac{\delta(\kk_1+\ldots+\kk_N)}{\sum_{n=1}^{N} \hbar^2 k_n^2/2m_n} \sum_{i=1}^{N_\uparrow} \sum_{j=N_\uparrow+1}^{N} (-1)^{i+j} D((\kk_n)_{n\neq i,j})$ where $D$ is the only unknown function (each $A_{ij}$ is a function of the $(\rr_k-\RR_{ij})_{k\neq i,j}$ of which $D((\kk_n)_{n\neq i,j})$ is the Fourier transform up to a factor).} which leads to a Skorniakov-Ter-Martirosian integral equation - here in the unitary limit at zero energy, in which we inject the Fourier equivalent of ansatz (\ref{eqe1_14}); the resulting transcendental equation (\ref{eqe1_15}) can be written explicitly for $N=3$ \cite{Tignone}, and is written as the determinant of an operator for $N>3$, this operator being given explicitly for $N=4$ in sector $(3,1)$ by reference \cite{MoraCastinPricoupenko} and in sector $(2,2)$ by reference \cite{Endopas4}). We must now distinguish between two cases. 

\paragraph{First case: $s^2>0$}
 There are two possible values for the scaling exponent, one value $>0$ which we agree to call $s$, and the opposite value $-s<0$. By a phenomenon similar to that of Equation (\ref{eqe4_11}), $\psi$ is in general a linear combination of two solutions, one containing a factor $R^s$, the other containing a factor $R^{-s}$, the relative amplitudes being univocally fixed by a length $\ell$ (the equivalent of $a$ in Equation (\ref{eqe4_11})) determined by the microscopic details of the interaction at short distance $O(b)$:\footnote{\label{noteeN2} Scattering state (\ref{eqe4_11}) corresponds to the case $N=2$; then $s=1/2$ and $(3N-5)/2=1/2$, and $\psi$ in Equation (\ref{eqe2_15}) is indeed a linear combination of $R^0$ and $R^{-1}$; in this case, $R\propto r_{12}$ and $\Phi(\Omega)=\mbox{const}$ in $s$ wave. Explicit calculation of expression (\ref{eqe2_15}) for $N=2$ gives $\psi\propto (r_{12}/\bar{\ell})^{s-1/2}-(r_{12}/\bar{\ell})^{-s-1/2}$ with $\bar{\ell}=(m_1+m_2)\ell/(m_1m_2)^{1/2}$, which must be proportional to the scattering state (\ref{eqe4_11}), hence the announced exponent value $s=1/2$; the two-body parameter $\bar{\ell}$ is none other than the scattering length $a$.} \footnote{A minus sign has been placed between the two bracketed terms in Equation (\ref{eqe2_15}); a plus sign would also be possible, depending on the microscopic model.} 
\begin{equation}\label{eqe2_15}
\psi=\left[(R/\ell)^{s}-(R/\ell)^{-s}\right]R^{-\frac{3N-5}{2}}\Psi(\Omega)
\end{equation} However, in the absence of $N$-body scattering resonance, $\ell=O(b)$ is expected, so that $\ell\to 0$ when $b\to 0$: the solution $R^{-s}$ becomes negligible, length $\ell$ disappears from the problem and we keep the following scale invariant $N$-body contact condition in the channel of scaling exponent $s$ \cite{theseFelix}: 
\begin{equation}\label{eqe3_15}
\psi\underset{R\to 0}{\approx} R^{s-\frac{3N-5}{2}}
\end{equation} The wave function $\psi$, considered as a function of $R$, is nodeless so energy $E=0$ corresponds to the ground state: there is no bound state (an eigenenergy $E<0$ would tend to $-\infty$ when $b\to 0$).\footnote{\label{notee13} The special case of a $(N_\uparrow,N_\downarrow)$-body scattering resonance, where $\ell$ remains finite in the $b\to 0$ limit, is treated in detail in reference \cite{LudovicN}, which explains which $N$-body contact condition to use to correctly describe the resulting low-energy bound state. Indeed, condition (\ref{eqe2_15}) already proposed in \cite{FWsep} is only satisfactory for $s$ small enough (for $s>1$, we can see that state (\ref{eqe2_15}) is no longer square integrable near $R=0$ and that a second length - a cutoff - must be introduced).}~\footnote{\label{notee13bis} On a narrow Feshbach resonance, see note \ref{notee1_10}, the effective range $r_{\rm e}$ is - for $1/a=0$ - the only relevant length scale when the true range $b$ tends towards zero, so that the length $\ell$ is of the order of $|r_{\rm e}|\gg b$. The wave function $\psi$ in Equation (\ref{eqe2_15}) then admits a node far “outside” the interaction potential: the solution at $E=0$ would not be of minimum energy, and the system would admit a bound state of $(N_\uparrow,N_\downarrow)$ fermions (with $N>2$)! However, a specific study of the case $(N_\uparrow=2,N_\downarrow=1)$ shows that this is not true (as long as the mass ratio $m_\uparrow/m_\downarrow$ remains low enough for $s^2\geq 0$ of course) \cite{Tignone,epls}. Should we be surprised? Let's argue by contradiction. If there really were a bound state, it would lead to a relative wave number $k_{\rm rel}\approx 1/|r_{\rm e}|$ between the fermions, the effective range term would not be negligible in the denominator of the scattering amplitude (\ref{eqefk}) and we would lose scale invariance and therefore separability in hyperspherical coordinates. Equation (\ref{eqe2_15}) would be inapplicable and the prediction of a bound state would be invalid.  More generally, in order to believe in (\ref{eqe2_15}) - this is a necessary condition, the hyperradius $R$ must be much larger than any length scale appearing in any subsystem $(n_\uparrow, n_\downarrow)$ [with $n_\uparrow\leq N_\uparrow$, $n_\downarrow\leq N_\downarrow$ and $n_\uparrow+n_\downarrow<N_\uparrow+N_\downarrow=N$], in particular $R\gg b$ and $R\gg |r_{\rm e}|$ for $(n_\uparrow=1,n_\downarrow=1)$. Needless to say, the $(N_\uparrow,N_\downarrow)$-body resonance of note \ref{notee13} does not call into question (\ref{eqe2_15}) and the existence of a node at hyperdistance $\ell$ since the abnormally large $\ell\gg b$ length that appears does not pre-exist in any subsystem.}

\paragraph{Second case: $s^2<0$}
 Here too, there are two possible values for the scaling exponent, one $s=\ii|s|$ in $\ii\mathbb{R}^+$, which we'll call $s$, and the other, its conjugate complex $-\ii |s|$, or its opposite $-s$, in $\ii\mathbb{R}^-$. As in the first case, we conclude that there is a length $\ell$, a function of the microscopic details of the interaction, setting the relative amplitude of the two solutions: 
\begin{equation}\label{eqe1_16}
\psi=\left[(R/\ell)^{\ii|s|}-(R/\ell)^{-\ii|s|}\right]R^{-\frac{3N-5}{2}}\Psi(\Omega)=2\ii\sin[|s|\ln(R/\ell)]R^{-\frac{3N-5}{2}}\Psi(\Omega)
\end{equation} This time, the two solutions have the same modulus, so we must keep them both (neither outweighs the other in the limit $b\to 0$)! The length $\ell$ does not disappear from the problem, but defines in the limit $b\to 0$ a $N$-body contact condition (\ref{eqe1_16}) which explicitly breaks the continuous scale invariance of the unitary limit. Since $\psi$ has an infinite number of nodes at an arbitrarily large hyperdistance $R$ - arbitrarily larger than the interaction range $b$ - (see the third expression in Equation (\ref{eqe1_16})), there is an infinite number of $N$-body bound states under the $E=0$ energy solution; since the boundary condition (\ref{eqe1_16}) is invariant by changing $\ell$ into $\exp(\pm\pi/|s|)\ell$, we pass from one $N$-body bound state to the other by a homothety of ratio $\exp(\pi/|s|)$: the corresponding spectrum forms a geometric sequence of zero limit but it is not bounded below in the zero range limit,\footnote{The precise expression of $E_{\rm glob}$ as a function of $s$ and $\ell$, and of the order of $\hbar^2/Mb^2$ for $\ell\approx b$, is given, for example, in reference \cite{virtrans}. In Equation (\ref{eqe2_16}) we take $n\geq 1$ (assuming $\exp(-2\pi/|s|)\ll 1$ - otherwise the spectrum would not be entirely geometric \cite{Tignone}) because the Wigner-Bethe-Peierls model can only be applied to a bound state of size $\gg b$. We can however have $\ell\approx |r_{\rm e}|\gg b$ and $E_{\rm glob}\approx \hbar^2/M r_{\rm e}^2\ll \hbar^2/M b^2$ on a narrow Feshbach resonance, see our note \ref{notee1_10}; even if it is not obvious, the exclusion of $n=0$ in (\ref{eqe2_16}) remains correct in this case \cite{Tignone,epls}. This exclusion of $n=0$ is consistent with the absence of a bound state when $s^2\geq 0$, see our note \ref{notee13bis}: on the Efimovian side, the whole discrete spectrum must tend to zero when $|s|\to 0$ knowing that $E_{\rm glob}$ has a finite, non-zero limit.} 
\begin{equation}\label{eqe2_16}
E_n=-E_{\rm glob} \eee^{-2\pi n/|s|} \quad ,\quad n\in\mathbb{N}^*, \quad \mbox{with} \quad E_{\rm glob}\approx\hbar^2/Mb^2
\end{equation} These $N$-body bound states, historically predicted by Efimov for $N=3$, are said to be Efimovian. It would be very interesting to stabilize them in a cold atom experiment (we unfortunately verify that, if $s\in\ii\mathbb{R}^{+*}$ in the problem with $(2,1)$ fermions, which occurs for $m_\uparrow/m_\downarrow>13.6069\ldots$ as we shall see, the three-body loss rate of Equation (\ref{eqe1_7}) does not tend to zero when $b\to 0$, but rather to a quantity proportional to $\hbar k_{\rm F}^2/m$ in the homogeneous gas at $T=0$, which is considerable), for example using note \ref{notee1_10}. 

But back to our mathematical problem: we conclude that the gas with $(N_\uparrow,N_\downarrow)$ fermions is stable for a zero-range $\uparrow\downarrow$ interaction, and that the Hamiltonian of the Wigner-Bethe-Peierls model is self-adjoint and bounded from below, if and only if the scaling exponents are all real: \begin{equation}\label{eqe1_17}
s\in\mathbb{R}^* \quad \forall s\ \mbox{solution of}\ \Lambda_{N_\uparrow,N_\downarrow}(s)=0
\end{equation}
Of course, no gas subsystem $(n_\uparrow,n_\downarrow)$ must exhibit an Efimov effect either, otherwise (i) a $n_\uparrow+n_\downarrow < N_\uparrow+N_\downarrow$-body parameter would have to be introduced as in Equation (\ref{eqe1_16}) and the scale invariance at the origin of separability (\ref{eqe1_14}) and of result (\ref{eqe1_17}) would be broken - we would lose separability at all distances, and (ii) the subsystem could collapse in the $b\to 0$ limit of a zero range interaction and the energy would not be bounded below. Without saying so, we have done a proof by induction and the equivalent of the stability condition (\ref{eqe1_17}) must be satisfied for any number $n_\uparrow\leq N_\uparrow$ and any number $n_\downarrow\leq N_\downarrow$.

\paragraph{Remark 1}
We could add $s=0$ in the third case: this is in fact precisely the threshold for an Efimov effect.  Expanding Equation (\ref{eqe1_16}) to first order in $|s|$, we find that
\begin{equation}\label{eqe2_17}
\psi\propto \ln(R/\ell) R^{-\frac{3N-5}{2}} \Phi(\Omega)
\end{equation} i.e.\ the length $\ell$ sets the relative amplitude of the solutions $R^{-(3N-5)/2}$ and $(\ln R)R^{-(3N-5)/2}$. This bears a striking resemblance to the definition of the scattering length of two particles in dimension two, see for example reference \cite{mon_cours} and Jan Solovej's contribution \cite{Solov}. However, in the $\ell=O(b)$ case where we are, the length $\ell$ tends towards zero when $b\to 0$, the first solution - with coefficient $\ln(1/\ell)\to +\infty$ - wins out over the second and we keep the $N$-body contact condition in continuity with (\ref{eqe3_15}):
\be
\psi\underset{R\to 0}{\approx} R^{-(3N-5)/2}
\ee
In the zero-range model, therefore, there is no breaking of scale invariance and no bound state at the Efimovian threshold.\footnote{The considerations of note \ref{notee13bis} apply at threshold. In particular, we must not believe in the bound state of energy $\propto -\hbar^2/M\ell^2$ that Equation (\ref{eqe2_17}) would lead us to predict: it would be of spatial extension $\ell$ and could not be described by our zero-range model when $\ell\approx b$; nor would it exist for $(N_\uparrow=2,N_\downarrow=1)$ in the - apparently favourable - case of a narrow Feshbach resonance where $\ell\approx |r_{\rm e}|\gg b$, see note \ref{notee13bis} and references \cite{Tignone,epls}.}~\footnote{The fact that $\ln(1/\ell)$ tends slowly towards infinity when $b\to 0$ is not without practical consequences: if we want to compare with experiments, it is better to keep the contribution $\ln R$ in (\ref{eqe2_17}) [and the term $(R/\ell)^{-s}$ in (\ref{eqe2_15}) for $s>0$ close enough to zero] to form the $N$-body contact condition. Thus, we find that the third cluster coefficient $b_{2,1}$ defined in Section \ref{sece3.4} is actually a regular function of the mass ratio $m_\uparrow/m_\downarrow$ and therefore of $s^2$, whereas it has an infinite derivative at $s^2=0$ in the zero-range model \cite{daily,virtrans}, hence in Equation (\ref{eqe1_20}).}

\paragraph{Remark 2}
We've taken $a=\infty$ here for simplicity, but if the system is unstable for $a=\infty$, it will remain so for finite $a$ (all else being equal): the Efimovian $N$-mers of the unitary gas of size $\ll |a|$ (there are as many as you like for $b\to 0$) make no difference between an infinite and a finite scattering length $a$.

\paragraph{Complement}
 In the unitary limit, the present analysis generalizes to nonzero energy $E$ (still with the center of mass of the system at rest). Equation (\ref{eqe1_14}) becomes 
\begin{equation}\label{eqe1_17p}
\psi=F(R) R^{-\frac{3N-5}{2}} \Phi(\Omega)
\end{equation} (satisfying the Wigner-Bethe-Peierls contact conditions under note \ref{notee1_14}) with 
\begin{equation}\label{eqe2_17p}
E F(R) = -\frac{\hbar^2}{2M}\Delta_{\rm 2D} F(R) + \frac{\hbar^2 s^2}{2MR^2} F(R)
\end{equation} and $\Delta_{\rm 2D}$, the 2D Laplacian for variable $R$, reduces here (in the absence of angular dependence) to $\dd^2/\dd R^2+R^{-1}\dd/\dd R$. The  Efimovian case $s^2<0$ therefore simply corresponds to the known “fall to the center” problem in an attractive $1/R^2$ potential \cite{Landaumec}. Separability in hyperspherical coordinates (\ref{eqe1_17p}) even extends to the trapped case \cite{FWsep,livre}, simply by adding the trapping term $(1/2)M\omega^2R^2F(R)$ to the right-hand side of Equation (\ref{eqe2_17p}), and making the substitution $E\to E-E_{\rm com}$, the eigenvalue of the equation for $F(R)$ being the internal energy as opposed to that $E_{\rm com}$ of the center of mass. For $s^2>0$, this leads to the spectrum 
\begin{equation}\label{eqe3_17p}
E-E_{\rm com}=(s+1+2q)\hbar\omega \quad,\quad q\in\mathbb{N}
\end{equation}
 
\subsection{What is known on the stability domain}
\label{sece3.3}

The problem of whether the stability condition (\ref{eqe1_17}) is satisfied can be tackled from two opposite ends.

\begin{figure}[t]
\includegraphics[width=12cm,clip=]{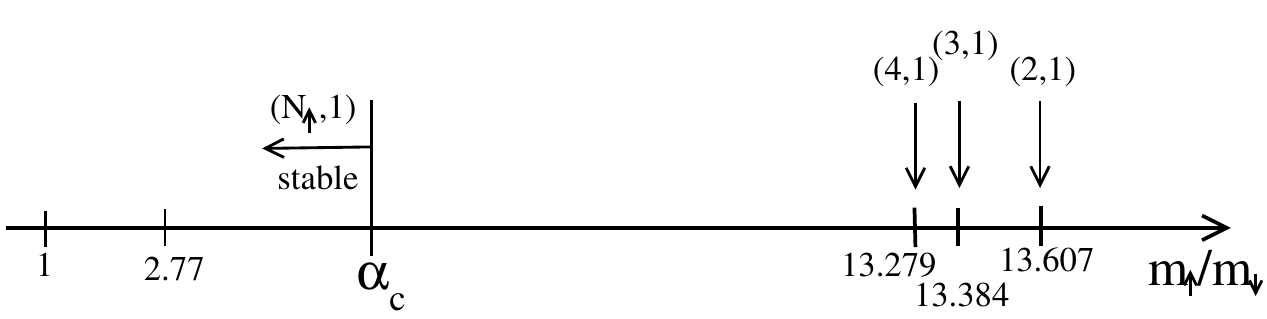}
\caption{Known results on the stability of the $(N_\uparrow,1)$ system of $N_\uparrow$ spin $\uparrow$ fermions and one spin $\downarrow$ fermion, for a zero-range interaction, as a function of the $m_\uparrow/m_\downarrow$ mass ratio between a $\uparrow$ particle and the $\downarrow$ particle. Vertical arrows: critical values of $m_\uparrow/m_\downarrow$ (thresholds) for the $(N_\uparrow,1)$-body Efimov effect obtained by solving the corresponding problem, see references \cite{Efimov} for $N_\uparrow=2$, \cite{MoraCastinPricoupenko} for $N_\uparrow=3$, \cite{BazakPetrov} for $N_\uparrow=4$; when $m_\uparrow/m_\downarrow$ exceeds these values, the energy of the system is no longer bounded below. Vertical bar with horizontal arrow: critical mass ratio $\alpha_{\rm c}$, whose existence is established by the Moser-Seiringer theorem \cite{MoserSeiringer}, below which the system $(N_\uparrow,1)$ is stable $\forall N_\uparrow$; the theorem does not give the exact value of $\alpha_{\rm c}$ but a more restrictive lower bound, $\alpha_{\rm c}>2.77$, than that, $m_\uparrow/m_\downarrow=1$, of the cold-atom experiments (a superiority of the latter, however, is that the unitary Fermi gas appears to be stable, without collapse or significant three-body losses, for all values of $N_\uparrow$ and $N_\downarrow$).}
\label{fige3}
\end{figure}

The first involves solving the problem with $(N_\uparrow,N_\downarrow)$ fermions at $E=0$ in the Wigner-Bethe-Peierls model and calculating the scaling exponents $s$ (we proceed analytically as far as possible, but there is a final numerical step, at least for $N>3$). To the best of our knowledge, this program has been completed in fermionic problem $(N_\uparrow>1,N_\downarrow=1)$ up to $N_\uparrow=4$, see Figure \ref{fige3}: in each case, an Efimov effect is found to appear above a critical mass ratio $m_\uparrow/m_\downarrow$ (impurity $\downarrow$ must be light enough), obviously a decreasing function of $N_\uparrow$.\footnote{Once we have an Efimov effect in the $(N_\uparrow,N_\downarrow)$ fermionic problem, as we said below Equation (\ref{eqe1_17}), we lose scale invariance and can no longer apply the reasoning behind Equations (\ref{eqe1_14},\ref{eqe3_14}) to the $(N_\uparrow+1,N_\downarrow)$ or $(N_\uparrow,N_\downarrow+1)$ fermionic problem; in the latter case, there is no separability in $(R,\Omega)$ coordinates as in Equation (\ref{eqe1_16}), no geometric spectrum (\ref{eqe2_16}) and, strictly speaking, no possible Efimov effect!} We note that the successive critical mass ratios get closer and closer together; mathematically, however, we don't know whether this sequence continues (is there an Efimov effect for (5,1) bodies? for (6,1) bodies? etc.). The $(N_\uparrow>1,N_\downarrow=2)$ case has been studied for $N_\uparrow=2$ by reference \cite{Endopas4}, which predicts stability as long as that the subsystems (2,1) and (1,2) are stable.

At the second extreme, the aim is to constrain (rather than calculate) critical mass ratios, by lower bounding the Hamiltonian spectrum. This is what reference \cite{MoserSeiringer} has done for the $(N_\uparrow,N_\downarrow=1)$ fermion problem: it demonstrates the magnificent 
\paragraph{Theorem}
\textit{There is a critical mass ratio $m_\uparrow/m_\downarrow=\alpha_{\rm c}$ below which the fermionic system $(N_\uparrow,N_\downarrow=1)$ is stable $\forall N_\uparrow$ for a contact interaction, and $\alpha_{\rm c}>1/0.36=2.77$.}
 
\subsection{Cluster or virial expansion}
\label{sece3.4}

Some would argue that it is the macroscopic many-body problem that should ultimately be the focus of our attention, rather than the few-body system. To which we reply that the latter can say something about the former by means of cluster expansion, an expansion of the pressure $P$ of the homogeneous gas at grand-canonical thermal equilibrium in powers of the fugacities of the components $\sigma$ (the temperature $T=1/k_{\rm B}\beta$ is fixed and the chemical potentials $\mu_\sigma$ tend to $-\infty$, which corresponds to a quantum non-degenerate limit): 
\begin{equation}\label{eqe1_19}
\frac{P\bar{\lambda}^3}{k_{\rm B}T}=\sum_{(n_\uparrow,n_\downarrow)\in\mathbb{N}^{2*}} b_{n_\uparrow,n_\downarrow} \eee^{\beta \mu_\uparrow n_\uparrow} \eee^{\beta\mu_\downarrow n_\downarrow}
\end{equation} where the thermal de Broglie wavelength is taken at the arbitrary reference mass $\bar{m}$, $\bar{\lambda}=(2\pi\hbar^2/\bar{m}k_{\rm B}T)^{1/2}$.\footnote{Reference \cite{EndoCastin2022} uses the natural choice $\bar{m}^{3/2}=(m_\uparrow^{3/2}+m_\downarrow^{3/2})/2$ leading to a first total cluster coefficient $b_1=(b_{1,0}+b_{0,1})/2$ equal to one.} The expansion (\ref{eqe1_19}), in recent literature, is often confused with the virial expansion, which expands in powers of the phase space densities $\rho_\sigma\lambda_\sigma^3$ ($\lambda_\sigma=(2\pi\hbar^2/m_\sigma k_{\rm B}T)^{1/2}$).

How to calculate the cluster coefficients $b_{n_\uparrow,n_\downarrow}$? At the unitary limit $a^{-1}=0$, the simplest way is to use the harmonic regulator method of reference \cite{ComtetOuvry}, which first places the system in the harmonic traps $U_\sigma(\rr)$ of Equation (\ref{eqe1_5}) and then, all calculations done, opens the traps to obtain the homogeneous case from the local density approximation (exact in the limit $\omega\to 0^+$). The trapped problem is then separable in hyperspherical coordinates as in Section \ref{sece3.2} and, if we know the zero-energy free space scaling exponents $s$ for all $n_\uparrow\leq n_\uparrow^{\rm target}, n_\downarrow\leq n_\downarrow^{\rm target}$, we also know the energy levels of the trapped system as in Equation (\ref{eqe3_17p}), hence all the canonical partition functions $Z_{n_\uparrow,n_\downarrow}$ and ultimately the coefficient $b_{n_\uparrow^{\rm target},n_\downarrow^{\rm target}}$. As a result, the cluster coefficients must be functionals of the $\Lambda_{n_\uparrow,n_\downarrow}$ of Equation (\ref{eqe1_15}). This is indeed what is predicted by the conjecture in reference \cite{EndoCastin}, claiming that, at the unitary limit, 
\begin{equation}\label{eqe1_20}
b_{n_\uparrow,n_\downarrow}=\frac{(n_\uparrow m_\uparrow+n_\downarrow m_\downarrow)^{3/2}}{\bar{m}^{3/2}} \left[\int_{-\infty}^{+\infty} \frac{\dd S}{4\pi} \, S \frac{\dd}{\dd S} \left(\ln\Lambda_{n_\uparrow,n_\downarrow}(\ii S)\right) + \mbox{CorrStat}_{n_\uparrow,n_\downarrow}\right]
\end{equation} for any $(n_\uparrow,n_\downarrow)\in\mathbb{N}^{*2}\setminus\{1,1\}$.\footnote{Case $(n_\uparrow=1,n_\downarrow=1)$ is different and must be excluded; it corresponds, contrary to what we have assumed, to a $N$-body scattering resonance with $N=2$ in $s$ wave: in the unitary limit, we must keep only the solution $R^{-s}, s=1/2$, in Equation (\ref{eqe2_15}), as if the right root of $\Lambda$ to keep was $-s$; in fact, the solution $R^s$ corresponds to the regular part $\propto 1/a$ of the zero-energy scattering state (see note \ref{noteeN2} for further details).}
Here, the prefactor relates the homogeneous case to the trapped case, and the first contribution in square brackets is modelled on the $N=3$ result of reference \cite{revuecanad}; the second contribution in square brackets is an ideal-gas quantum statistical correction originating from the indistinguishable non-monoatomic subclusters into which the internal eigenstates of the trapped system $(n_\uparrow,n_\downarrow)$ decouple at high energy (the center of mass of the system remains in its ground state). 

\begin{figure}[t]
\includegraphics[width=8cm,clip=]{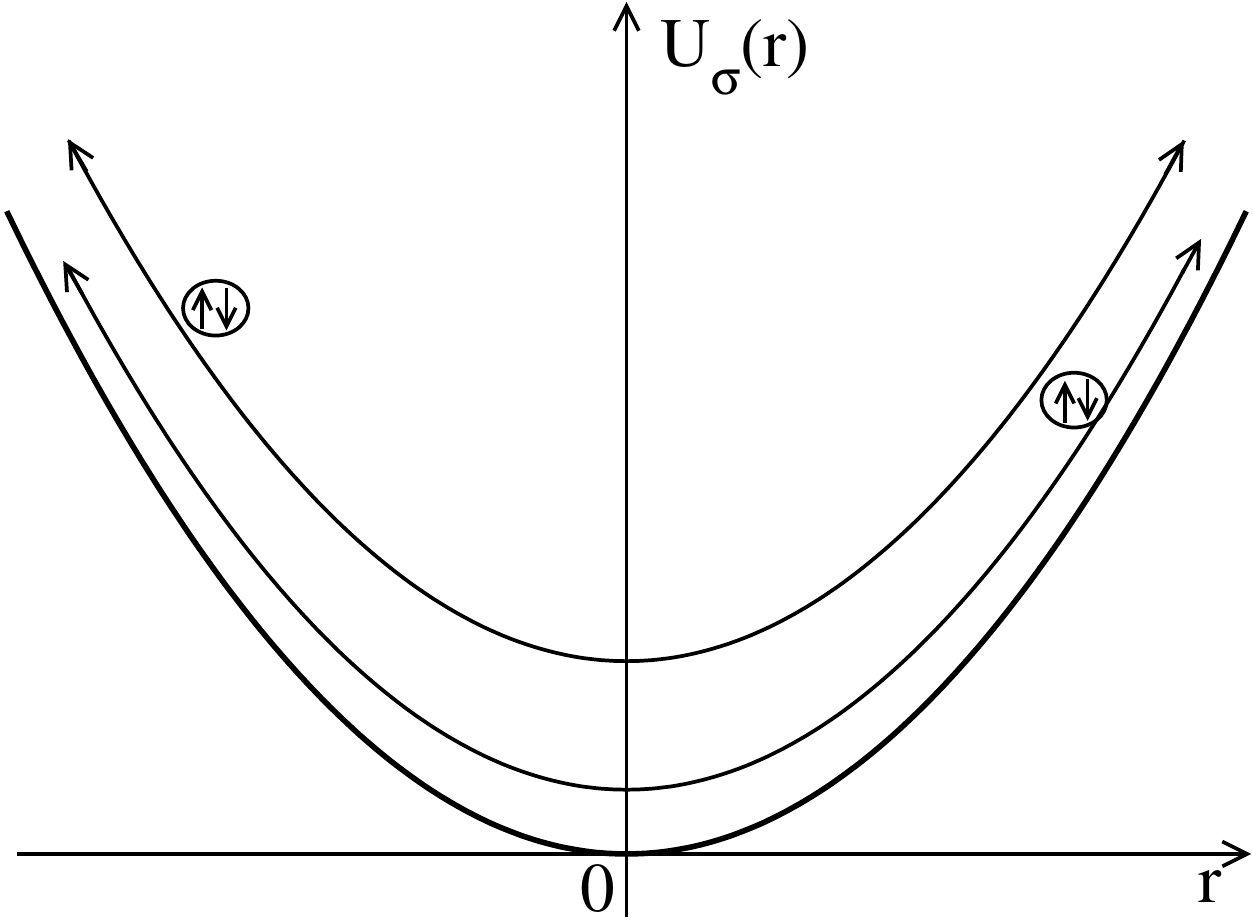}
\caption{A possible asymptotic behavior of the unitary four-fermion problem $(N_\uparrow=2,N_\downarrow=2)$ in harmonic potentials $U_\sigma(\rr)$: two pairons oscillate furiously (with large-amplitude motions); fermions $\uparrow$ and $\downarrow$ in each pairon have relative energy $O(1)$ (in $\hbar\omega$ units if dimension is to be respected) and remain strongly correlated; the two pairons have relative energy $\to +\infty$ and are decoupled, so they can be seen as two identical bosons that have an internal structure (that of relative motion $\uparrow\downarrow$ within a pairon) and do not interact. The quantum statistical correction $\mbox{CorrStat}_{2,2}$ is then non-zero in Equation (\ref{eqe1_20}).}
\label{fige3bis}
\end{figure}

Let's explain this subcluster decoupling story a little better with some examples. If $(n_\uparrow,n_\downarrow)=(1,1)$, the eigenstates asymptotically (for arbitrarily large energy values) take the form of two uncorrelated fermions $\uparrow$ and $\downarrow$ in large-amplitude oscillatory levels;\footnote{If the relative wavenumber $k_{\rm rel}\to+\infty$, the scattering amplitude $f_{k_{\rm rel}}\to 0$ in Equation (\ref{eqe2_10}) so even at the unitary limit, interactions become negligible.} the subclusters being monoatomic, we have $\mbox{CorrStat}=0$. If $(n_\uparrow,n_\downarrow)=(2,1)$, a new decoupling is possible, alongside that into three decorrelated fermions: the particles can separate into an atom $\uparrow$ and a “pairon” $\uparrow\downarrow$ of strongly correlated fermions (the relative motion within the pairon remaining of low amplitude), with highly excited oscillatory levels for the atom $\uparrow$ and for the center of mass of the pairon $\uparrow\downarrow$; the pairon is alone in its category, and we again have $\mbox{CorrStat}=0$. The conclusion remains the same for $(n_\uparrow,n_\downarrow)=(3,1)$, except that the triplon $\uparrow\uparrow\downarrow$ appears as a new decoupled subcluster. On the other hand, for $(n_\uparrow,n_\downarrow)=(2,2)$, there is possible decoupling into two $\uparrow\downarrow$ pairons of very high relative energy, see Figure \ref{fige3bis}; they no longer interact but are indistinguishable and lead, like identical bosons in an ideal gas, to a quantum statistical correction ignored by the integral over $S$ in Equation (\ref{eqe1_20}); the calculation gives $\mbox{CorrStat}=1/32$ \cite{EndoCastin}.

Conjecture (\ref{eqe1_20}) is well established for $N=3$, by inverse application of the residue theorem, which converts the sum over the spectra (\ref{eqe3_17p}), i.e.\ over the roots $s$ of $\Lambda_{n_\uparrow,n_\downarrow}$, into an integral \cite{revuecanad}. For $N=4$, the analytical properties of function $\Lambda_{n_\uparrow,n_\downarrow}$ in the complex plane are not sufficiently known to apply Cauchy's theorem;\footnote{It must be possible to unfold the integration path surrounding the roots and poles of $\Lambda_{n_\uparrow,n_\downarrow}$ on the real axis onto the imaginary axis without crossing any singularity - pole or branch cut - in the upper and lower half-planes.} for the special case $m_\uparrow/m_\downarrow=1$, however, the conjecture has been confirmed by a very accurate few-body quantum Monte Carlo calculation \cite{HouDrut} (on the other hand, the experimental values \cite{ens,mit} are not confirmed, the problem arising from the impossibility of obtaining the correct polynomial of degree 4 in $z=\exp(\beta\mu)$ by fitting the pressure $P$ or density $\rho$ measured on the experimentally accessible fugacity interval \cite{halaug}).\footnote{References \cite{ens,mit} have access only to the fourth total cluster coefficient $b_4=(b_{4,0}+b_{3,1}+b_{2,2}+b_{1,3}+b_{0,4})/2$, which prohibits a comparison with the conjecture (\ref{eqe1_20}) sector by sector.} 
 The proof of expression (\ref{eqe1_20}) in the general case therefore remains open. Another interesting question concerns the behavior of the cluster coefficients $b_{n_\uparrow,n_\downarrow}$ at large orders $n_\sigma\to +\infty$, which is required to perform an efficient summation of the series (\ref{eqe1_19}) after calculation of its first terms, in order to extend its applicability to the degenerate regime $T\lesssim T_{\rm F}$ (for example, if the radius of convergence is zero, one could implement a conformal Borel-type resummation as in references \cite{felixborel,zinnborel}).

\section{Open questions from a macroscopic point of view}
\label{sece4}

In this section, the interacting Fermi gas, considered in the thermodynamic limit and at non-zero but arbitrarily low temperature, is described by an effective low-energy Hamiltonian theory, Landau and Khalatnikov quantum hydrodynamics \cite{LK}.\footnote{\label{notee1_23} The term “quantum hydrodynamics”, in non-linear physics, is understood in contrast to classical fluid hydrodynamics and refers to an Euler equation for a real-valued rather than operator-valued velocity field $\vv(\rr)$ with a $\propto\hbar^2$ quantum pressure term, which allows to describe the motion of quantum vortices - with quantized circulation - in the superfluid (like the Gross-Pitaevskii equation for the wave function of a Bose condensate written in terms of density and phase gradient). Here, the name is to be taken in the sense of the second quantization, the velocity field now being operator-valued $\hat{\vv}(\rr)$.}

\subsection{Overview of the considered superfluid regime}
\label{sece4.1}

The three-dimensional system of fermions is here spatially homogeneous (in a quantization volume $[0,L]^3$ close to the thermodynamic limit, which is taken at the end of the calculations), with particles of equal mass $m_\uparrow=m_\downarrow=m$ in both internal states, spin-unpolarized (there are the same number of particles in both components, $N_\uparrow=N_\downarrow$, to allow complete pairing) and at canonical thermal equilibrium in a low-temperature limit ($T\neq 0$ but $T\to 0$). 

Under these conditions, (i) fermions assemble into bound pairs $\uparrow\downarrow$ in the $s$ wave; in the presence of Fermi seas in both internal states, this is what the attractive interaction between $\uparrow$ and $\downarrow$ in Section \ref{sece3} leads to, via the famous Cooper mechanism; this is true even for a negative scattering length $a$, where there is no $\uparrow\downarrow$ bound state in free space, although the size of a pair tends to $+\infty$ as $a\to 0^-$ (in the $a>0$ case, a dimer state does exist and, not surprisingly, this is what the bound pair state reduces to in the low-density limit $\rho\to 0$);\footnote{It is not entirely obvious that the Wigner-Bethe-Peierls contact interaction is attractive. To see this, we obtain it as the continuous limit $b\to 0$ of a model on a cubic lattice $b\mathbb{Z}^3$ with a coupling $\propto \hbar^2/mb^2$ between neighboring sites (to represent the kinetic energy) and an on-site interaction $g_0/b^3$; at a fixed scattering length $a\neq 0$, we find that $g_0\approx -\hbar^2b/m<0$ as $b\to 0$ (the bare coupling constant $g_0$ is therefore, in the resonant scattering regime $b\ll|a|$, quite different from the effective coupling constant $g=4\pi\hbar^2a/m$) \cite{varenna}.} (ii) these bound pairs, being composite bosons of sorts, form a condensate in the mode of wave vector $\KK_{\rm pair}=\zero$ of their center of mass (of coherence length limited by the size of the box, infinite at the thermodynamic limit) and a superfluid. 

\begin{figure}[tb]
\includegraphics[width=6cm,clip=]{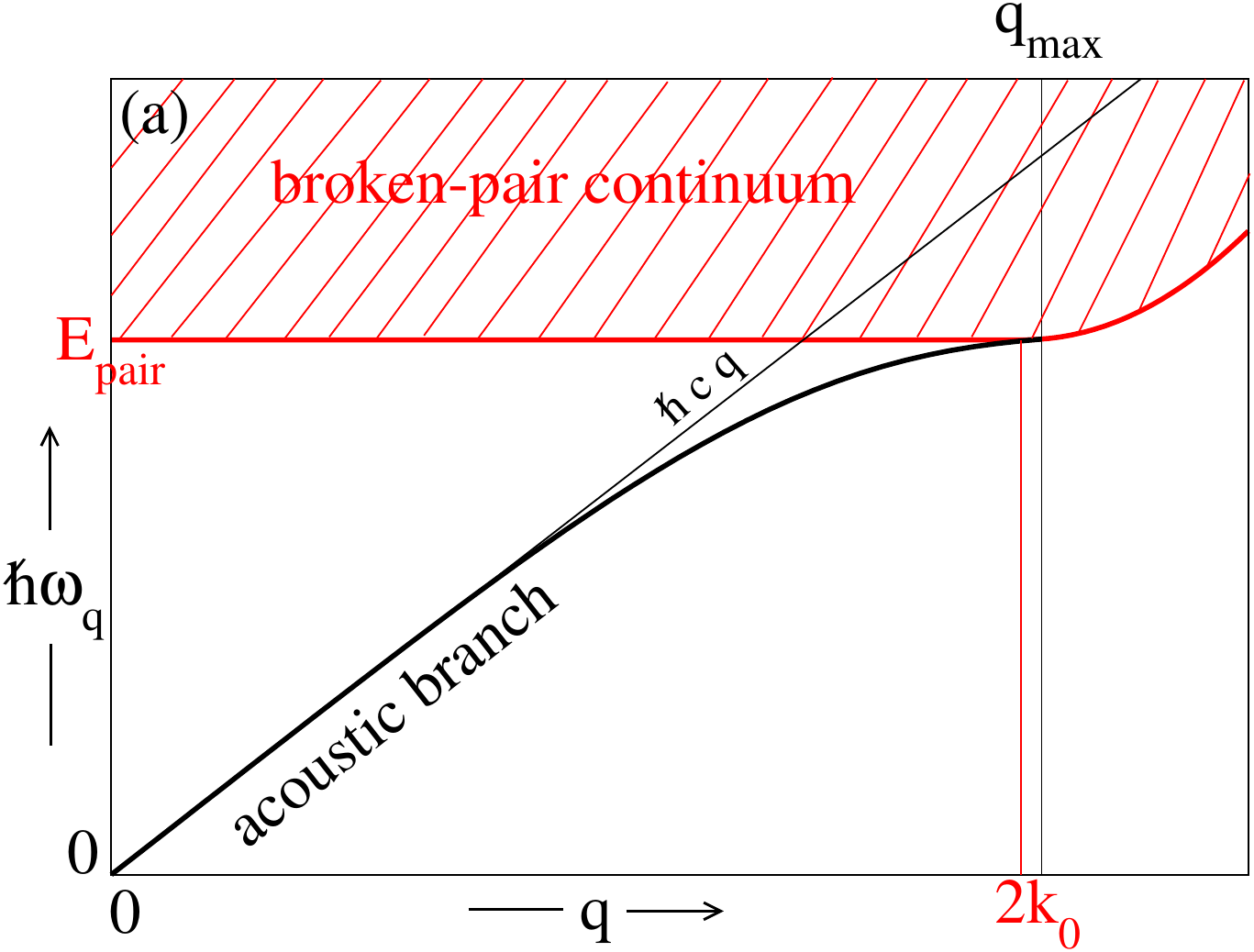}\hspace{1cm}\includegraphics[width=6cm,clip=]{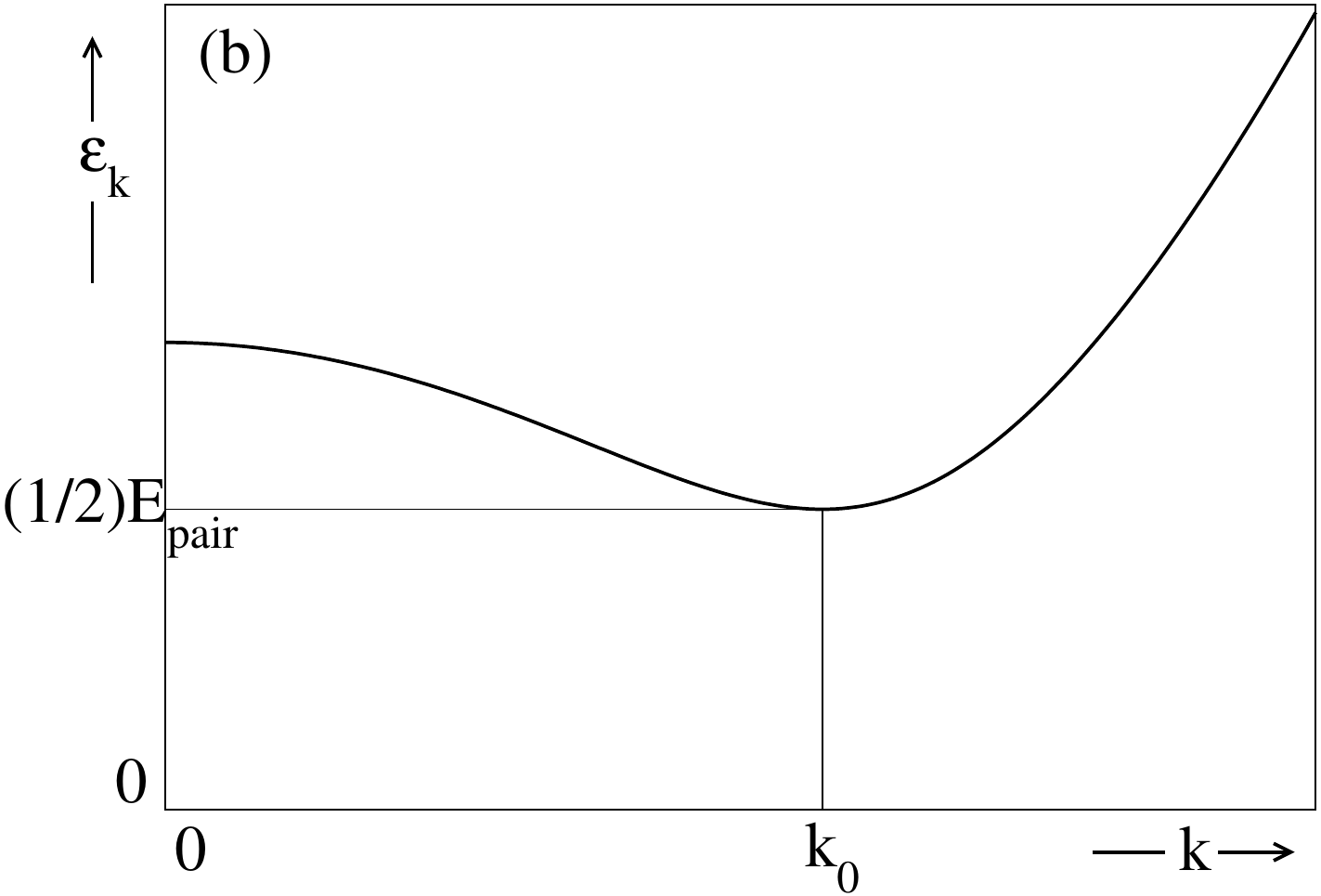}
\caption{Energy of different types of excitations of an unpolarized homogeneous two-component Fermi gas at zero temperature, as a function of their wave number $q$ or $k$. (a) Acoustic excitation branch $\hbar\omega_\qq$ of linear departure $\hbar c q$ ($c$ is the speed of sound), bounded at the top by the broken-pair continuum (hatched area). Here, the branch tangentially reaches the lower edge of the continuum at the terminal point of wave number $q_{\rm max}$. (b) Dispersion relation $\veps_\kk$ of a fermionic quasiparticle $\chi$ (see text). Under the effect of a percussive excitation of wave vector $\qq$, a bound pair $\uparrow\downarrow$ of the pair condensate, initially at rest, dissociates into two fermionic quasiparticles $\chi$ of opposite spins, wave vectors $\kk$ and $\kk'=\qq-\kk$ and energies $\veps_\kk$ and $\veps_{\kk'}$; as vector $\kk$ is not constrained (no conservation of energy for a percussive excitation), a continuum of final energies $\{\veps_\kk+\veps_{\qq-\kk},\kk\in\mathbb{R}^3\}$ appears. For the interaction strength chosen in the figure ($|\Delta|/\mu=0.84$ or $1/k_{\rm F}a\simeq -0.16$ according to BCS theory, with $\mu$ the chemical potential of the gas and $\Delta$ the complex order parameter of the pair condensate), the acoustic branch is concave at low $q$ ($\gamma<0$ in Equation (\ref{eqe1_24})) and the dispersion relation $\veps_\kk$ has a minimum $E_{\rm pair}/2$ at $k=k_0>0$; the lower edge of the continuum is therefore exactly $E_{\rm pair}$, at least as long as one can have $k=k'=k_0$, i.e.\ as long as $q=|\kk+\kk'|\leq 2 k_0$. The approximate dispersion relations shown are derived from BCS theory for $\veps_\kk$ and Anderson's RPA for $\hbar\omega_\qq$. Depending on the interaction strength, the domain of existence in $q$ of the acoustic branch may also be non-compact, connected $q\in\,]0,+\infty[$ or not $q\in\,]0,q_{\rm max}[\bigcup]q_{\rm min},+\infty[$ ($q_{\rm min}> q_{\rm max}$) \cite{CKS}; the concavity of the branch is also variable \cite{PRAconcav}; finally, $k_0=0$ and $E_{\rm pair}=2(\mu^2+|\Delta|^2)^{1/2}$ if $\mu<0$, $k_0=(2m\mu)^{1/2}/\hbar$ and $E_{\rm pair}=2|\Delta|$ otherwise.}
\label{fige4}
\end{figure}

As a result, we expect the system to have, in its ground state, a branch of acoustic excitation (by sound waves) that starts out linear in wavenumber $q$ with a cubic correction, 
\begin{equation}\label{eqe1_24}
\omega_\qq\underset{q\to 0}{=} cq\left[1+\frac{\gamma}{8}\left(\frac{\hbar q}{mc}\right)^2+O(q^4\ln q)\right]
\end{equation} and this branch to be energy-limited at the top by the broken-pair continuum, with a lower edge of the continuum given by the pair binding energy $E_{\rm pair}$, as in Figure \ref{fige4}a. Here $\omega_\qq$ is the angular eigenfrequency at wave vector $\qq$, $c$ is the speed of sound at zero temperature and the curvature parameter $\gamma$ is scaled so that it tends to one in the limit $k_{\rm F}a\to 0^+$ of a weakly interacting condensed gas of dimers (in agreement with Bogolioubov theory).\footnote{Our sign convention on $\gamma$ differs from that used in liquid helium-4, see reference \cite{Maris}.} The acoustic branch is often referred to as Goldstone \cite{Goldstone}, because it is associated with $U(1)$ symmetry breaking in pair condensation; its Higgs counterpart is discussed in Section \ref{sece5}.

In the following, our low-temperature regime satisfies two conditions:
\begin{equation}\label{eqe2_24}
0< k_{\rm B}T\ll mc^2 \quad \mbox{and}\quad 0<k_{\rm B}T\ll E_{\rm pair}
\end{equation} The first ensures that only the linear part of the acoustic branch is thermally populated, the second that there is a negligible density of broken pairs 
(according to Boltzmann's law, this density has an activation factor $\exp(-E_{\rm pair}/2 k_{\rm B}T)$, the fragments resulting from the dissociation of a bound pair - the $\chi$ fermionic quasiparticles in Figure \ref{fig4}b - individually having a minimum energy $E_{\rm pair}/2$). Our system then reduces to a thermal gas of phonons, if we agree to call the quanta of the acoustic branch as such (in view of its linear departure).\footnote{Equation (\ref{eqe1_24}) holds only for a short-range interaction $V(\rr_{ij})$, decreasing fast enough when $r_{ij}\to +\infty$. In the case of a dipolar interaction, as in gases of cold magnetic atoms, the speed of sound is anisotropic \cite{Axel}, see also Wilhelm Zwerger's contribution to this special issue \cite{Zwergerdt} and Jean Dalibard's 2023-2024 lecture at the Collège de France \cite{jd}. In the case of a Coulomb interaction, as in superconducting electron gases, the acoustic branch gives way to a gapped plasmon branch ($\omega_\qq$ has a positive limit at $q=0$) \cite{Anderson}. Here, our atoms are neutral and of negligible dipole moment.}

This raises three types of partially open questions: 
\begin{enumerate}\item as we shall see, phonons (abbreviated to $\phi$) interact, the underlying Fermi superfluid constituting a nonlinear medium for sound. What are the effects of these interactions on phonons of wave vector $\qq$? In particular, they are expected to damp with a rate $\Gamma_\qq(T)$ and to undergo a thermal angular frequency shift $\Delta_\qq(T)$ (we don't count the shift at zero temperature, which by definition gives rise to the spectrum (\ref{eqe1_24}) - the neglected term $q^5\ln q$ arises precisely from the cross-effect of interactions and quantum fluctuations in the phonon field \cite{espagnols,insuffisance}).\footnote{In the convex case $\gamma>0$, it is accompanied - still at $T=0$ - by a non-zero imaginary part $\approx q^5$, corresponding to the Beliaev damping mechanism $\qq\to\kk,\kk'$, see below.} \item what are the consequences of phonon collisional dynamics on the evolution of a particularly interesting macroscopic gas variable, the phase operator $\hat{\phi}_0(t)$ of the pair condensate?
\item we enrich the problem by considering the partially polarized case $N_\uparrow\neq N_\downarrow$. In the weakly polarized case, e.g. $N_\uparrow-N_\downarrow=O(1)$, the unpaired supernumerary fermions of the majority spin component form, in the interacting gas, fermionic quasiparticles (abbreviated to $\chi$) with a dispersion relation $\veps_\kk$ different from those of the free fermions: it has a non-zero minimum, i.e.\ a gap, given by half the binding energy of a pair $\uparrow\downarrow$, where it varies quadratically with the wave number $k$ (see Figure \ref{fige4}b). The question then arises of the $\phi-\chi$ and of the $\chi-\chi$ interactions; in particular, disagreement persists over the expression of the low-energy $\phi-\chi$ scattering amplitude (references \cite{cras} and \cite{italiens} differ).\footnote{This problem is relevant for superfluid helium-4, whose excitation branch also has a quadratic minimum, the roton minimum; our $\phi-\chi$ scattering problem is therefore formally equivalent to the roton-phonon scattering already studied in reference \cite{LK}, with the difference that rotons are bosons. The predictions of \cite{LK} are, however, incomplete and at odds with \cite{cras,italiens}.} In the highly polarized case, where $N_\uparrow-N_\downarrow$ is extensive like $N$, it is expected that $T=0$ pair condensation may take place in a plane-wave superposition of their center of mass (rather than in $\KK_{\rm pair}=\zero$ as assumed here), see references \cite{ff,lo}, giving rise to a spatially modulated superfluid (a supersolid following the fashionable terminology), with an imperfectly known domain of existence in parameter space $((N_\uparrow-N_\downarrow)/N,1/k_{\rm F}a,T/T_{\rm F})$, the problem being complicated by its high sensitivity to thermal fluctuations \cite{Leo} (there are as yet no experimental results in three-dimensional cold atom gases \cite{Hulet}). 
\end{enumerate}
 The most interesting case for points 1 and 2 is that of a concave acoustic branch, $\gamma<0$ in Equation (\ref{eqe1_24}), very different from the fairly well-known weakly interacting Bose gas (where $\gamma\simeq 1>0$ as we said).\footnote{Qualitatively, the case $\gamma<0$ is obtained when the binding energy $E_{\rm pair}$ is low enough: if we reduce $E_{\rm pair}$, the broken pair continuum in Figure \ref{fige4}a lowers, pushes on the acoustic branch and eventually bends it downwards. This is what happens in the BCS limit $k_{\rm F}a\to 0^-$ where $E_{\rm pair}/mc^2=O(\exp(-\pi/2k_{\rm F}|a|))$ tends rapidly to zero; it no longer happens in the BEC limit $k_{\rm F}a\to 0^+$ where $E_{\rm pair}\sim E_{\rm dim}=\hbar^2/ma^2\gg mc^2$. It is not clear on which side of the unitary limit $1/k_{\rm F}a=0$ (i.e.\ for which sign of the scattering length $a$) the curvature parameter $\gamma$ changes sign, see Section \ref{sece5}. Given the shape of the lower continuum edge in Figure \ref{fige4}a - regardless of the sign of $\gamma$, the repulsion effect on the acoustic branch is strongest at large $q$ and weakest at small $q$ (where the energy difference between the continuum edge and the branch is greatest). We therefore expect to have an interval of values of $1/k_{\rm F}a$ over which the branch is convex at small $q$ and concave at large $q$ \cite{PRAconcav}.} In particular, phonon damping for $\gamma<0$ can only occur at $T\neq 0$, since the decay of a phonon into any number $n>1$ of phonons is forbidden by energy-momentum conservation for a concave acoustic branch; at the leading order in temperature, it results for the same reason from the $\phi\phi\to\phi\phi$ four-phonon processes of Landau and Khalatnikov \cite{LK} (for $\gamma>0$ it results from the Beliaev $\phi\to\phi\phi$ or Landau $\phi\phi\to\phi$ three-phonon processes \cite{Belyaev,Martin}). To our knowledge, this four-phonon damping has not yet been observed experimentally in any system. It could be observed in a Fermi gas of cold atoms in a box trap \cite{epl}. It could also be observed in superfluid helium-4 (a liquid of bosons) if the pressure is increased sufficiently to make $\gamma<0$ (the roton minimum is lowered, which ends up making the acoustic branch concave at low $q$) and if the temperature is lowered sufficiently to reduce the density of rotons (through the activation factor $\exp(-E_{\rm roton}/k_{\rm B}T)$) and make the parasitic damping of phonons by rotons negligible \cite{prlrevi}.\footnote{In liquid helium-4, four-phonon scattering processes between intentionally produced (non-thermal) phonon beams have already been the subject of theoretical and experimental studies \cite{Adamenko}.}

\subsection{Which macroscopic theory to use?}
\label{sece4.2}

An effective low-energy theory gives up describing the system below a certain length scale $\ell$; on the other hand, the theory is expected to be accurate at long wavelengths, in this case at leading order in temperature.  
\begin{figure}[t]
\includegraphics[width=4cm,clip=]{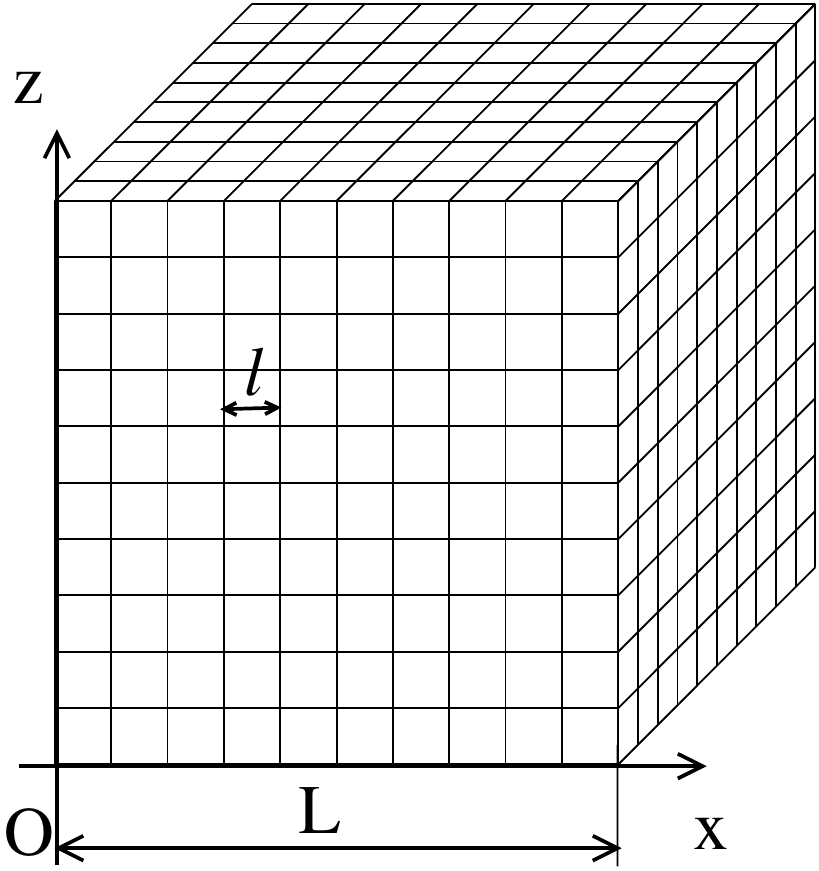}
\caption{Cutting the gas into mesoscopic cubic portions of side $\ell$, in the quantum hydrodynamics of Landau and Khalatnikov (this effective theory does not describe length scales $<\ell$). See text for choice of $\ell$.}
\label{fige5}
\end{figure}
 Under these conditions, it is legitimate to cut the gas into portions of size $\ell$, for example into cubic boxes of side $\ell$ centered on the cubic lattice $\ell\mathbb{Z}^3$, see Figure \ref{fige5}. Let's set out some constraints on the choice of $\ell$: 
\begin{enumerate}\item we must have $\ell\gg\xi$ (here $\xi=\hbar/mc$ is the so-called healing or correlation length of the superfluid) and $\rho\ell^3\gg 1$ (there are a large number of fermions per site) so that (i) each cubic portion can be considered mesoscopic with a well-defined equation of state linking pressure or chemical potential to density (as is the case in the thermodynamic limit), and (ii) the lattice spacing $\ell$ provides a wave number cutoff $\pi/\ell\ll mc/\hbar$ to the phononic excitations of the gas, restricting them to the quasi-linear part of the branch (\ref{eqe1_24}), which is universal because it is described by two parameters, $c$ and $\gamma$. \item we must have $k_{\rm B}T\ll \hbar c \pi/\ell$ (this is the energy of the ground phonon mode in a portion) so that we can consider that (i) each portion is at zero temperature, and (ii) each portion is spatially homogeneous on the scale of the typical wavelength $q_{\rm th}^{-1}=\hbar c/k_{\rm B}T$ of thermal sound waves.
\item it is also necessary that $\ell\ll\ell_{\rm coh}$ where $\ell_{\rm coh}$ is the coherence length of the fermion pairs, so that the notion of global phase $\hat{\phi}$ makes sense in each portion (as for a condensate). This constraint is inoperative here since $\ell_{\rm coh}\approx L$ (bound pairs are condensed in 3D). 
\end{enumerate}
We can then represent the system by two field operators, the density field $\hat{\rho}(\rr)$ and the phase field $\hat{\phi}(\rr')$, with $\rr,\rr'\in\ell\mathbb{Z}^3$; these are canonically conjugate variables,
\begin{equation}\label{eqe1_28}
[\hat{\rho}(\rr)\ell^3,\hat{\phi}(\rr')]=\ii\,\delta_{\rr,\rr'}
\end{equation} as if $\hat{\phi}$ were a momentum operator and $\hat{\rho}\ell^3$ a position operator in ordinary quantum mechanics \cite{LK,tome9landau}.\footnote{These historical references use a continuous space description for simplicity. The need to discretize the space to avoid infinities and make the theory renormalizable is emphasized in publication \cite{brouilfer}. Here, we sweep these difficulties under the carpet; for example, we do not distinguish in (\ref{eqe1_29}) between the notion of bare equation of state $e_{0,0}(\rho)$ - which enters the Hamiltonian - and true or effective equation of state $e_0(\rho)$ - which is observed in an experiment.} The phase field gives access to the velocity field by simple differentiation (this is a discrete gradient):\footnote{\label{notee_passuper}
The fact that the velocity field operator is a gradient vector in no way implies that the flow is entirely superfluid (this would be physically false even at thermal equilibrium, at non-zero temperature). Let's explain this in two points. (i) Don't confuse the $\hat{\vv}(\rr)$ operator of quantum hydrodynamics (which contains all possible quantum and thermal fluctuations) with the $\vv(\rr)$ mean velocity field of ordinary hydrodynamics; in particular, whether or not the flow is superfluid depends on whether or not $\vv(\rr)$ (without hat) is irrotational. (ii) In general, we have $\vv(\rr)\neq\langle\hat{\vv}(\rr)\rangle$ where the expectation value is taken in the quantum state of the system, since $\vv(\rr)$ is defined in terms of the mean matter current density, $\vv(\rr)=\langle\hat{\mathbf{j}}(\rr)\rangle/\langle\hat{\rho}(\rr)\rangle$ with here $\mathbf{\hat{j}}(\rr)=[\hat{\rho}(\rr)\hat{\vv}(\rr)+\hat{\vv}(\rr)\hat{\rho}(\rr)]/2$ (by definition, the evolution equation of $\hat{\rho}$ in the Heisenberg picture is written $\partial_t\hat{\rho}+\mathrm{div}\,\hat{\mathbf{j}}=0$ and that for the mean density $\rho(\rr)=\langle\hat{\rho}(\rr)\rangle$ is written $\partial_t\rho+\mathrm{div}\,(\rho \vv)=0$); it would therefore be wrong to believe that $\vv(\rr)=(\hbar/m)\mathbf{grad}\langle[\hat{\phi}(\rr)-\hat{\phi}(\rr_0)]\rangle$ (where $\rr_0$ is an arbitrary reference position) and deduce that $\vv(\rr)$ is necessarily a gradient vector.}
\begin{equation}\label{eqe2_28}
\hat{\vv}(\rr)=\frac{\hbar}{m}\mathbf{grad}\,\hat{\phi}(\rr)
\end{equation} The Hamiltonian is obtained by summing the internal energy and the kinetic energy associated with the local fluid velocity in each portion: 
\begin{equation}\label{eqe1_29}
H=\sum_\rr \frac{1}{2} m\hat{\vv}(\rr)\cdot \hat{\rho}(\rr)\ell^3 \hat{\vv}(\rr)+\ell^3 e_0(\hat{\rho}(\rr))
\end{equation} Here, $e_0(\rho)$ is the zero-temperature energy density of the homogeneous Fermi gas with density $\rho$, and $m\hat{\rho}(\rr)\ell^3$ is the amount of matter (the mass) in the portion centered at $\rr$. The equations of motion for $\hat{\rho}$ and $\hat{\vv}$ in the Heisenberg picture, derived from the Hamiltonian $H$, take the form of a continuity equation and an operator-valued Euler equation (without viscosity term),\footnote{See note \ref{note_passuper} for the equation for $\hat{\rho}(\rr)$. The equation for $\hat{\vv}(\rr)$ is obtained by taking the gradient of that for $\hat{\phi}(\rr)$, $\hbar\partial_t\hat{\phi}=-\mu_0(\hat{\rho})-m\hat{\vv}^2/2$, where $\mu_0(\rho)$ is the zero-temperature chemical potential function, as in Equation (\ref{eqe2_30}).} hence the name quantum hydrodynamics given to the theory (with the risk of confusion pointed out in note \ref{notee1_23}). 

As you will have gathered, the great strength of this effective theory is that it does not depend on the nature of the bosonic or fermionic particles constituting the underlying superfluid, nor on their interactions (strong or weak, in liquid or gas phase) as long as they remain short-range, except through the equation of state $e_0(\rho)$ and the curvature parameter $\gamma$ at zero temperature. It therefore applies equally well to weakly interacting Bose gases, strongly interacting Fermi gases and liquid helium-4 (an extremely dense system that defies microscopic theory). 

However, the formalism can only be trusted in a low-temperature limit, $T\to 0$, where spatial density fluctuations and phase gradients are small; Equation (\ref{eqe1_29}) must therefore be expanded to the relevant order (here, fourth order) in powers of 
\begin{equation}\label{eqe2_29}
\delta\hat{\rho}(\rr)\equiv\hat{\rho}(\rr)-\hat{\rho}_0 \quad \mbox{and}\quad \delta\hat{\phi}(\rr)\equiv\hat{\phi}(\rr)-\hat{\phi}_0
\end{equation} where $\hat{\rho}_0$ and $\hat{\phi}_0$ are the zero-wave-vector Fourier components of the fields $\hat{\rho}(\rr)$ and $\hat{\phi}(\rr)$ (physically, $\hat{\rho}_0=\hat{N}/L^3$, where $\hat{N}$ is the total number of fermions operator, $L^3$ is the volume of the quantization box $[0,L]^3$ and $\hat{\phi}_0$ is the phase operator of the pair condensate \cite{brouilfer}). The expanded Hamiltonian is formally written as 
\begin{equation}\label{eqe3_29}
H=H_0+H_2+H_3+H_4+\ldots
\end{equation} where $H_n$ is the contribution of total degree $n$ in $\delta\hat{\rho}$ and $\delta\hat{\phi}$.

The 0-order contribution $H_0$ is a constant of little interest; the 1-order contribution is exactly zero (because $\sum_\rr\ell^3\delta\hat{\rho}(\rr)=0$ by construction) and has been omitted directly from Equation (\ref{eqe3_29}). The quadratic contribution $H_2$ is diagonalized by a Bogolioubov transformation:\footnote{\label{notee32} This transformation corresponds to the modal expansions $\delta\hat{\rho}(\rr)=L^{-3/2}\sum_{\kk\neq\zero} \rho_\kk (\hat{b}_\kk+\hat{b}_{-\kk}^\dagger) \exp(\ii\kk\cdot\rr)$ and $\delta\hat{\phi}(\rr)=L^{-3/2}\sum_{\kk\neq\zero}\phi_\kk (\hat{b}_\kk-\hat{b}_{-\kk}^\dagger)\exp(\ii\kk\cdot\rr)$ where $\rho_{\kk}=(\hbar\rho k/2 m c)^{1/2}$ and $\phi_\kk=(-\ii)(mc/2\hbar\rho k)^{1/2}$ are the amplitudes of the quantum fluctuations of density and phase in the phonon mode of wave vector $\kk$. Note the relation $-\ii\omega_\kk\delta\rho_\kk-\rho(\hbar/m)k^2\phi_\kk=0$ imposed by the linearized continuity equation.} 
\begin{equation}\label{eqe4_29}
H_2=\mbox{const}+\sum_{\kk\neq\zero} \hbar\omega_\kk \hat{b}_\kk^\dagger\hat{b}_\kk
\end{equation} where the creation $\hat{b}_\kk^\dagger$ and annihilation $\hat{b}_\kk$ operators of an elementary excitation (a phonon) of wave vector $\kk$ obey the usual bosonic commutation relations 
\begin{equation}\label{eqe1_30}
[\hat{b}_\kk,\hat{b}_{\kk'}]=0 \enspace\mbox{and}\enspace[\hat{b}_\kk,\hat{b}_{\kk'}^\dagger]=\delta_{\kk,\kk'}
\end{equation} The spectrum obtained here is exactly linear, $\omega_\kk=c k$, with the speed of sound given by 
\begin{equation}\label{eqe2_30}
mc^2=\rho\frac{\dd^2}{\dd\rho^2} e_0(\rho) = \rho \frac{\dd}{\dd\rho}\mu_0(\rho)
\end{equation} where $\mu_0(\rho)$ is the zero-temperature chemical potential of the Fermi gas with density $\rho$. The relation (\ref{eqe2_30}) is exact (it's well known from superfluid hydrodynamics \cite{tome9landau}), but the systematic absence of curvature in the spectrum is not physically realistic: this pathology stems from the fact that we have omitted so-called gradient corrections \cite{SonWingate} from Hamiltonian $H$; to simplify, as illustrious predecessors \cite{LK} have done, we replace $\omega_\kk$ in $H_2$ by hand with its cubic approximation (\ref{eqe1_24}), which is justified by reference \cite{Annalen}. 

The approximation $H_2$ corresponds to an ideal gas of phonons, and cannot describe sound attenuation. The interaction between phonons that causes their damping comes from the cubic contribution $H_3$ and the quartic contribution $H_4$. For the sake of simplicity, we give here only the expression of the most useful part of $H_3$, simplified to make the physics clear: 
\begin{equation}\label{eqe3_30}
H_3|_{\rm simpl} = \frac{\mathcal{A}}{2L^{3/2}}\sum_{\kk,\kk',\qq} (k k' q)^{1/2} \hat{b}_\kk^\dagger\hat{b}_{\kk'}^\dagger\hat{b}_\qq \delta_{\kk+\kk',\qq} + (k' k q)^{1/2} \hat{b}_{\kk'}^\dagger\hat{b}_\kk \hat{b}_\qq \delta_{\kk',\kk+\qq}+\ldots
\end{equation} with the constant amplitude (independent of the wave numbers) as a factor,
\begin{equation}\label{eqe4_30}
\mathcal{A}=(\xi/2)^{3/2}\rho^{-1/2}\left[3 mc^2+\rho^2 \frac{\dd^3}{\dd\rho^3}e_0(\rho)\right]
\end{equation} and $\xi=\hbar/mc$ as before.\footnote{The true coupling amplitude depends on the angles between the three wave vectors $\kk_i$ involved; as damping is actually dominated at low temperatures by processes at small angles between wave vectors, due to the small denominator effect described below, we have written the amplitude directly at zero angles $\kk_i\cdot\kk_j/k_ik_j=1$.} The ellipse in Equation (\ref{eqe3_30}) contains terms $\hat{b}^\dagger\hat{b}^\dagger\hat{b}^\dagger$ and $\hat{b}\hat{b}\hat{b}$ of no great importance, as they do not conserve the energy $H_2$. The terms $\hat{b}^\dagger\hat{b}^\dagger\hat{b}$ and $\hat{b}^\dagger\hat{b}\hat{b}$, on the other hand, are central to damping: they correspond respectively to the Beliaev process (phonon $\qq$ decays into two phonons $\kk$ and $\kk'$) and to the Landau process (phonon $\qq$ merges with a phonon $\kk$ to form a single phonon $\kk'$), which can be represented diagrammatically as follows, where the phonon whose damping we are studying plays a privileged role: 
\begin{equation}\label{eqe1_31}
\begin{tabular}{c}
\includegraphics[width=5cm,clip=]{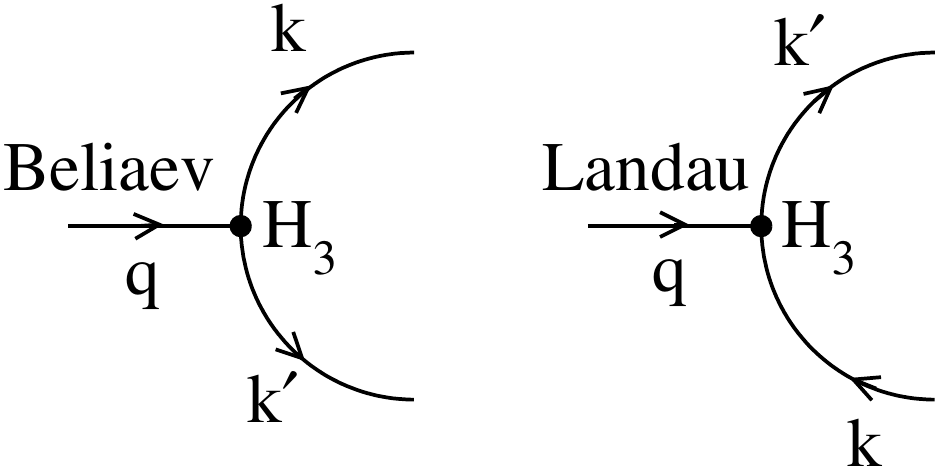}
\end{tabular}
\end{equation} As $H_3$ is cubic, each vertex of the diagram representing its action is the meeting point of three phonon lines. We could proceed in the same way with $H_4$ (the vertices would be four lines), but we won't do so, as quartic processes generally play a sub-dominant role in damping compared with cubic processes (for example, in the four-phonon damping $\phi\phi\to\phi\phi$ for $\gamma<0$, the amplitude of the direct process $\kk,\qq\to\kk',\kk''$ induced by $H_4$ in first-order perturbation theory is in practice negligible compared to that of the indirect process of the same initial and final state $\kk,\qq\to\kk+\qq\to\kk',\kk''$ induced by $H_3$ treated in second-order, due to the appearance in the latter of a very small energy denominator at small angles between $\kk$ and $\qq$).\footnote{This energy denominator $\hbar\omega_\qq+\hbar\omega_\kk-\hbar\omega_{\kk+\qq}$ would be exactly zero at zero angle without the curvature terms in the dispersion relation (\ref{eqe1_24}), which incidentally shows the singularity of a $\gamma=0$ theory (to which we'll return in Section \ref{sece4.3}).}

\subsection{How to calculate phonon damping?}
\label{sece4.3}

Let's imagine that we apply a short pulse of Bragg excitation to the gas, initially at thermal equilibrium, inducing a small coherent Glauber displacement of amplitude $\alpha\in\mathbb{C}^*$ in the phonon mode of wave vector $\qq$ without affecting the other modes, which corresponds to the unitary evolution operator $U_{\rm exc}=\exp(\alpha\hat{b}_\qq^\dagger-\alpha^*\hat{b}_\qq)$.\footnote{In a cold atom experiment, Bragg excitation is induced by the superposition of two far off-resonance laser beams of wave vectors $\kk_1$ and $\kk_2$ with $\kk_1-\kk_2=\qq$; even if the $\pm \qq$ acoustic modes are initially empty ($\hbar\omega_\qq\gg k_{\rm B} T$) and can only receive phonons, Raman (two-photon) processes - absorption of a photon in one laser beam, stimulated emission in the other - induce $\pm\hbar(\kk_1-\kk_2)=\pm\hbar\qq$ momentum changes in the Fermi gas and generally excite the two modes; however, the duration of the Bragg excitation can be adjusted so that the $-\qq$ mode emerges intact from the excitation procedure \cite{Cartago}.} Immediately after excitation, we have a non-zero mean for the corresponding annihilation operator: 
\begin{equation}\label{eqe1_32}
\langle\hat{b}_\qq(0^+)\rangle=\alpha\neq 0
\end{equation} This leads to an observable modulation of the mean gas density at wave vectors $\pm\qq$ since $\langle\delta\hat{\rho}(\rr,0^+)\rangle$, a linear combination of $\langle\hat{b}_\kk(0^+)\rangle$ and $\langle\hat{b}_\kk^\dagger(0^+)\rangle$ as in note \ref{notee32}, is then $\neq 0$. 

\begin{figure}[t]
\includegraphics[width=10cm,clip=]{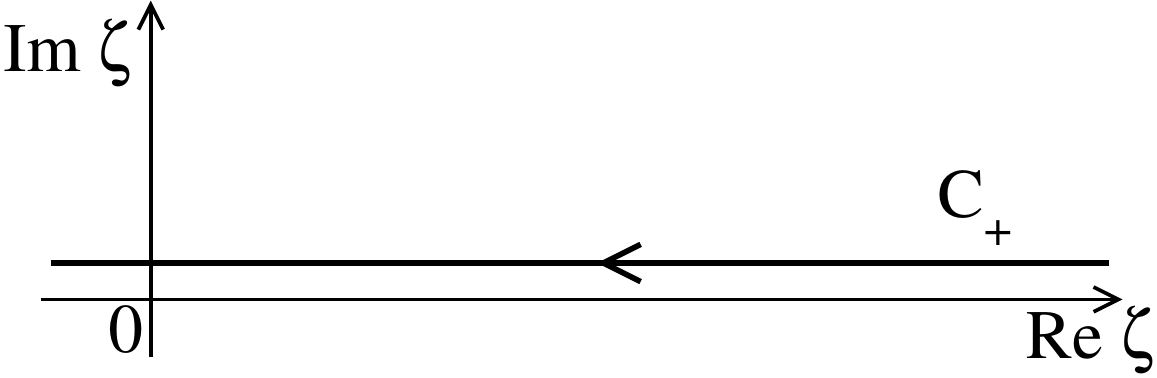}
\caption{Integration path in the complex plane followed by Equation (\ref{eqe2_32}).}
\label{fige6}
\end{figure}

In the limit $\alpha\to 0$, i.e.\ in the linear response regime, the many-body Green's function formalism applied to the effective low-energy theory, hence to the phonon Hamiltonian (\ref{eqe3_29}) [rather than to a microscopic description of the Fermi gas with interaction potential $V(\rr_{ij})$ as references \cite{tome9landau,FetterW} for example do], leads to the exact expression 
\begin{equation}\label{eqe2_32}
\langle\hat{b}_\qq(t)\rangle \stackrel{t>0}{=} \alpha \eee^{-\ii\omega_\qq t} \int_{C_+} \frac{\dd\zeta}{2\ii\pi} \frac{\eee^{-\ii\zeta t/\hbar}}{\zeta-\Sigma_\qq(\zeta)}
\end{equation} In this expression, $C_+$ is the integration path parallel to the real axis in the upper complex half-plane, followed from right to left (from $\re\zeta=+\infty$ to $\re\zeta=-\infty$), see Figure \ref{fige6}, and $\Sigma_\qq(\zeta)$ is the self-energy at wave vector $\qq$ and complex energy $\zeta$,\footnote{\label{notee1_33p}In comparison with the usual energy variable $z$ (of reference \cite{FetterW}, for example), the energy variable used here is shifted by the unperturbed energy of mode $\qq$, $\zeta=z-\hbar\omega_\qq$. This explains why we were able to take out the unperturbed phase factor in (\ref{eqe2_32}) and why $\Sigma_\qq(\zeta)$ is taken at $\zeta=\ii 0^+$ in the approximation (\ref{eqe1_34}) to come (it actually corresponds to $z =\hbar\omega_\qq+\ii 0^+$).} which is not known explictly but is defined by its perturbative expansion to all orders in the phonon-phonon interaction. Here, we restrict ourselves to the cubic interaction $H_3$ (see Section \ref{sece4.2}) and the expansion takes the following diagrammatic form 
\begin{equation}\label{eqe1_33}
\begin{tabular}{c}
\includegraphics[width=8cm,clip=]{diagb.pdf}
\end{tabular}
\end{equation} where the integer $n$ gives the order in $H_3$ \cite{insuffisance}. We only indicate the topology; it remains to sum over all possible non-redundant orientations of the internal lines, see example of diagram section (\ref{eqe1_31});\footnote{In the first diagram of (\ref{eqe1_33}) (with one loop), (i) orienting the top line to the right and the bottom line to the left and (ii) orienting the top line to the left and the bottom line to the right correspond to the same contribution, by invariance of the diagram by angle $\pi$ rotation around its horizontal axis. The same applies to the inner loop of the second diagram in (\ref{eqe1_33}), which is locally rotationally symmetric. For the same reason of rotational symmetry (this time global) of the diagrams of order $n=4$, we decide, in order to avoid double counting, to put the inner loop in the upper branch and, in the third diagram of (\ref{eqe1_33}), to orient the bridge downwards.} a precise value can then be assigned to each diagram, involving a sum over the wave vectors and Matsubara frequencies of the internal lines \cite{FetterW}. 

\begin{figure}[tb]
\includegraphics[width=8cm,clip=]{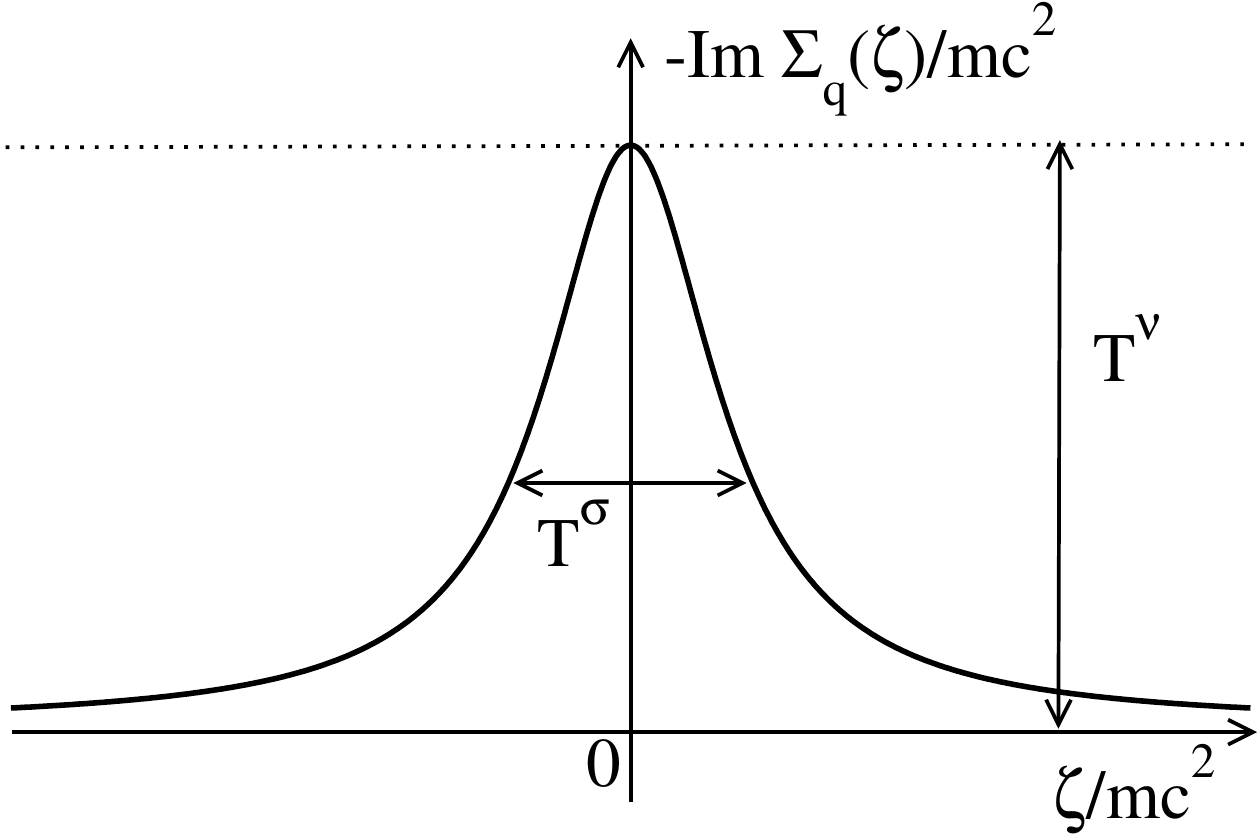}
\caption{In the limit $T\to 0$, the order of magnitude and typical width of function $\im\Sigma_\qq(\zeta)$ near $\zeta=\ii 0^+$ are assumed to be characterized by two power laws in temperature, with exponents $\nu$ and $\sigma$ (taking into account a possible temperature dependence of the curvature parameter $\gamma$).}
\label{fige7}
\end{figure}

In order to obtain explicit results for the damping, we traditionally perform the following two approximations:\footnote{\label{notee2_33}Our effective theory being exact to the leading order in temperature, we use it only in the $T\to 0$ limit, fixing the ratio $\bar{q}=\hbar c q/k_{\rm B}T$ so that the mode $\qq$ is also described exactly. We then have $\omega_\qq/\Gamma_{\rm th}\to +\infty$ where $\Gamma_{\rm th}=\Gamma_{q=k_{\rm B} T/\hbar c}$ is the thermalization rate of the phonon gas and $\Gamma_q$ is the function (\ref{eqe3_34}), since the exponent $\nu$ introduced in Figure \ref{fige7} on page \pageref{fige7} is always $>1$, see Table \ref{tablee1} on page \pageref{tablee1}: the thermalization rate $\Gamma_{\rm th}$ tends towards zero faster than the angular eigenfrequency $\omega_\qq$ and the mode by definition enters the collisionless regime. In the opposite, hydrodynamic regime $\omega_\qq\ll\Gamma_{\rm th}$, the phonon gas has time to reach local thermal equilibrium at each oscillation point of the sound wave $\qq$, and damping is described by viscosity-type coefficients in the classical hydrodynamic equations of a two-fluid model \cite{Khalatlivre}.} 
\begin{enumerate}\item the Markov approximation (the phonon gas seen by mode $\qq$ constitutes a reservoir with no memory, i.e.\ with negligible correlation time): we ignore the energy dependence of the self-energy as follows (see note \ref{notee1_33p}), 
\begin{equation}\label{eqe1_34}
\Sigma_\qq(\zeta)\simeq \Sigma_\qq(\ii 0^+)
\end{equation} The integral in Equation (\ref{eqe2_32}) is then calculated by the residue theorem (one closes the contour with an infinite semicircle in the lower complex half-plane), 
\begin{equation}\label{eqe2_34}
\langle\hat{b}_\qq(t)\rangle|_{\rm Markov} \stackrel{t\geq 0}{=} \alpha \eee^{-\ii\omega_\qq t} \eee^{-\ii \Sigma_\qq(\ii 0^+)t/\hbar}
\end{equation} The signal decay in this case is exponential, with a rate corresponding to the imaginary part of $\Sigma_\qq(\ii 0^+)$ (the real part gives the change in the mode angular frequency).
\item Born approximation: $\im\Sigma_\qq(\ii 0^+)$ is calculated perturbatively to leading order $n$ in $H_3$. The damping rate of phonon $\qq$ is then\footnote{To shed further light on the approximation (\ref{eqe3_34}), let's point out for $n=2$ that it can be obtained by the quantum master equation method well known in quantum optics (we obtain a closed evolution equation for the density operator $\hat{\rho}_S(t)$ of a small system $S$ - here the phonon mode $\qq$ - coupled to a large reservoir $R$ - here the other phonon modes $\kk\neq\qq$, by resorting to the Born-Markov approximation) \cite{CCTbordeaux,mon_cours} or, more simply, using Fermi's golden rule (we calculate $\dd\langle\hat{n}_\qq\rangle/\dd t$, where $\langle\hat{n}_\qq\rangle=\langle\hat{b}_\qq^\dagger\hat{b}_\qq\rangle$ is the mean number of phonons in the non-equilibrium mode $\qq$, by summing the incoming fluxes - population processes $\kk,\kk'\to \qq$ and $\kk'\to \kk,\qq$ - and outgoing fluxes - inverse depopulation processes $\qq\to\kk,\kk'$ and $\qq,\kk\to\kk'$ - and linearizing just after the Bragg excitation, as $(\dd/\dd t)\delta n_\qq(t=0^+)=-\Gamma_\qq \delta n_\qq(t=0^+)$, where $\delta n_\qq(t)\equiv \langle n_\qq\rangle(t)-\bar{n}_\qq$ is the deviation from thermal equilibrium). For $n=4$, we find the same result (\ref{eqe3_34}) by extending Fermi's golden rule to higher orders \cite{Landaumec}.}\ \footnote{\label{notee_sub} There's a little mathematical subtlety here: if the leading order $n$ is $\geq 4$, the Markov approximation must no longer simply replace $\Sigma_\qq(\zeta)$ with $\Sigma_\qq(\ii 0^+)$ but must approximate it by a Taylor expansion around $\zeta=\ii 0^+$. For example, for $n=4$, we take $\Sigma_\qq(\zeta)\simeq \Sigma_\qq(\ii 0^+)+\zeta\frac{\dd}{\dd\zeta}\Sigma_\qq(\ii 0^+)$ so that at order 4 in $H_3$, $\zeta-\Sigma_\qq(\zeta)\simeq[1-\frac{\dd}{\dd\zeta}\Sigma_\qq(\ii 0^+)][\zeta-\Sigma_\qq^{(2)}(\ii 0^+)-\Sigma_\qq^{(4){\rm eff}}(\ii 0^+)]$ with $\Sigma_\qq^{(4){\rm eff}}(\ii 0^+)=\Sigma_\qq^{(4)}(\ii 0^+)+\Sigma_\qq^{(2)}(\ii 0^+)\frac{\dd}{\dd\zeta}\Sigma_\qq^{(2)}(\ii 0^+)$. Then replace $\Sigma_\qq^{(4)}(\ii 0^+)$ with $\Sigma_\qq^{(4){\rm eff}}(\ii 0^+)$ in the damping rate expression (\ref{eqe3_34}). In the three-dimensional concave case, this makes no difference to $\Gamma_\qq|_{\rm Born-Markov}$, since $\frac{\dd}{\dd\zeta}\Sigma_\qq^{(2)}(\ii 0^+)$ is a real quantity, as is $\Sigma_\qq^{(2)}(\ii 0^+)$; in the two-dimensional concave case, the conclusion is less obvious but remains the same, see note \ref{notee_scott}. On the other hand, this substitution must be made in the calculation of the thermal angular frequency shift $\Delta_\qq$ of the mode, $\hbar\Delta_\qq|_{\rm Born-Markov}=\re[\Sigma_\qq^{(2)}(\ii 0^+)+\Sigma_\qq^{(4){\rm eff}}(\ii 0^+)]-\mbox{idem at }T=0$.}
\begin{equation}\label{eqe3_34}
\Gamma_\qq|_{\rm Born-Markov} = -\frac{2}{\hbar} \im\Sigma_\qq^{(n)}(\ii 0^+)
\end{equation} where the exponent gives the order in $H_3$. In the case of a convex acoustic branch ($\gamma>0$) in 3D, it's sufficient to go to order $n=2$: this is the three-phonon Beliaev-Landau damping, much studied theoretically and observed in liquid helium-4 \cite{Roach,hel4} and, to a lesser extent, in bosonic cold-atom gases, only Beliaev damping having been seen there \cite{NirDav}. In the concave case ($\gamma<0$), we need to go to order $n=4$ (the first contribution in Equation (\ref{eqe1_33}) is purely real for $\zeta=\ii 0^+$ because its energy denominators, of the form $\hbar\omega_\qq+\hbar\omega_\kk-\hbar\omega_{\kk+\qq}$ and $\hbar\omega_\qq-(\hbar\omega_\kk+\hbar\omega_{\qq-\kk})$, cannot vanish, but the next two are not in $\phi\phi\to\phi\phi$ processes); this case has been little studied theoretically (reference \cite{Annalen} noted and corrected an error in the original calculation \cite{LK}, and reference \cite{diffpferm} obtained a much more explicit expression of the result, even generalizing it to a non-zero phonon chemical potential $\mu_\phi$)\footnote{If we restrict ourselves at low temperature to the leading collisional processes $\phi\phi\rightarrow\phi\phi$, the phonon number becomes a conserved quantity, allowing us to take $\mu_\phi<0$.} and it has, to our knowledge, never been observed experimentally (no precise measurement of $\Gamma_\qq^{\gamma <0}$ has been made in any system). 
\end{enumerate}

 Let's determine the validity of the Born approximation by means of the estimate of order $n \in 2\mathbb{N}^*$ given in reference \cite{insuffisance}: 
\begin{equation}\label{eqe1_35}
\Sigma_{\qq}^{(n)}(\ii 0^+)\approx \int\left(\prod_{i=1}^{n/2}\dd^dk_i\right) \frac{\langle\ \ |L^{d/2}H_3|\ \ \rangle^n}{(\Delta E)^{n-1}}\approx |\gamma|T^3 \left(\epsilon_{d{\rm D}}=\frac{T^{2d-4}}{|\gamma|^{(5-d)/2}}\right)^{n/2}
\end{equation} where $d\geq 2$ is the dimension of space. The writing in the second expression symbolically represents the product of $n$ matrix elements of the cubic phonon interaction in the numerator and the product of $n-1$ energy denominators (associated with $n-1$ intermediate states) in the denominator, and the integral is taken over the independent phonon wave vectors $\kk_i$. The order of magnitude in the third expression is obtained as in reference \cite{insuffisance} by restricting to the small $O(|\gamma|^{1/2}T)$ angles between $\kk_i$ and $\qq$, which is legitimate when $T\to 0$ \cite{LK}; we omit here the dependence in the density $\rho$, in $\xi=\hbar/mc$ and in the coupling constant $\mathcal{A}$ of Equation (\ref{eqe4_30}), unlike reference \cite{insuffisance}, but we keep that in the curvature parameter $\gamma$, as we will soon link $\gamma$ and $T$. In short, the Born expansion is legitimate in dimension $d = 3$ if its small parameter tends to zero at low temperature: 
\begin{equation}\label{eqe2_35}
\epsilon_{{\rm 3D}}=\frac{T^2}{|\gamma|}\underset{T\to 0}{\to} 0
\end{equation} To discuss the validity of the Markov approximation, we assume that the behavior of the self-energy in the vicinity of $\zeta=\ii 0^+$ is characterized by two exponents, $\nu$ and $\sigma$, the one giving the typical values $\propto T^\nu$ of its imaginary part and the one giving its typical scale of variation $\propto T^\sigma$, as in Figure \ref{fige7}.\footnote{The exponents introduced here differ by one unit from those in reference \cite{insuffisance} due to a different choice of convention.} \footnote{For the sake of clarity, we have assumed in Figure \ref{fige7} that the function shown has a maximum at the origin. This is not necessarily true (the two-dimensional convex case of reference \cite{insuffisance} provides a counter-example, see its Equation (114)). The true definition of exponents $\nu$ and $\sigma$ is that the scaled function $\im \Sigma_\qq\left(\zeta=\bar{\zeta}mc^2(k_{\rm B}T/mc^2)^\sigma\right)/\left[mc^2(k_{\rm B}T/mc^2)^\nu\right]$ has a finite, non-zero limit as $T\to 0$ with fixed reduced complex energy $\bar{\zeta}$ ($\im\bar{\zeta}>0$).} 
From Equation (\ref{eqe3_34}), we therefore have 
\begin{equation}\label{eqe3_35}
\Gamma_\qq \underset{T\to 0}{\approx} T^\nu
\end{equation} The function $\Sigma_\qq(\zeta)$ then has a slow (negligible) energy variation on the scale of the damping rate (which is indeed the inverse of the characteristic time in Equation (\ref{eqe2_32})) if it is wider than high, which imposes 
\begin{equation}\label{eqe1_36}
\nu>\sigma
\end{equation} The exponent $\nu$ is obtained by an explicit calculation of the right-hand side of Equation (\ref{eqe3_34}) in the limit $T\to 0$, as was done in reference \cite{Annalen} for the three-dimensional case\footnote{The historical reference \cite{LK} for fixed $\gamma<0$ also finds $\nu = 7$ but the $q$ dependence of $\Gamma_\qq$ is different, e.g.\ $\Gamma_\qq\approx q T^6$ in \cite{LK} instead of $q^3T^4$ in \cite{Annalen} as $q\to 0$.} and in reference \cite{insuffisance} for the two-dimensional convex case; more simply, we can use the estimate (\ref{eqe1_35}) with $n=2$ if $\gamma>0$ and $d\in\{2,3\}$, $n=4$ if $\gamma<0$ and $d=3$.\footnote{\label{notee_scott} For fixed $\gamma<0$ and $d=2$, Alice Sinatra obtained in 2021, in the formulation of references \cite{epl,Annalen}, the unpublished result that $\im\Sigma_\qq^{(n=4)}(\ii 0^+)=0$ at order $T^3$ (the expected leading order in temperature). To see this, it is actually simpler to use expressions (84) and (85) in reference \cite{insuffisance}: (i) in (84), we can ignore processes $\phi\leftrightarrow\phi\phi\phi$ and restrict ourselves to process $\phi\phi\to\phi\phi$ (second contribution), the only one conserving energy-momentum; (ii) in the integrand of (85), we are entitled to replace $\zeta$ by $0$ in the numerator of the large fraction for the similar reason that processes $\phi\leftrightarrow\phi\phi$ do not conserve energy-momentum - this makes the numerator real ; (iii) we then check that, if the scaled energy difference $\Delta E/\left[k_{\rm B}T (k_{\rm B}T/mc^2)^2\right]$ next to $\ii 0^+$ vanishes in the denominator of the large fraction, as required by the Dirac distribution $\delta(\Delta E)$ of the generalized Fermi golden rule \cite{Landaumec}, the numerator also vanishes (we show this by formally replacing $\gamma$ in the numerator by its $\Delta E$-cancelling expression, a rational function of the moduli and angles of the phonon wavevectors). 
In other words, the limit at $\zeta=0$ (but also at $\zeta=-\Delta E$) of the transition amplitude in the numerator of the large fraction, considered as a rational fraction of the angles, can be written $\Delta E\times P/Q$, where the polynomials $\Delta E$, $P$ and $Q$ are two by two coprime.

However, this reasoning neglects possible edge effects in the integral on phonon wave numbers, in the sense of reference \cite{fermi1D}, where one of the wave numbers tends to zero, which makes one of the energy denominators of the $\phi\leftrightarrow\phi\phi$ processes tend to zero. Including these edge effects, we find that the processes $\phi\to\phi\phi\phi$, $\phi\phi\to\phi\phi$ and $\phi\phi\phi\to\phi$, abbreviated as $1\to 3$, $2\to 2$ and $3\to 1$ in reference \cite{insuffisance}, each make a non-zero contribution to $\im\Sigma_\qq^{(n=4)}(\ii 0^+)$ at order $T^3$, but the sum of these contributions is exactly zero (the contribution of the edge $q_1'+q_2'=q$ in $2\to 2$, i.e.\ $C\int_0^{\bar{q}}\dd\bar{k}\,(\bar{n}_k^{\rm lin}+\bar{n}_{q-k}^{\rm lin}+1)\bar{k}(\bar{q}-\bar{k})/\bar{q}$, is exactly offset by $1\to 3$, and that of the edges $q_1'=0$ and $q_2'=0$ as a whole, i.e.\ $2C\int_{\bar{q}}^{+\infty}\dd\bar{k}\, (\bar{n}_{k-q}^{\rm lin}-\bar{n}_k^{\rm lin})\bar{k}(\bar{k}-\bar{q})/\bar{q}$, is exactly offset by $3\to 1$; here, $\bar{k}=\hbar c k/k_{\rm B}T$, $\bar{n}_k^{\rm lin}=1/(\exp\bar{k}-1)$, $\Lambda=\rho^2\left(\frac{\dd^3}{\dd\rho^3}e_0(\rho)\right)/(3 mc^2)$ and $C=k_{\rm B}T (k_{\rm B}T/mc^2)^2\left[9(1+\Lambda)^2/8\rho\xi^2\right]^2/\left[\pi(3\gamma)^2\right]$). 

The conclusion is not changed by the correction $\Sigma_\qq^{(2)}(\ii 0^+)\frac{\dd}{\dd\zeta}\Sigma_\qq^{(2)}(\ii 0^+)$ of our note \ref{notee_sub}, as we find that factors $\Sigma_\qq^{(2)}(\ii 0^+)$ and $\frac{\dd}{\dd\zeta}\Sigma_\qq^{(2)}(\ii 0^+)$ are both real. This was obvious for the first factor (the edge effects it presents in 1D \cite{fermi1D} are suppressed in 2D by a lowering of the phonon density of states at low wavenumber). This was not the case for the second factor: because of edge effects in the integration over $\kk$ (see Equation (39) in reference \cite{insuffisance}), the Beliaev and Landau processes each give a non-zero contribution to $\im\frac{\dd}{\dd\zeta}\Sigma_\qq^{(2)}(\ii 0^+)$ (it is $2[9(1+\Lambda)^2/8\rho\xi^2]/[\bar{q}(3|\gamma|)^{3/2}]$ for Landau at temperature leading order) but these contributions are exactly opposite, in particular because the energy denominators $\veps_\qq-(\veps_\kk+\veps_{\qq-\kk})$ and $\veps_\qq+\veps_\kk-\veps_{\qq+\kk}$ are opposite $\sim\mp\hbar c k[1-(v_q/c)\cos\theta]$ at leading order in $k$ ($v_q=\dd\veps_\qq/\hbar\dd q$ is the group velocity and $\theta$ is the angle between $\kk$ and $\qq$) and enter the derivative of the Dirac distribution $\delta'(\veps)$, which is an odd function of its argument.

The two-dimensional concave case is therefore special: the quantity $\im\Sigma_\qq^{(4){\rm eff}}(\ii 0^+)$ of note \ref{notee_sub} - considered at all temperature orders - does not give the correct temperature scaling law $\propto T^3$ of the self-energy at fourth order in $H_3$ on a $O(T^3)$ neighborhood of $\zeta=\ii 0^+$; nor does it give the damping rate $\Gamma_\qq$, since the Born and Markov approximations fail, as in the two-dimensional convex case (see the last line of our Table \ref{tablee1}). Note, however, still for $\gamma<0$, that the limiting case $\rho\xi^2\to +\infty$ of a very weakly interacting underlying superfluid must be set apart as there we have the additional small parameter $1/\rho\xi^2$ helping validity of Markov (as in Section 3.2 of reference \cite{insuffisance}) and Born (as in Equation (17) of the same reference); this limiting case is inaccessible in a gas of spin-$1/2$ fermions with contact interaction - we have $\rho\xi^2=O(1)$ when $\gamma<0$ \cite{csdm} - but it is in a Bose gas with an interaction range $\gtrsim\xi$ as considered in reference \cite{Annalen}.} The exponent $\sigma$ is obtained by generalizing the previous estimate to the case $\zeta\neq 0$, i.e.\ by adding $\zeta$ to $\Delta E$ in (\ref{eqe1_35}); however, in an expansion at small angles between $\kk_i$ and $\qq$, the part of $\Delta E$ linear in the wave numbers vanishes, leaving only the cubic contributions $\approx\gamma T^3$, so that, independently of the dimension of space $d$, 
\begin{equation}\label{eqe2_36}
\frac{1}{\zeta+\Delta E} \approx \frac{1}{\zeta+\gamma T^3} \quad\mbox{and therefore}\quad T^\sigma\approx|\gamma|T^3
\end{equation} taking into account the dependence on the parameter $\gamma$, which - as we said - may vary with temperature. The resulting validity condition $T^\nu=o(\gamma T^3)$ in (\ref{eqe1_36}) can be given a simple interpretation: in the limit $T\to 0$ taken with the scaling law $q\approx T$, the damping rate $\Gamma_\qq$ must tend to zero faster than the $q^3$ term in $\omega_\qq$, 
\be
\Gamma_q\stackrel{\hbar c q/k_{\rm B}T\,\mbox{\scriptsize fixed}}{\underset{T\to 0}{=}}o(\omega_\qq^{(3)})\quad\mbox{with}\quad \omega_\qq^{(3)}=\gamma mc^2(q\xi)^3/8\hbar
\ee
which is the true mark of the Markovian nature of damping (rather than the naive perturbative condition $\Gamma_\qq=o(\omega_\qq)$).

The situation is summarized in Table \ref{tablee1} below.\footnote{In the third row of the table, possible logarithmic factors $\ln(1/T)$ are omitted for simplicity. These factors arise from the fact that, for scaling laws $\gamma\propto T^2$ and $q\propto T$, the terms $q^3$ and $q^5\ln q$ are of the same order of magnitude in the dispersion relation (\ref{eqe1_24}): in this low curvature regime, the logarithmico-quintic contribution to $\omega_\qq$ is no longer a small correction and must be kept.} The Born-Markov approximation is therefore usable in dimension 3, except over a narrow interval of values of $\gamma$, of width $\approx (k_{\rm B}T/mc^2)^2$ around $\gamma = 0$; for $\gamma = 0$, the phonon dispersion relation (\ref{eqe1_24}) deviates quintically from the linear law $cq$, which is obviously a special case. \begin{table}[h]
\begin{tabular}{|l||c|c|c||c|c|}
\hline
& $\nu$ & $\sigma$ & Markov & $\epsilon_{d{\rm D}}$ &  Born \\
\hline
$d=3,\gamma>0$ \mbox{fixed}& 5 & 3 & yes & $\approx T^2\to 0$ & yes \\
\hline
$d=3,\gamma<0$ \mbox{fixed}& 7 & 3 & yes & $\approx T^2\to 0$ & yes \\
\hline
$d=3,\gamma=O(T^2)$ & 5 & 5 & no & $\approx T^0\not\to 0$ & no \\
\hline
$d=2,\gamma>0$ \mbox{fixed}& 3 & 3 & no & $\approx T^0\not\to 0$ & no \\
\hline
\end{tabular}
\medskip
\caption{In the study of phonon damping in a superfluid, validity of the Born-Markov approximation in the low-temperature limit $T\to 0$ depending on the dimension of space $d$ and the curvature parameter $\gamma$ of the acoustic branch (more precisely, its sign and its temperature dependence, the third line holding regardless of the sign of $\gamma$). The exponents $\nu$ and $\sigma$ used in the definition (\ref{eqe1_36}) of the Markovian regime are shown in Figure \ref{fige7}, and the small parameter of the Born expansion $\epsilon_{d{\rm D}}$ is given in Equation (\ref{eqe1_35}).}
\label{tablee1}
\end{table}

The precise calculation of the damping rate $\Gamma_\qq$ (or what takes its place for non-exponential decay, such as the inverse of the width of $|\langle\hat{b}_\qq(t)\rangle|^2$ at relative height $1/\eee$) for these small values of curvature is, to our knowledge, an open question; it is of great experimental relevance, as the interaction strength leading to $\gamma=0$ seems to be close to the unitary limit (see reference \cite{PRAconcav} and our Section \ref{sece5.1}), the preferred point for cold atom experiments in a regime of fairly high values of $T_{\rm c}/T_{\rm F}$ and collisional properties conducive to evaporative cooling \cite{Shlyap}.

For good measure, we have also considered the two-dimensional convex case in Table \ref{tablee1}: the Born-Markov approximation fails here, and reference \cite{insuffisance} had to resort to a non-perturbative heuristic approximation on the self-energy $\Sigma_\qq(\zeta)$ to reach good agreement with classical field simulations (quantum hydrodynamics operators $\hat{b}_\qq,\hat{b}_\qq^\dagger$ replaced by complex numbers $b_\qq,b_\qq^*$) in the weakly interacting regime $\rho\xi^2\gg 1$ of the underlying bosonic superfluid, where a small parameter was thought to be available and to ensure the success of Fermi's golden rule even in the $k_{\rm B}T/mc^2\to 0$ limit (this reasonable expectation, confirmed in Section 3.2 of reference \cite{insuffisance} to order two in $H_3$, is invalidated in Section 4.3 of the same reference by a calculation to order four).\footnote{We haven't even mentioned the very special case of dimension $d=1$, where two wave vectors make a very small angle (zero!) as soon as they are in the same direction. Let's just say that the Born small parameter is still given by Equation (\ref{eqe1_35}), even though this equation was obtained under the assumption $d\geq 2$. Restoring the density dependence as in \cite{insuffisance}, we find more precisely $\epsilon_{1\rm D}=1/[\gamma^2\rho\xi(k_{\rm B}T/mc^2)^2]$, the prefactor in (\ref{eqe1_35}) being written $\gamma (k_{\rm B}T)^3/(mc^2)^2$. In the low-temperature limit $k_{\rm B}T/mc^2\to 0$ at $\rho\xi$ fixed considered here, $\epsilon_{1\rm D}\to +\infty$ and we must immediately resort to non-perturbative approximations on $\Sigma_\qq(\zeta)$ and $\Gamma_\qq$, such as the self-consistent calculation of references \cite{auto1,auto2}. In the opposite weakly-interacting limit $\rho\xi\to+\infty$ at $k_{\rm B}T/mc^2$ fixed, $\epsilon_{1\rm D}\to 0$ and we can use Fermi's golden rule as in reference \cite{fermi1D}; more precisely, we expect the validity condition of the golden rule to be written as $\rho\xi(k_{\rm B}T/mc^2)^2\gg \phi_{1\rm D}(\bar{q})$ where $\phi_{1\rm D}$ is some function of $\bar{q}=\hbar c q/k_{\rm B}T$, forgetting the $\gamma$ dependence for simplicity (at a fixed $\bar{q}$, the Born approximation imposes this condition, but the Markov approximation is then also satisfied because we have $\hbar\Gamma_\qq|^{\scriptsize\mbox{golden}}_{\scriptsize\mbox{rule}}\approx\gamma[(k_{\rm B}T)^3/(mc^2)^2]\epsilon_{1\rm D}\approx k_{\rm B}T/(\gamma\rho\xi)\ll\Delta E\approx\gamma(k_{\rm B}T)^3/(mc^2)^2$ where $\Delta E$, the typical Beliaev-Landau energy denominator, gives the width in $\zeta$ of the self-energy as in Equation (\ref{eqe2_36})). In 2D, as shown in reference \cite{insuffisance}, the perturbative $H_3$-expansion of the self-energy is subject to a similar validity condition, $\rho\xi^2(k_{\rm B}T/mc^2)^2\gg \phi_{2\rm D}(\bar{q})$, see its Equation (96), which, unlike the 1D case, is not obtained by simple power counting.}

\subsection{Phase diffusion of the pair condensate}
\label{sece4.4}

A fundamental and practical question concerns the coherence time of the pair condensate at thermal equilibrium in a Fermi gas perfectly isolated from its environment.

For an infinite unpolarized system of $\uparrow$ and $\downarrow$ fermions, the coherence time is infinite, as asserted by the $U(1)$ symmetry-breaking phenomenon: in the grand canonical ensemble (term $-\mu\hat{N}$ added to the Hamiltonian of the Fermi gas where $\mu$ is the chemical potential and $\hat{N}$ the total number of particles operator), the complex order parameter $\Delta(\rr,t)$ is uniform and constant; in the canonical ensemble, it therefore evolves with the undamped phase factor $\exp(-2\ii \mu t/\hbar)$,\footnote{There is a factor of 2 under the exponential because $\Delta$ is a pair-order parameter whereas $\mu$ is the chemical potential of the fermions. There is no factor of 2 in Equation (\ref{eqe1_37}) because the phase operator $\hat{\phi}_0$ is conjugate to the fermion density.} advancing at the immutable angular frequency $2\mu/\hbar$; in any case, the coherence time is infinite. 

What about a finite size system (quantization box $[0,L]^3$, fixed total number of fermions $N$)? To find out, let's follow reference \cite{brouilfer} and write the evolution equation for the condensate phase operator, which we denote $\hat{\phi}_0$ as in Equation (\ref{eqe2_29}), in the quantum hydrodynamics regime:\footnote{To obtain this equation, we had to eliminate $\hat{b}_\qq\hat{b}_{-\qq}$ and $\hat{b}^\dagger_\qq\hat{b}^\dagger_{-\qq}$ terms by temporal smoothing; this is of no consequence, as they oscillate with a period $\approx\hbar/k_{\rm B}T$ much shorter than the collisional timescales of interest here (see below) and automatically average to zero.}
\begin{equation}\label{eqe1_37}
-\hbar\frac{\dd}{\dd t}\hat{\phi}_0 = \mu_0(\rho) + \sum_{\qq\neq\zero} \hat{b}_\qq^\dagger\hat{b}_\qq \frac{\dd}{\dd N}(\hbar\omega_\qq) \equiv \hat{\mu}
\end{equation} In the second expression, $\mu_0(\rho)=\dd E_0/\dd N$ is the chemical potential of the $N$ fermions in the ground state of energy $E_0$ at density $\rho$, and the sum over $\qq$ can be interpreted as the adiabatic derivative (meaning at fixed number operators $\hat{b}_\qq^\dagger\hat{b}_\qq$ of the phonon modes) with respect to $N$ of the corresponding sum in the phonon Hamiltonian $H_2$ (\ref{eqe4_29}). The second expression as a whole is therefore the isentropic derivative of the Hamiltonian with respect to the total number of particles. In this sense, it is a chemical potential operator for the fermions, hence the notation $\hat{\mu}$ in the third expression, and Equation (\ref{eqe1_37}) is a quantum version of the famous second Josephson relation, linking the time derivative  of the (classical) phase of the order parameter to the equilibrium chemical potential $\mu$. 

In a given realization of the experiment, which we assume to correspond to a $N$-body eigenstate $|\psi_\lambda\rangle$ of energy $E_\lambda$ sampling the canonical ensemble, the occupation numbers $\hat{b}_\qq^\dagger\hat{b}_\qq$ fluctuate and decorrelate under the effect of incessant collisions between phonons due in particular to $H_3$, see Equation (\ref{eqe3_30}). At times long enough for a large number of collisions to have taken place, we therefore expect a diffusive spreading of the condensate phase, the variance of the random phase shift increasing linearly with time: 
\begin{equation}\label{eqe1_38}
\mbox{Var}_\lambda[\hat{\phi}_0(t)-\hat{\phi}_0(0)] \underset{\Gamma_{\rm coll}^{\phi}t\gg 1}{\sim} 2 D_\lambda t
\end{equation} 
with a subintensive diffusion coefficient $D_\lambda$, i.e. $\approx 1/N$ in the thermodynamic limit. Here $\Gamma_{\rm coll}^{\phi}=\Gamma_{q=k_{\rm B}T/\hbar c}$ is the typical collision rate between thermal phonons (the function $\Gamma_q$ is that of Equation (\ref{eqe3_34})). The spreading (\ref{eqe1_38}) induces an exponential loss of temporal coherence of rate $D_\lambda$, by virtue of Wick's relation (the phase shift statistic in $|\psi_\lambda\rangle$ is expected to be approximately Gaussian \cite{vraiediff}), 
\begin{multline}
\label{eqe2_38}
\left\langle\exp\left\{-\ii\left[\hat{\phi}_0(t)-\hat{\phi}_0(0)\right]\right\}\right\rangle_\lambda \simeq \exp\left[-\ii\langle\hat{\phi}_0(t)-\hat{\phi}_0(0)\rangle_\lambda\right] \exp\left\{-\frac{1}{2}\mathrm{Var}_\lambda[\hat{\phi}_0(t)-\hat{\phi}_0(0)]\right\} \\
\underset{\Gamma_{\rm coll}^{\phi}t\gg 1}{\simeq} \exp\left[-\ii\langle\hat{\mu}\rangle_\lambda t/\hbar\right] \exp(-D_\lambda t)
\end{multline}
which is confirmed by the resolvent analysis of reference \cite{brouilfer}.\footnote{\label{notee1_38} If we assume that the interacting phonon gas is an ergodic quantum system \cite{eth1,eth2}, the mean $\langle\hat{\mu}\rangle_\lambda$ in the steady state $|\psi_\lambda\rangle$ depends only on the two conserved quantities, the energy $E$ and the number of particles $N$, and coincides for a large system with the microcanonical chemical potential $\mu_{\rm mc}(E=E_\lambda,N)$. If the energy $E$ fluctuates from one realization of the experiment to the next around the mean value $\bar{E}$, as in the canonical ensemble, the phase factor in the third expression of Equation (\ref{eqe2_38}) fluctuates and leads to a Gaussian-in-time loss of coherence: linearization of $\mu_{\rm mc}(E,N)$ around $\bar{E}$ gives $\mbox{Var}\,[\hat{\phi}_0(t)-\hat{\phi}_0(0)]\sim[\partial_E \mu_{\rm mc}(\bar{E},N)]^2(\mbox{Var}\, E)t^2/\hbar^2$, a parasitic effect $\approx t^2/N$ rapidly masking the phase diffusion (\ref{eqe1_38}) $\approx t/N$ \cite{superdiff}.}.

In the case $\gamma>0$ of a convex acoustic branch, the situation resembles that of weakly interacting Bose condensates well studied in reference \cite{praphase}: the dominant collisions are the three-phonon Beliaev-Landau ones $\phi\leftrightarrow\phi\phi$, and $D_\lambda$ has been calculated for the low-temperature Fermi pair condensate in reference \cite{diffpferm}; we give here a simplified expression, keeping only the scaling laws in $N$, $T$ and $\gamma$ (under the assumption $\gamma=O(1)$): 
\begin{equation}\label{eqe3_38}
D_\lambda^{\gamma>0} \approx N^{-1} T^4 \gamma^0
\end{equation} There has been no experimental verification yet in cold atomic gases (the only quantum fluids sufficiently well isolated for the condensate loss of coherence to be intrinsic), even for bosons.

In the concave case $\gamma<0$, on the other hand, the question remains largely open. An attempt to calculate $D_\lambda$ in reference \cite{diffpferm}, taking into account only the Landau-Khalatnikov four-phonon collision processes $\phi\phi\to\phi\phi$ at small angles, of typical rate $\Gamma_{\rm coll}^{\phi}\propto (k_{\rm B}T/mc^2)^7mc^2/\hbar|\gamma|\approx T^7$, led to an infinite diffusion coefficient, 
\begin{equation}\label{eqe1_39}
D_\lambda^{\gamma<0} = +\infty
\end{equation} more precisely to a superdiffusive spreading law (simplified as in (\ref{eqe3_38}))\footnote{In all cases, see Equations (\ref{eqe3_38}) and (\ref{eqe2_39}), we find that there is no phase spreading at the thermodynamic limit $N\to +\infty$: in an isolated gas, the limited coherence time of the condensate is a finite-size effect.} 
\begin{equation}\label{eqe2_39}
\mbox{Var}_\lambda^{\gamma<0} [\hat{\phi}_0(t)-\hat{\phi}_0(0)] \approx N^{-1} T^{20/3} |\gamma|^{1/3} t^{5/3}
\end{equation} in particular because collisions $\phi\phi\to\phi\phi$ preserve the total number $N_\phi$ of phonons (unlike $\phi\leftrightarrow\phi\phi$).\footnote{$N_\phi$ should then be added to the list of constants of motion, alongside $E$ and $N$, in note \ref{notee1_38}.}\ \footnote{The fact that, for $\gamma<0$, the phonon damping rate $\Gamma_\qq$ tends towards zero as $q^3$ (instead of $q$ for $\gamma>0$) also plays a role; however, without the conservation of $N_\phi$, it would lead to a marginally superdiffusive $t\,\ln t$ spreading law (see Equation (C.20) of reference \cite{diffpferm} and the morality stated below its Equation (72)).} To go beyond this and obtain the true (a priori finite) value of $D_\lambda$ remains an open question: one would have to take into account the subdominant five-phonon processes $\phi\phi\leftrightarrow\phi\phi\phi$ which change $N_\phi$ and occur at a rate $\approx T^9$ \cite{Khalat}, of the same order of magnitude as that of large-angle $\phi\phi\to\phi\phi$ processes \cite{LK,Annalen}, which is not easy.\footnote{Publication \cite{diffpferm}, misunderstanding reference \cite{Khalat}, had seen there a five-phonon damping rate scaling as $T^{11}$. The error has been corrected here. Indeed, reference \cite{Khalat}, considering a quasi-thermal equilibrium with a small non-zero phonon chemical potential $\mu_\phi\to 0^-$, obtains the evolution equation $L^{-3}\dd N_\phi/\dd t=-\Gamma_\phi\mu_\phi$ for the average phonon number, where $\Gamma_\phi\approx T^{11}$ is not the rate sought despite appearances; as $L^{-3}\dd N_\phi/\dd t\approx T^2\dd\mu_\phi/\dd t$ for Bose's law $\bar{n}_\qq=1/\{\exp[(\hbar c q -\mu_\phi)/k_{\rm B}T]-1\}$, we actually have $-\dd\mu_\phi/\dd t\propto (\Gamma_\phi/T^2)\mu_\phi$, of rate $\approx T^9$.}

\section{Open questions requiring a microscopic theory of the many-body problem}
\label{sece5}

The quantum hydrodynamics of Section \ref{sece4} is only a low-energy effective theory. It therefore has limitations of two kinds, which we briefly review here, and which make it impossible to dispense with a many-body microscopic calculation. 

\subsection{Determining the ingredients of quantum hydrodynamics}
\label{sece5.1}

Quantum hydrodynamics involves two quantities that are external to it, the equation of state of the unpolarized Fermi gas at zero temperature (through the energy density $e_0(\rho)$ or the chemical potential $\mu_0(\rho)$ - its derivative - at density $\rho$) and the curvature parameter $\gamma$ of the acoustic branch (\ref{eqe1_24}). 

In the present case of equal mass $m_\uparrow=m_\downarrow=m$, the equation of state has been measured experimentally \cite{Salomon,mit} and various approximate calculation methods give satisfactory results, such as fixed-node diffusive quantum Monte Carlo \cite{Giorgini,Gezerlis} or the Gaussian fluctuations approximation in a path integral formulation \cite{Randeria,Drummond}. 

The situation is much more open for the curvature parameter $\gamma$. The Anderson random phase approximation (RPA) \cite{Anderson}, equivalent for this problem to the eigenfrequency calculation of linearized time-dependent BCS equations or even to the more powerful Gaussian fluctuations approximation \cite{Strinati,Randeria},\footnote{These different approaches lead to exactly the same implicit equation linking angular eigenfrequency $\omega_\qq$, chemical potential $\mu$ and order parameter $\Delta$, and exactly the same equation linking $\mu, \Delta$ and $s$-wave scattering length $a$ \cite{PRAconcav}; they differ only in the equation of state $\mu=\mu_0(\rho)$ linking $\mu$ to $\rho$ in the ground state, the one of Gaussian fluctuations being the most accurate. For example, at the unitary limit $a^{-1}=0$, the approaches all give $mc^2/\mu=2/3$ (this is exact by scale invariance, $\mu_0(\rho)\propto \rho^{2/3}$ in Equation (\ref{eqe2_30})), $|\Delta|/\mu\simeq 1.16$ (close to the experimental value $0.44 E_{\rm F}/0.376 E_{\rm F}\simeq 1.17$ knowing that $|\Delta|=E_{\rm pair}/2$ in these theories and that $E_{\rm pair}/2E_{\rm F}\simeq 0.44$ in the experiment \cite{Ketterle}) but the ratio $\mu/E_{\rm F}\simeq 0.376$ in the experiment \cite{mit}, very poorly reproduced $\simeq 0.59$ by RPA and BCS, is much better $\simeq 0.40$ in Gaussian fluctuations.} leads to a fairly simple analytical expression of $\gamma$ in terms of $\mu/|\Delta|$ and $(\partial\mu/\partial|\Delta|)_a$, exact in the limit $k_{\rm F}a\to 0^+$ of a condensate of dimers ($\gamma\to 1$), reasonable in the BCS limit $k_{\rm F}a\to 0^-$ ($\gamma\to -\infty$ exponentially with $1/k_{\rm F}|a|$ since the acoustic branch is crushed by the broken-pair continuum) and changing sign for $|\Delta|/\mu\simeq 0.87$, i.e.\ $1/k_{\rm F}a\simeq -0.14$ for the rather approximate BCS equation of state, not far from the unitary limit in any case \cite{PRAconcav}.

In particular, $\gamma$ has the same positive value at the unitary limit in all three approaches (convex acoustic branch at low $q$): 
\begin{equation}\label{eqe1_41}
\gamma_{a^{-1}=0}^{\rm RPA}\simeq 0.084
\end{equation} The error, however, is uncontrolled, and we're not even sure of the sign. 

A completely different method proceeds by extending the problem to an arbitrary spatial dimension $d$ and expanding around dimension four, in powers of the small parameter $\epsilon=4-d$. At the unitary limit, it too predicts a convex branch at low $q$ \cite{Rupak}:\footnote{We obtained expression (\ref{eqe2_41}) by directly inserting Equation (50) of reference \cite{Rupak} into the dispersion relation (48) of the same reference and using the exact property $mc^2=2\mu/3$ due to scale invariance. Proceeding differently, i.e.\ via its Equation (52) and its result $c_2/c_1=O(\epsilon^2)\simeq 0$ with $d=3$ in its Equation (48), we find the fairly close value $\gamma_{a^{-1}=0}^{\rm dimension}=8/45\simeq 0.18$.} 
\begin{equation}\label{eqe2_41}
\gamma_{a^{-1}=0}^{\rm dimension} = \frac{1}{3}\left[1-\frac{1}{4}\epsilon+O(\epsilon^2)\right]\underset{d=3}{\stackrel{\epsilon=1}{\simeq}}\frac{1}{4}>0
\end{equation}

Experimentally, a recent measurement of the acoustic branch by Bragg excitation in a cold-atom Fermi gas leads, on the contrary, to a concave branch at the unitary limit \cite{Moritz}:
\begin{equation}\label{eqe1_42}
\gamma_{a^{-1}=0}^{\rm exp}=\frac{8\mu}{3E_{\rm F}}\zeta\simeq \zeta\quad\mbox{with}\quad\zeta=-0.085(8)<0
\end{equation} where the ratio $\mu/E_{\rm F}$ is $\simeq 3/8$ for the ground-state unitary gas \cite{mit} and $\zeta$ is the acoustic-branch curvature parameter for a rescaling of $q$ by $k_{\rm F}$, $\omega_\qq=cq(1+\zeta q^2/k_{\rm F}^2+\ldots)$. The result (\ref{eqe1_42}) suffers from two limitations \cite{comment}: (i) a cubic fit of the branch over an interval of rather high $q$ values, $q/k_{\rm F}\in[0.29,1.63]$, rather than over a narrow neighborhood of $q=0$ (for the RPA dispersion relation, for example, which has an inflection point at $q\simeq 0.5 k_{\rm F}$, such a fit, blindly mixing convex and concave parts, would not give the right sign of $\gamma_{a^{-1}=0}^{\rm RPA}$), and (ii) a relatively high temperature, $T=0.128(8)T_{\rm F}\simeq 0.8\!T_{\rm c}$: even if we start from the RPA branch of parameter $\gamma>0$ in the ground state, quantum hydrodynamics predicts a thermal change $\delta\gamma_{\rm th}$ in curvature (by interaction of the mode $\qq$ with thermal phonons) negative enough to change its sign: 
\begin{equation}\label{eqe2_42}
\delta\gamma_{\rm th}\sim -\frac{8\pi^2}{9(3\mu/E_{\rm F})^{1/2}}\left(\frac{T}{T_{\rm F}}\right)^2\simeq -0.14<-\gamma_{a^{-1}=0}^{\rm RPA}
\end{equation}
The question of the sign of $\gamma$ at the unitary limit, which crucially determines the three-phonon ($\gamma>0$) or four-phonon ($\gamma<0$) nature of sound damping in the low-temperature collisionless regime, therefore remains largely open.\footnote{The damping studied experimentally in reference \cite{Zwierlein_visco} is in the hydrodynamic regime, in the sense of note \ref{notee2_33}. This reference therefore does not allow us to resolve the problem.}

\subsection{Describing high-frequency modes}
\label{sece5.2}

Quantum hydrodynamics, with its almost linear acoustic branch, cannot reliably describe the sound waves of angular frequency $\omega_\qq>mc^2/\hbar$ in the Fermi superfluid. Ignoring the composite nature of bound pairs $\uparrow\downarrow$, it is totally inapplicable to angular frequencies $\omega\approx E_{\rm pair}/\hbar$, where $E_{\rm pair}$ is the binding energy of a pair: at these frequencies, pairs can break into two fermionic excitations $\chi$ (conservation of energy no longer prohibits this), see Figure \ref{fige4}a.

A microscopic description of the Fermi gas is then required. At zero temperature, the main method available is that of the time-dependent BCS variational theory \cite{BlaizotRipka}. Its specialization to the linear response regime yields the following eigenvalue equation for the energy $z$ of the modes of wave vector $\qq$: 
\begin{equation}\label{eqe1_43}
\mathrm{det}\, M(\qq,z)=0\quad\mbox{with}\quad M(\qq,z)=\begin{pmatrix} M_{|\Delta||\Delta|}(\qq,z) & M_{|\Delta|\theta}(\qq,z)\\ M_{\theta|\Delta|}(\qq,z) & M_{\theta\theta}(\qq,z)\end{pmatrix}
\end{equation} where the coefficients of the $2\times 2$ matrix correspond to a response in the modulus $|\Delta|$ or in the phase $\theta$ of the complex order parameter $\Delta(\rr,t)$. In the weakly-interacting BCS limit $k_{\rm F}a\to 0^-$, off-diagonal elements are usually (rightly) neglected and the dynamics decouples into modulus and phase modes; in the general case, this distinction no longer applies. 

The exploration of solutions to Equation (\ref{eqe1_43}) has begun. At a fixed wavenumber $q$, we find at most one root under the edge $\veps_\qq^{\rm bord}$ of the broken-pair continuum, the root $\hbar\omega_\qq$ of the acoustic branch. On the interval $z\in\,]\veps_\qq^{\rm bord},+\infty[$, the function $\mathrm{det}\, M(\qq,z)$ has a branch cut,\footnote{The matrix elements of $M(\qq,z)$ involve an integral over the wave vector $\kk$ of one of the dissociation fragments of a bound pair of the condensate, and the integrand contains the corresponding energy denominator $z-(\veps_\kk+\veps_{\qq-\kk})$; by definition, the denominator vanishes when $z$ belongs to the broken-pair continuum, see the legend to Figure \ref{fige4}.} it is necessary to add an infinitesimal shift $\ii 0^+$ to $z$ to make sense of it; the function then acquires an imaginary part, which cannot vanish simultaneously with the real part, and Equation (\ref{eqe1_43}) has no solution. On the other hand, a complex $z_\qq$ with a non-infinitesimal $<0$ imaginary part can be found, by performing an analytic continuation of function $z\mapsto\mathrm{det}\, M(\qq,z)$ from the upper half-plane to the lower half-plane through its branch cut (indicated by arrow $\downarrow$ in the subscript): 
\begin{equation}\label{eqe1_44}
\mathrm{det}M_{\downarrow}(\qq,z_\qq)=0\quad\mbox{with}\quad \im z_\qq<0
\end{equation} There is therefore a collective mode in the continuum, which decays exponentially in time through the emission of broken pairs. The calculation was first performed in the BCS limit $k_{\rm F}a\to 0^-$, both for neutral fermions and for electrons in a superconductor, in reference \cite{AndrianovPopov}. It was then generalized to fermionic cold-atom gases for arbitrary values of $k_{\rm F}a$, no longer neglecting the off-diagonal elements $M_{|\Delta|\theta}$ and $M_{\theta|\Delta|}$ \cite{prlhiggs,crashiggs}. The Andrianov-Popov branch persists up to $1/k_{\rm F}a=0.55$ (point of zero chemical potential $\mu=0$ in BCS theory) and always starts at $2|\Delta|$ quadratically in $q$ with a complex coefficient: 
\begin{equation}\label{eqe1_45}
z_\qq\stackrel{\mu>0}{\underset{q\to 0}{=}}2|\Delta|+\zeta\frac{\hbar^2q^2}{4 m_*}+O(q^3) \quad (\im\zeta<0)
\end{equation} where $m_*$ is the effective mass of a fermionic quasiparticle $\chi$ at the location $k=k_0$ of its energy minimum.\footnote{The effective mass is such that $\veps_\kk-E_{\rm pair}/2\sim\hbar^2(k-k_0)^2/2m_*$ as $k\to k_0$. Scaling by $m_*$ in Equation (\ref{eqe1_45}) ensures that $\zeta$ has a finite, non-zero limit when $k_{\rm F}a\to 0^-$ \cite{AndrianovPopov}. In this regime, the well-known reference \cite{LV} predicts an incorrect behavior for $z_\qq$ at low $q$, with an imaginary part tending linearly to zero $\propto q$, see its Equation (2.38). The quantity $\zeta$ here has nothing to do with that in Equation (\ref{eqe1_42}), there is an unfortunate coincidence of notations.} We have written $2|\Delta|$ here rather than $E_{\rm pair}$, where $\Delta$ is the equilibrium order parameter, even though BCS theory is unable to distinguish (we have exactly $E_{\rm pair}=2|\Delta|$ for all $\mu>0$ in this theory), in order to evoke the Higgs mechanism \cite{Higgs} which we think the collective mode of the continuum comes under \cite{Varma};\footnote{As this mechanism results from $U(1)$ symmetry breaking, here by condensation of bound pairs, it must be characterized by the energy scale associated with the order parameter, i.e.\ $|\Delta|$ up to a factor; this is indeed what reference \cite{Higgs} finds, see its Equation (2b). On the other hand, the energy scale $E_{\rm pair}$ is related to pair breaking, not pair condensation, and therefore has no a priori connection with the Higgs branch. Having $E_{\rm pair}=2|\Delta|$ is a source of confusion and prevents the two phenomena from being decoupled. It would also be interesting to see whether property $E_{\rm pair}=2|\Delta|$ remains rigorously true at zero temperature in a theory more elaborate than BCS or in experiments.} moreover, in the opposite limit $k_{\rm F}a\to 0^+$ of a bosonic dimer condensate, where $2|\Delta|\ll E_{\rm pair}\sim 2|\mu|\sim\hbar^2/ma^2$ (this time we have $\mu<0$), we do indeed find a collective excitation branch starting quadratically at $2|\Delta|$ and not at $E_{\rm pair}$ \cite{crashiggs}. The extension of Equation (\ref{eqe1_43}) to non-zero temperature (beyond a simple BCS-type mean-field generalization, perhaps insufficient\footnote{This generalization is not a panacea, as can be seen on the imaginary part of the acoustic branch. For $\gamma>0$ (but not for $\gamma<0$), this criticism can be leveled at RPA already at zero temperature, since it wrongly predicts a purely real angular eigenfrequency $\omega_\qq$. However, this doesn't seem too serious, since the imaginary part $(-1/2)\Gamma_\qq(T=0)\approx q^5$ obtained by quantum hydrodynamics (Beliaev damping) gets lost in the neglected sub-subdominant terms in Equation (\ref{eqe1_24}). This problem is most visible at low $q$ at non-zero temperature, where $\Gamma_\qq(T>0)$ starts linearly in $q$ with a coefficient $\propto T^4$ (exponent $\nu$ is 5 in Table \ref{tablee1} for the scaling law $q\propto T$ of note \ref{notee2_33}), which the BCS-type mean-field theory cannot account for (it predicts a coefficient $O[-\exp(E_{\rm pair}/2k_{\rm B}T)]$ \cite{Kulik,Tempere} since the only thermal occupation numbers it brings out are those $\bar{n}_\kk=1/[\exp(\veps_\kk/k_{\rm B}T)+1]$ of fermionic quasiparticles $\chi$). In other words, the linearization of the time-dependent BCS equations or, what amounts to the same thing, the Gaussian fluctuations approximation takes into account the $\phi-\chi$ coupling but not the $\phi-\phi$ coupling.}) remains to our knowledge an open question. 

From an experimental point of view, in cold atoms or superconductors, excitation at angular frequencies $\omega>E_{\rm pair}/\hbar$ has only been carried out at zero wavenumber, where there is, according to zero temperature theories, no collective mode in the continuum, the spectral weight of the mode tending to 0 as $q\to 0$ \cite{prlhiggs}; at long times, we simply observe oscillations of the order parameter at angular frequency $E_{\rm pair}/\hbar$ (this is the effect of the non-zero edge of the continuum), which attenuate with a power law $t^{-\alpha}$ \cite{Shimano,ValePRL} by the same mechanism as the spreading of the Gaussian wave packet of a free particle in ordinary quantum mechanics (percussive excitation creates a “wave packet” of broken pairs $(\kk,-\kk)$ in the continuum, whose evolution governed by the dispersion relation $2\veps_\kk$ is effectively one-dimensional for $k_0>0$ ($\mu>0$), in which case $\alpha=1/2$ \cite{Volkov}, and three-dimensional for $k_0=0$ ($\mu<0$), in which case $\alpha=3/2$ \cite{Gurarie}).\footnote{The observations of reference \cite{ValePRL} in a unitary Fermi gas at $T\neq 0$ raise several questions: (i) the measured exponent $\alpha\simeq 1\pm 0.15$ is very different from the theoretically predicted value ($\alpha=1/2$ at the unitary limit), (ii) it cannot be ruled out that the decay of the oscillation amplitude is in fact exponential, (iii) in contrast to the amplitude, the angular frequency of the oscillations shows no observable reduction as $T$ approaches the transition temperature $T_{\rm c}$ (where there is no more $U(1)$ symmetry breaking and $|\Delta|$ tends to zero), which seems incompatible with the qualification of Higgs oscillations used in this reference (the measured angular frequency is not proportional to $|\Delta|/\hbar$), but also suggests a rather troubling lack of dependence of $E_{\rm pair}$ with temperature (the measured angular frequency should be given by $E_{\rm pair}/\hbar$ since the excitation is done at $q=0$). Incidentally, the temperature is never very low in the experiment, $T\gtrsim 0.1T_{\rm F}$, see our Section \ref{sece2}, which makes the theory at $T=0$ stricto sensu inapplicable.} The observation of the continuum mode (at $q>0$) and the precise measurement of its dispersion relation $z_\qq$ therefore remain to be done (hints are given in references \cite{prlhiggs,Bragg}). 
\selectlanguage{french}

\bibliographystyle{crunsrt}

\nocite{*}


\end{document}